\documentclass[12pt]{revtex4}
\usepackage{color}
\usepackage{graphicx}
      \def\di{\displaystyle}
      
      \def\bS{{\bf S}}
      \def\bl{{\bf l}}
      \def\bp{{\bf p}}
      \def\bq{{\bf q}}
      \def\br{{\bf r}}
      \def\bs{{\bf s}}

      \def\E{{\cal E}}
      \def\F{{\cal F}}
      \def\H{{\cal H}}

      \def\L{{\cal L}}
      
      \def\P{{\cal P}}
      \def\R{{\cal R}}

      \def\u{{{\uparrow\downarrow}}}
      \def\d{{{\downarrow\uparrow}}}
\begin{document}

\title{Nuclear scissors modes and hidden angular momenta}
\author{\firstname{E.~B.}~\surname{Balbutsev}}
\email{balbuts@theor.jinr.ru}
\author{\firstname{I.~V.}~\surname{Molodtsova}}
\affiliation{Joint Institute for Nuclear Research, 141980 Dubna, Moscow Region, Russia}
\author{\firstname{P.}~\surname{Schuck}}
\affiliation{
Institut de Physique Nucl\'eaire, IN2P3-CNRS, Universit\'e Paris-Sud,
F-91406 Orsay C\'edex, France;\\
Laboratoire de Physique et Mod\'elisation des Milieux Condens\'es,
CNRS and Universit\'e Joseph Fourier,
25 avenue des Martyrs BP166, F-38042 Grenoble C\'edex 9, France
}

\begin{abstract}
The coupled dynamics of low lying modes and various giant resonances 
are studied with the help of the Wigner Function Moments method 
generalized to take into account spin degrees of freedom and pair correlations simultaneously. The method is based on Time Dependent Hartree-Fock-Bogoliubov equations. The model of the harmonic oscillator including 
spin-orbit potential plus quadrupole-quadrupole and spin-spin interactions 
is considered. New low lying spin dependent modes are analyzed. Special 
attention is paid to the scissors modes.
A new source of nuclear magnetism, connected with
counter-rotation of spins up and down around the symmetry axis
(hidden angular momenta), 
is discovered. Its inclusion into
the theory allows one to improve substantially the agreement with experimental data in the 
description of energies and transition probabilities of scissors modes.
\end{abstract}


\maketitle

\section{Introduction}

The idea of the possible existence of the collective motion in deformed nuclei
similar to the scissors motion continues to attract the attention of physicists who extend it to various 
kinds of objects, not necessarily nuclei, (for example, magnetic traps, see the review 
by Heyde at al~\cite{Heyd}) and invent new sorts of scissors, 
for example, the rotational oscillations of neutron skin against a proton-neutron core~\cite{Pena}.

The nuclear scissors mode was predicted~\cite{Hilt}--\cite{Lo} 
as a counter-rotation of protons against neutrons in deformed nuclei.  
However, its collectivity turned out to be small. From RPA results which
 were in qualitative agreement with experiment, it was even questioned 
 whether this mode is collective at all~\cite{Zaw,Sushkov}. 
Purely phenomenological models (such as, e.g., 
the two rotors model~\cite{Lo2000} and the sum rule approach~\cite{Lipp})
did not clear up the situation in this 
respect. Finally in a very recent review~\cite{Heyd} it is concluded 
that the scissors mode is "weakly collective, but strong 
on the single-particle scale" and further: "The weakly
collective scissors mode excitation has become an ideal test of models
-- especially microscopic models -- of nuclear vibrations. Most models
are usually calibrated to reproduce properties of strongly collective
excitations (e.g. of $J^{\pi}=2^+$ or $3^-$ states, giant resonances,
...). Weakly-collective phenomena, however, force the models to make
genuine predictions and the fact that the transitions in question are
strong on the single-particle scale makes it impossible to dismiss
failures as a mere detail, especially in the light of the overwhelming
experimental evidence for them in many nuclei~\cite{Kneis,Richt}."

The Wigner Function Moments (WFM) or phase space moments method turns out to be very 
useful in this situation. On the one hand it is a purely microscopic method, because
it is based on the Time Dependent Hartree-Fock (TDHF) equation. On the
other hand the method works with average values (moments) of operators
which have a direct relation to the considered phenomenon and, thus, make a 
natural bridge with the macroscopic description. This 
makes it an ideal instrument to describe the basic characteristics 
(energies and excitation probabilities) of collective excitations such as,
in particular, the scissors mode. Our investigations have shown that 
already the minimal set of collective variables, i.e. phase space 
moments up to quadratic order,
is sufficient to reproduce the most important property of the
scissors mode: its inevitable coexistence with the IsoVector Giant
Quadrupole Resonance (IVGQR) implying a deformation of the Fermi surface.

Further developments of the WFM
method, namely, the switch from \mbox{TDHF}
to Time Dependent Hartree-Fock Bogoliubov \mbox{(TDHFB)} equations, i.e. taking 
into account pair correlations, allowed
us to improve considerably the quantitative description of the 
scissors mode~\cite{Malov,Urban}: for rare earth nuclei the energies were reproduced with
$\sim 10\%$ accuracy and $B(M1)$ values were reduced by about a factor of two  
with respect to their non superfluid values. 
However, they remain about two times too high with respect to experiment.
We have suspected, that the reason of this last discrepancy is hidden in the spin 
degrees of freedom, which were so far ignored by the WFM method. One cannot exclude,
that due to spin dependent interactions some part of the 
force of M1 transitions is shifted to the energy region of 5-10~MeV,
where a 1$^+$ resonance of spin nature is observed~\cite{Zaw}. 

\begin{figure}
\centering\includegraphics[width=0.5\textwidth]{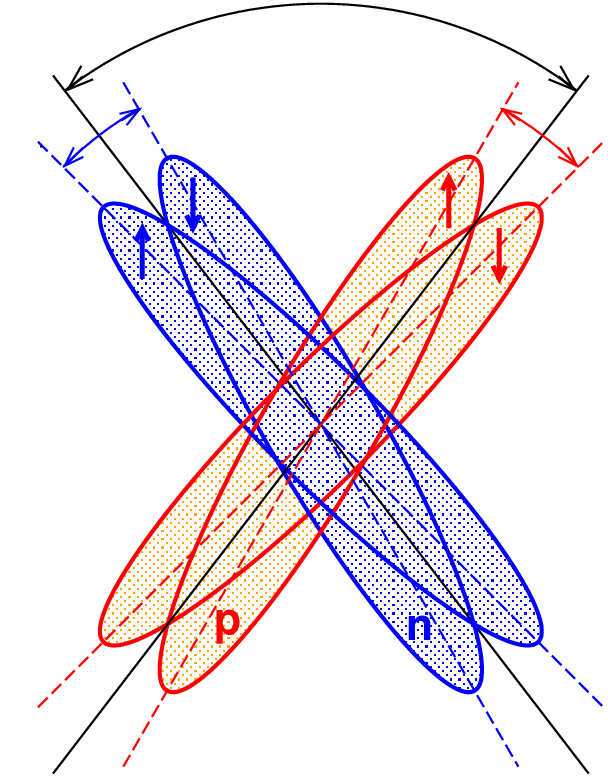}
\caption{Pictorial representation of two intermingled scissors: the orbital (neutrons 
versus protons) scissors + spin (spin-up nucleons versus spin-down nucleons)
scissors. Arrows inside of ellipses show the direction of spin projections.
 {\bf p} - protons, {\bf n} - neutrons.}
\label{fig0}\end{figure}

  In a recent paper~\cite{BaMo} the WFM method was 
applied for the first time to solve the TDHF equations including spin
dynamics.
As a first step, only the spin-orbit interaction was included in the
consideration, as the most important
one among all possible spin dependent interactions because it enters 
into the mean field. This allows one to understand the structure
of necessary modifications of the method avoiding  cumbersome 
calculations.
 The most remarkable result was the discovery of a new type
of nuclear collective motion: rotational oscillations of "spin-up"
nucleons with respect of "spin-down" nucleons (the spin scissors mode). 
It turns out that the experimentally 
observed group of peaks in the energy interval 2-4~MeV corresponds 
very likely to
two different types of motion: the conventional (orbital) scissors mode and this new kind 
of mode, i.e. the spin scissors mode.
 The pictorial view of these two intermingled scissors
is shown on Fig.~\ref{fig0}, which is just the modification (or generalization) of the 
classical picture for 
the orbital scissors (see, for example,~\cite{Lo2000,Heyd}).

Three low lying excitations of a new nature were found: isovector and 
isoscalar spin scissors and the excitation generated by the relative 
motion of the orbital angular momentum and the spin of the nucleus (they 
can change their absolute values and directions keeping the total spin 
unchanged). 
In the frame of the same approach ten high lying excitations were also 
obtained: well known isoscalar and isovector Giant Quadrupole Resonances (GQR), 
two resonances of
a new nature describing isoscalar and isovector quadrupole vibrations
of "spin-up" nucleons with respect of "spin-down" nucleons, and six
resonances which can be interpreted as spin flip modes of various 
kinds and multipolarity. 

The next step was done in the paper~\cite{BaMoPRC}, where the influence of the spin-spin
interaction on the scissors modes was studied. There was hope
that, due to spin dependent interactions, some part of the 
force of $M1$ transitions will be shifted to the area of a
spin-flip resonance,
decreasing in such a way the $M1$ force of scissors. However, these expectations were not realised. It turned out that the spin-spin interaction does not change the general
picture of the positions of excitations described in~\cite{BaMo} pushing all levels
up proportionally to its strength without changing their order. The most interesting
result concerns the $B(M1)$ values of both scissors
-- the spin-spin interaction strongly redistributes $M1$ strength in  favour
of the spin scissors mode practically without changing their summed strength.
One of the main points of this work was, indeed, that we were
able to give a tentative explanation of a recent experimental finding 
\cite{Siem} where the $B(M1)$ values in $^{232}$Th of the two low lying 
magnetic states are inverted in strength in favour of the lowest, i.e.,
the spin scissors mode, when cranking up the spin-spin interaction. 
Indeed, the explanation with respect to a triaxial deformation given in~\cite{Siem} 
yields a stronger $B(M1)$ value for the higher lying state, contrary to 
observation, as remarked by the authors themselves.

In the last work~\cite{BaMoPRC15} we suggested a generalization of the WFM method which takes into account
spin degrees of freedom and pair correlations simultaneously. These two factors, working together, improve considerably the agreement 
between the theory and experiment in the description of nuclear scissors modes.

The paper is organized as follows.
 In Sec. 2 the TDHFB equations for the $2\times 2$ normal and anomalous density matrices are
formulated and their Wigner transform is found.
In Sec. 3 the model Hamiltonian and the mean field are analyzed.
In Sec. 4 the collective variables are defined and the respective 
dynamical equations are derived.
In Sec. 5 the choice of parameters and the results of calculations of energies and $B(M1)$
values of two scissors modes are discussed.
The phenomenon of counter-rotating angular momenta with spin up/down,
which can be considered also as a phenomenon of hidden angular momenta, is analysed in Sec. 6.
Results of calculations for 26 nuclei in the rare earth region 
with the hidden angular 
momenta taken into account 
are discussed in Sec. 7.
The summary of main results is given in the conclusion section. 
The mathematical details are concentrated in Appendices~\ref{AppA},~\ref{AppB},~\ref{AppC},~\ref{AppD},~\ref{AppE}.

\section{WIGNER TRANSFORMATION OF TIME-DEPENDENT HARTREE-FOCK-BOGOLIUBOV EQUATIONS}

\hspace{5mm} The Time-Dependent Hartree--Fock--Bogoliubov (TDHFB)
equations in matrix formulation are
\cite{Solov,Ring}
\begin{equation}
i\hbar\dot\R=[\H,\R]
\label{tHFB}
\end{equation}
with
\begin{equation}
\R={\hat\rho\qquad-\hat\kappa\choose-\hat\kappa^{\dagger}\;\;1-\hat\rho^*},
\quad\H={\hat
h\quad\;\;\hat\Delta\choose\hat\Delta^{\dagger}\quad-\hat h^*}
\end{equation}
The normal density matrix $\hat \rho$ and Hamiltonian $\hat h$ are
hermitian whereas the abnormal density $\hat\kappa$ and the pairing
gap $\hat\Delta$ are skew symmetric: $\hat\kappa^{\dagger}=-\hat\kappa^*$, 
$\hat\Delta^{\dagger}=-\hat\Delta^*$.

The detailed form of the TDHFB equations is
\begin{eqnarray}
&& i\hbar\dot{\hat\rho} =\hat h\hat\rho -\hat\rho\hat h
-\hat\Delta \hat\kappa ^{\dagger}+\hat\kappa \hat\Delta^\dagger,
\nonumber\\
&&-i\hbar\dot{\hat\rho}^*=\hat h^*\hat\rho ^*-\hat\rho ^*\hat h^*
-\hat\Delta^\dagger\hat\kappa +\hat\kappa^\dagger\hat\Delta ,
\nonumber\\
&&-i\hbar\dot{\hat\kappa} =-\hat h\hat\kappa -\hat\kappa \hat h^*+\hat\Delta
-\hat\Delta \hat\rho ^*-\hat\rho \hat\Delta ,
\nonumber\\
&&-i\hbar\dot{\hat\kappa}^\dagger=\hat h^*\hat\kappa^\dagger
+\hat\kappa^\dagger\hat h-\hat\Delta^\dagger
+\hat\Delta^\dagger\hat\rho +\hat\rho^*\hat\Delta^\dagger .
\label{HFB}
\end{eqnarray}
It is easy to see that the second and fourth equations are complex 
conjugate to the first and third ones respectively.
Let us consider their matrix form in coordinate 
space keeping all spin indices $s, s', s''$~\cite{BaMoPRC15}:
\begin{eqnarray}
i\hbar\langle \br,s|\dot{\hat\rho}|\br'',s''\rangle  =&&
\sum_{s'}\int\!d^3r'\left(
\langle \br,s|\hat h|\br',s'\rangle \langle \br',s'|\hat\rho|\br'',s''\rangle  
-\langle \br,s|\hat\rho|\br',s'\rangle \langle \br',s'|\hat h|\br'',s''\rangle \right.
\nonumber\\
&&\left.
-\langle \br,s|\hat\Delta|\br',s'\rangle \langle \br',s'|\hat\kappa^{\dagger}|\br'',s''\rangle  
+\langle \br,s|\hat\kappa|\br',s'\rangle \langle \br',s'|\hat\Delta^{\dagger}|\br'',s''\rangle 
\right),
\nonumber\\
i\hbar\langle \br,s|\dot{\hat\kappa}|\br'',s''\rangle  =&& -\langle \br,s|\hat\Delta|\br'',s''\rangle 
+\sum_{s'}\int\!d^3r'\left(
\langle \br,s|\hat h|\br',s'\rangle \langle \br',s'|\hat\kappa|\br'',s''\rangle  \right.
\nonumber\\
&&\left.
+\langle \br,s|\hat\kappa|\br',s'\rangle \langle \br',s'|\hat h^*|\br'',s''\rangle \right.
\nonumber\\
&&\left.
+\langle \br,s|\hat\Delta|\br',s'\rangle \langle \br',s'|\hat\rho^*|\br'',s''\rangle  
+\langle \br,s|\hat\rho|\br',s'\rangle \langle \br',s'|\hat\Delta|\br'',s''\rangle 
\right),
\nonumber\\
i\hbar\langle \br,s|\dot{\hat\rho}^*|\br'',s''\rangle  =&&
\sum_{s'}\int\!d^3r'\left(
 \langle \br,s|\hat\rho^*|\br',s'\rangle \langle \br',s'|\hat h^*|\br'',s''\rangle 
-\langle \br,s|\hat h^*|\br',s'\rangle \langle \br',s'|\hat\rho^*|\br'',s''\rangle \right. 
\nonumber\\
&&\left.
+\langle \br,s|\hat\Delta^{\dagger}|\br',s'\rangle \langle \br',s'|\hat\kappa|\br'',s''\rangle  
-\langle \br,s|\hat\kappa^{\dagger}|\br',s'\rangle \langle \br',s'|\hat\Delta|\br'',s''\rangle 
\right),
\nonumber\\
i\hbar\langle \br,s|\dot{\hat\kappa}^{\dagger}|\br'',s''\rangle  =&& \langle \br,s|\hat\Delta^{\dagger}|\br'',s''\rangle 
-\sum_{s'}\int\!d^3r'\left(
\langle \br,s|\hat h^*|\br',s'\rangle \langle \br',s'|\hat\kappa^{\dagger}|\br'',s''\rangle \right. 
\nonumber\\
&&\left.
+\langle \br,s|\hat\kappa^{\dagger}|\br',s'\rangle \langle \br',s'|\hat h|\br'',s''\rangle \right.
\nonumber\\
&&\left.
+\langle \br,s|\hat\Delta^{\dagger}|\br',s'\rangle \langle \br',s'|\hat\rho|\br'',s''\rangle  
+\langle \br,s|\hat\rho^*|\br',s'\rangle \langle \br',s'|\hat\Delta^{\dagger}|\br'',s''\rangle 
\right).
\label{HFmatr}
\end{eqnarray}
 We do not specify the isospin indices in order to make
formulae more transparent. They will be re-introduced at the end. 
Let us introduce the more compact notation
$\langle \br,s|\hat X|\br',s'\rangle =X_{rr'}^{ss'}$. Then
the set of TDHFB equations (\ref{HFmatr}) with specified spin indices reads
\begin{eqnarray}
&&i\hbar\dot{\rho}_{rr''}^{\uparrow\uparrow} =
\int\!d^3r'\left(
 h_{rr'}^{\uparrow\uparrow}\rho_{r'r''}^{\uparrow\uparrow} 
-\rho_{rr'}^{\uparrow\uparrow} h_{r'r''}^{\uparrow\uparrow}
+\hat h_{rr'}^{\uparrow\downarrow}\rho_{r'r''}^{\downarrow\uparrow} 
-\rho_{rr'}^{\uparrow\downarrow} h_{r'r''}^{\downarrow\uparrow}
-\Delta_{rr'}^{\uparrow\downarrow}{\kappa^{\dagger}}_{r'r''}^{\downarrow\uparrow}
+\kappa_{rr'}^{\uparrow\downarrow}{\Delta^{\dagger}}_{r'r''}^{\downarrow\uparrow}\right),
\nonumber\\
&&i\hbar\dot{\rho}_{rr''}^{\uparrow\downarrow} =
\int\!d^3r'\left(
 h_{rr'}^{\uparrow\uparrow}\rho_{r'r''}^{\uparrow\downarrow} 
-\rho_{rr'}^{\uparrow\uparrow} h_{r'r''}^{\uparrow\downarrow}
+\hat h_{rr'}^{\uparrow\downarrow}\rho_{r'r''}^{\downarrow\downarrow} 
-\rho_{rr'}^{\uparrow\downarrow} h_{r'r''}^{\downarrow\downarrow}\right),
\nonumber\\
&&i\hbar\dot{\rho}_{rr''}^{\downarrow\uparrow} =
\int\!d^3r'\left(
 h_{rr'}^{\downarrow\uparrow}\rho_{r'r''}^{\uparrow\uparrow} 
-\rho_{rr'}^{\downarrow\uparrow} h_{r'r''}^{\uparrow\uparrow}
+\hat h_{rr'}^{\downarrow\downarrow}\rho_{r'r''}^{\downarrow\uparrow} 
-\rho_{rr'}^{\downarrow\downarrow} h_{r'r''}^{\downarrow\uparrow}\right),
\nonumber\\
&&i\hbar\dot{\rho}_{rr''}^{\downarrow\downarrow} =
\int\!d^3r'\left(
 h_{rr'}^{\downarrow\uparrow}\rho_{r'r''}^{\uparrow\downarrow} 
-\rho_{rr'}^{\downarrow\uparrow} h_{r'r''}^{\uparrow\downarrow}
+\hat h_{rr'}^{\downarrow\downarrow}\rho_{r'r''}^{\downarrow\downarrow} 
-\rho_{rr'}^{\downarrow\downarrow} h_{r'r''}^{\downarrow\downarrow}
-\Delta_{rr'}^{\downarrow\uparrow}{\kappa^{\dagger}}_{r'r''}^{\uparrow\downarrow}
+\kappa_{rr'}^{\downarrow\uparrow}{\Delta^{\dagger}}_{r'r''}^{\uparrow\downarrow}\right),
\nonumber\\
&&i\hbar\dot{\kappa}_{rr''}^{\uparrow\downarrow} = -\hat\Delta_{rr''}^{\uparrow\downarrow}
+\int\!d^3r'\left(
 h_{rr'}^{\uparrow\uparrow}\kappa_{r'r''}^{\uparrow\downarrow} 
+\kappa_{rr'}^{\uparrow\downarrow} {h^*}_{r'r''}^{\downarrow\downarrow}
+\Delta_{rr'}^{\uparrow\downarrow}{\rho^*}_{r'r''}^{\downarrow\downarrow} 
+\rho_{rr'}^{\uparrow\uparrow}\Delta_{r'r''}^{\uparrow\downarrow}
\right),
\nonumber\\
&&i\hbar\dot{\kappa}_{rr''}^{\downarrow\uparrow} = -\hat\Delta_{rr''}^{\downarrow\uparrow}
+\int\!d^3r'\left(
 h_{rr'}^{\downarrow\downarrow}\kappa_{r'r''}^{\downarrow\uparrow} 
+\kappa_{rr'}^{\downarrow\uparrow} {h^*}_{r'r''}^{\uparrow\uparrow}
+\Delta_{rr'}^{\downarrow\uparrow}{\rho^*}_{r'r''}^{\uparrow\uparrow} 
+\rho_{rr'}^{\downarrow\downarrow}\Delta_{r'r''}^{\downarrow\uparrow}
\right).
\label{HFsp}
\end{eqnarray}
This set of equations must be complemented by the complex conjugated equations.
Writing these equations, we neglected the diagonal matrix elements in spin,
$\kappa_{rr'}^{ss}$ and $\Delta_{rr'}^{ss}$. It is shown in Appendix~\ref{AppA} that such 
approximation works very well in the case of monopole pairing considered here.

We will work with the Wigner transform~\cite{Ring} of 
equations (\ref{HFsp}). The relevant mathematical details can be found in
\cite{Malov}. The most essential relations are outlined in  Appendix~\ref{AppB}.
Let us remind of some essential details of the Wigner transform of equations
(\ref{HFsp}) on the example of the first of these equations. Its left hand side is
transformed with the help of formula (\ref{B.1}) without any approximations,
i.e. exactly. The right hand side of this equation contains the products of two
matrices which are transformed with the help of formula (\ref{B.4}), where the
exponent represents an infinite series of terms with increasing powers of 
$\hbar$. It was shown in~\cite{Malov,Urban} that after integration of the
obtained equation over the phase space with second order weights 
$x_ix_j, x_ip_j, p_ip_j$ only terms proportional to powers  in $\hbar$ less than
2 survive. That is why we will write out only these terms.
From now on, we will not write out the coordinate 
dependence $(\br,\bp)$ of all functions in order to make the formulae 
more transparent. We have
\begin{eqnarray}
      i\hbar\dot f^{\uparrow\uparrow} &=&i\hbar\{h^{\uparrow\uparrow},f^{\uparrow\uparrow}\}
+h^{\uparrow\downarrow}f^{\downarrow\uparrow}-f^{\uparrow\downarrow}h^{\downarrow\uparrow}
+\frac{i\hbar}{2}\{h^{\uparrow\downarrow},f^{\downarrow\uparrow}\}
-\frac{i\hbar}{2}\{f^{\uparrow\downarrow},h^{\downarrow\uparrow}\}
\nonumber\\
&-&\frac{\hbar^2}{8}\{\!\{h^{\uparrow\downarrow},f^{\downarrow\uparrow}\}\!\}
+\frac{\hbar^2}{8}\{\!\{f^{\uparrow\downarrow},h^{\downarrow\uparrow}\}\!\} 
+ \kappa\Delta^* - \Delta\kappa^* 
\nonumber\\
&+&\frac{i\hbar}{2}\{\kappa,\Delta^*\}-\frac{i\hbar}{2}\{\Delta,\kappa^*\}
- \frac{\hbar^2}{8}\{\!\{\kappa,\Delta^*\}\!\} + \frac{\hbar^2}{8}\{\!\{\Delta,\kappa^*\}\!\}
+...,
\nonumber\\
      i\hbar\dot f^{\downarrow\downarrow} &=&i\hbar\{h^{\downarrow\downarrow},f^{\downarrow\downarrow}\}
+h^{\downarrow\uparrow}f^{\uparrow\downarrow}-f^{\downarrow\uparrow}h^{\uparrow\downarrow}
+\frac{i\hbar}{2}\{h^{\downarrow\uparrow},f^{\uparrow\downarrow}\}
-\frac{i\hbar}{2}\{f^{\downarrow\uparrow},h^{\uparrow\downarrow}\}
\nonumber\\
&-&\frac{\hbar^2}{8}\{\!\{h^{\downarrow\uparrow},f^{\uparrow\downarrow}\}\!\}
+\frac{\hbar^2}{8}\{\!\{f^{\downarrow\uparrow},h^{\uparrow\downarrow}\}\!\} 
+ \bar\Delta^* \bar\kappa - \bar\kappa^* \bar\Delta
\nonumber\\
&+&\frac{i\hbar}{2}\{\bar\Delta^*,\bar\kappa\}-\frac{i\hbar}{2}\{\bar\kappa^*,\bar\Delta\}
- \frac{\hbar^2}{8}\{\!\{\bar\Delta^*,\bar\kappa\}\!\} + \frac{\hbar^2}{8}\{\!\{\bar\kappa^*,\bar\Delta\}\!\}
+...,
\nonumber\\
      i\hbar\dot f^{\uparrow\downarrow} &=&
f^{\uparrow\downarrow}(h^{\uparrow\uparrow}-h^{\downarrow\downarrow})
+\frac{i\hbar}{2}\{(h^{\uparrow\uparrow}+h^{\downarrow\downarrow}),f^{\uparrow\downarrow}\}
-\frac{\hbar^2}{8}\{\!\{(h^{\uparrow\uparrow}-h^{\downarrow\downarrow}),f^{\uparrow\downarrow}\}\!\}
\nonumber\\
&-&h^{\uparrow\downarrow}(f^{\uparrow\uparrow}-f^{\downarrow\downarrow})
+\frac{i\hbar}{2}\{h^{\uparrow\downarrow},(f^{\uparrow\uparrow}+f^{\downarrow\downarrow})\}
+\frac{\hbar^2}{8}\{\!\{h^{\uparrow\downarrow},(f^{\uparrow\uparrow}-f^{\downarrow\downarrow})\}\!\}+....,
\nonumber\\
      i\hbar\dot f^{\downarrow\uparrow} &=&
f^{\downarrow\uparrow}(h^{\downarrow\downarrow}-h^{\uparrow\uparrow})
+\frac{i\hbar}{2}\{(h^{\downarrow\downarrow}+h^{\uparrow\uparrow}),f^{\downarrow\uparrow}\}
-\frac{\hbar^2}{8}\{\!\{(h^{\downarrow\downarrow}-h^{\uparrow\uparrow}),f^{\downarrow\uparrow}\}\!\}
\nonumber\\
&-&h^{\downarrow\uparrow}(f^{\downarrow\downarrow}-f^{\uparrow\uparrow})
+\frac{i\hbar}{2}\{h^{\downarrow\uparrow},(f^{\downarrow\downarrow}+f^{\uparrow\uparrow})\}
+\frac{\hbar^2}{8}\{\!\{h^{\downarrow\uparrow},(f^{\downarrow\downarrow}-f^{\uparrow\uparrow})\}\!\}+...,
\nonumber\\
      i\hbar\dot \kappa &=& \kappa\,(h^{\uparrow\uparrow}+\bar h^{\downarrow\downarrow})
  +\frac{i\hbar}{2}\{(h^{\uparrow\uparrow}-\bar h^{\downarrow\downarrow}),\kappa\}    
  -\frac{\hbar^2}{8}\{\!\{(h^{\uparrow\uparrow}+\bar h^{\downarrow\downarrow}),\kappa\}\!\}   
 \nonumber\\ 
 &+&\Delta\,(f^{\uparrow\uparrow}+\bar f^{\downarrow\downarrow})
  +\frac{i\hbar}{2}\{(f^{\uparrow\uparrow}-\bar f^{\downarrow\downarrow}),\Delta\}    
  -\frac{\hbar^2}{8}\{\!\{(f^{\uparrow\uparrow}+\bar f^{\downarrow\downarrow}),\Delta\}\!\}
  - \Delta + ...,
 \nonumber\\     
    i\hbar\dot \kappa^* &=&  -\kappa^*(h^{\uparrow\uparrow}+\bar h^{\downarrow\downarrow}) 
  +\frac{i\hbar}{2}\{(h^{\uparrow\uparrow}-\bar h^{\downarrow\downarrow}),\kappa^*\}    
  +\frac{\hbar^2}{8}\{\!\{(h^{\uparrow\uparrow}+\bar h^{\downarrow\downarrow}),\kappa^*\}\!\}   
  \nonumber\\
  &-& \Delta^*(f^{\uparrow\uparrow}+\bar f^{\downarrow\downarrow}) 
  +\frac{i\hbar}{2}\{(f^{\uparrow\uparrow}-\bar f^{\downarrow\downarrow}),\Delta^*\}    
  +\frac{\hbar^2}{8}\{\!\{(f^{\uparrow\uparrow}+\bar f^{\downarrow\downarrow}),\Delta^*\}\!\}
  + \Delta^* +..., 
\label{WHF}
\end{eqnarray} 
where the functions  $f$, $\kappa$, $h$ and $\Delta$ are the Wigner
transforms of $\hat\rho$, $\hat\kappa$, $\hat h$ and $\hat\Delta$
respectively, $\bar f(\br,\bp)=f(\br,-\bp)$,
 $\{f,g\}$ is the Poisson
bracket of the functions $f(\br,\bp)$ and $g(\br,\bp)$ and
$\{\{f,g\}\}$ is their double Poisson bracket.
The dots stand for terms proportional to higher powers of $\hbar$ -- after 
integration over phase space these terms disappear and we arrive to the set of
exact integral equations.
This set of equations must be complemented by the dynamical equations for 
$\bar f^{\uparrow\uparrow}, \bar f^{\downarrow\downarrow}, \bar f^{\uparrow\downarrow}, 
\bar f^{\downarrow\uparrow},\bar\kappa,\bar\kappa^*$.
They are obtained by the change $\bp \rightarrow -\bp$ in arguments of functions and Poisson brackets. 
So, in reality we deal with the set of twelve equations. We introduced the notation
$\kappa \equiv \kappa^{\uparrow\downarrow}$ and $\Delta \equiv \Delta^{\uparrow\downarrow}$.
Symmetry properties of matrices $\hat\kappa, \hat\Delta$ and the properties of their Wigner 
transforms (see Appendix~\ref{AppB}) allow one to replace the functions 
$\kappa^{\downarrow\uparrow}(\br,\bp)$ and  $\Delta^{\downarrow\uparrow}(\br,\bp)$ by the functions $\bar\kappa^{\uparrow\downarrow}(\br,\bp)$ and  $\bar\Delta^{\uparrow\downarrow}(\br,\bp)$. 

Following the paper~\cite{BaMo} we will write above equations in terms of spin-scalar
$$f^+=f^{\uparrow\uparrow}+ f^{\downarrow\downarrow}$$
and spin-vector
$$f^-=f^{\uparrow\uparrow}- f^{\downarrow\downarrow}$$
functions. Furthermore, it is useful to 
rewrite the obtained equations in terms of even and odd functions 
$f_{e}=\frac{1}{2}(f+\bar f)$ and $f_{o}=\frac{1}{2}(f-\bar f)$
and real and imaginary parts of $\kappa$ and $\Delta$: $\kappa^r=\frac{1}{2}(\kappa+\kappa^*),\,
\kappa^i=\frac{1}{2i}(\kappa-\kappa^*),\,\Delta^r=\frac{1}{2}(\Delta+\Delta^*),\,
\Delta^i=\frac{1}{2i}(\Delta-\Delta^*)$. We have
\begin{eqnarray}
i\hbar\dot f^{+}_e &=&\frac{i\hbar}{2}\left(
\{h^+_o,f^+_e\}+\{h^+_e,f^+_o\}+\{h^-_o,f^-_e\}+\{h^-_e,f^-_o\} \right)
\nonumber\\
&+&{i\hbar}\left(
 \{h^{\uparrow\downarrow}_o,f^{\downarrow\uparrow}_e\}
+\{h^{\uparrow\downarrow}_e,f^{\downarrow\uparrow}_o\}
+\{h^{\downarrow\uparrow}_o,f^{\uparrow\downarrow}_e\}
+\{h^{\downarrow\uparrow}_e,f^{\uparrow\downarrow}_o\} \right)
\nonumber\\
&+&4i\left([\kappa^i_e\Delta^r_e]-[\kappa^r_e\Delta^i_e]
+[\kappa^i_o\Delta^r_o]-[\kappa^r_o\Delta^i_o]\right)
+...,
\nonumber\\
i\hbar\dot f^{+}_o &=&\frac{i\hbar}{2}\left(
\{h^+_o,f^+_o\}+\{h^+_e,f^+_e\}+\{h^-_o,f^-_o\}+\{h^-_e,f^-_e\} \right)
\nonumber\\
&+&{i\hbar}\left(
 \{h^{\uparrow\downarrow}_o,f^{\downarrow\uparrow}_o\}
+\{h^{\uparrow\downarrow}_e,f^{\downarrow\uparrow}_e\}
+\{h^{\downarrow\uparrow}_o,f^{\uparrow\downarrow}_o\}
+\{h^{\downarrow\uparrow}_e,f^{\uparrow\downarrow}_e\} \right)
\nonumber\\
&+&2i\hbar\left(
\{\kappa^r_e,\Delta^r_e\}+\{\kappa^i_e,\Delta^i_e\}
+\{\kappa^r_o,\Delta^r_o\}+\{\kappa^i_o,\Delta^i_o\}   \right)+...,
\nonumber\\
      i\hbar\dot f^{-}_e &=&
 2\left([h^{\uparrow\downarrow}_e f^{\downarrow\uparrow}_e]
+[h^{\uparrow\downarrow}_o f^{\downarrow\uparrow}_o]
-[h^{\downarrow\uparrow}_e f^{\uparrow\downarrow}_e]
-[h^{\downarrow\uparrow}_o f^{\uparrow\downarrow}_o]\right)
\nonumber\\  
&+&\frac{i\hbar}{2}\left(
\{h^+_o,f^-_e\}+\{h^+_e,f^-_o\}+\{h^-_o,f^+_e\}+\{h^-_e,f^+_o\} \right)
\nonumber\\
&+&2i\hbar\left(
\{\kappa^r_e,\Delta^r_o\}+\{\kappa^i_e,\Delta^i_o\}
+\{\kappa^r_o,\Delta^r_e\}+\{\kappa^i_o,\Delta^i_e\}   \right)
+...,
\nonumber\\
      i\hbar\dot f^{-}_o &=&
 2\left([h^{\uparrow\downarrow}_e f^{\downarrow\uparrow}_o]
+[h^{\uparrow\downarrow}_o f^{\downarrow\uparrow}_e]
-[h^{\downarrow\uparrow}_e f^{\uparrow\downarrow}_o]
-[h^{\downarrow\uparrow}_o f^{\uparrow\downarrow}_e]\right)
\nonumber\\  
&+&\frac{i\hbar}{2}\left(
\{h^+_o,f^-_o\}+\{h^+_e,f^-_e\}+\{h^-_o,f^+_o\}+\{h^-_e,f^+_e\} \right)
\nonumber\\&+&
4i\left(
[\kappa^i_e\Delta^r_o]-[\kappa^r_e\Delta^i_o]
+[\kappa^i_o\Delta^r_e]-[\kappa^r_o\Delta^i_e]\right)
+...,
\nonumber\\
      i\hbar\dot f^{\uparrow\downarrow}_e &=&
 [h^-_e f^{\uparrow\downarrow}_e] + [h^-_o f^{\uparrow\downarrow}_o] 
- [h^{\uparrow\downarrow}_e f^-_e]
- [h^{\uparrow\downarrow}_o f^-_o]
\nonumber\\
&+&\frac{i\hbar}{2}\left(
 \{h^{\uparrow\downarrow}_e,f^+_o\}
+\{h^{\uparrow\downarrow}_o,f^+_e\}
+\{h^+_e,f^{\uparrow\downarrow}_o\} + \{h^+_o,f^{\uparrow\downarrow}_e\} 
 \right)
+...,
\nonumber\\  
      i\hbar \dot f^{\downarrow\uparrow}_e &=&       
     - [h^-_e f^{\downarrow\uparrow}_e] 
-[h^-_o f^{\downarrow\uparrow}_o] 
+[h^{\downarrow\uparrow}_e f^-_e]
+[h^{\downarrow\uparrow}_o f^-_o ]
\nonumber\\
&+&\frac{i\hbar}{2}\left(
 \{h^{\downarrow\uparrow}_e,f^+_o\}
+\{h^{\downarrow\uparrow}_o,f^+_e\}
+\{h^+_e,f^{\downarrow\uparrow}_o\} 
+\{h^+_o,f^{\downarrow\uparrow}_e\} 
 \right)
+...,
\nonumber\\
      i\hbar\dot f^{\uparrow\downarrow}_o &=&
 [ h^-_e f^{\uparrow\downarrow}_o] + [h^-_o f^{\uparrow\downarrow}_e] 
- [h^{\uparrow\downarrow}_e f^-_o]
- [h^{\uparrow\downarrow}_o f^-_e]
\nonumber\\
&+&\frac{i\hbar}{2}\left(
 \{h^{\uparrow\downarrow}_e,f^+_e\}
+\{h^{\uparrow\downarrow}_o,f^+_o\}
+\{h^+_e,f^{\uparrow\downarrow}_e\}
+ \{h^+_o,f^{\uparrow\downarrow}_o\}  \right)
+...,
\nonumber\\  
      i\hbar \dot f^{\downarrow\uparrow}_o &=&       
 -[h^-_e f^{\downarrow\uparrow}_o]
-[h^-_o f^{\downarrow\uparrow}_e] 
+[h^{\downarrow\uparrow}_e f^-_o]
+[h^{\downarrow\uparrow}_o f^-_e]
\nonumber\\
&+&\frac{i\hbar}{2}\left(
 \{h^{\downarrow\uparrow}_e,f^+_e\}
+\{h^{\downarrow\uparrow}_o,f^+_o\}
+\{h^+_e,f^{\downarrow\uparrow}_e\} 
+\{h^+_o,f^{\downarrow\uparrow}_o\}  \right)
+...,
\nonumber\\
     i\hbar\dot \kappa^r_e &=& i[h^+_e\kappa^i_e]+i[h^-_o\kappa^i_o]
    +i[f^+_e\Delta^i_e]+ i[f^-_o\Delta^i_o] -i\Delta^i_e
\nonumber\\
&+& 
\frac{i\hbar}{2}\left(\{h^+_o,\kappa^r_e\}+\{h^-_e,\kappa^r_o\}+\{f^+_o,\Delta^r_e\}+\{f^-_e,\Delta^r_o\}\right)
  +...,
\nonumber\\
     i\hbar\dot \kappa^r_o &=& i[h^+_e\kappa^i_o]+i[h^-_o\kappa^i_e]
   + i[f^+_e\Delta^i_o]+ i[f^-_o\Delta^i_e]   -i\Delta^i_o
\nonumber\\
&+&    
  \frac{i\hbar}{2}\left(\{h^+_o,\kappa^r_o\}+\{h^-_e,\kappa^r_e\}
  +\{f^+_o,\Delta^r_o\}+ \{f^-_e,\Delta^r_e\}\right)
  +...,
\nonumber\\
     i\hbar\dot \kappa^i_e &=& -i[h^+_e\kappa^r_e]-i[h^-_o\kappa^r_o]-i[f^+_e\Delta^r_e]-i[f^-_o\Delta^r_o]
  +i\Delta^r_e
\nonumber\\
&+&  
   \frac{i\hbar}{2}\left(\{h^+_o,\kappa^i_e\}+\{h^-_e,\kappa^i_o\}
   +\{f^+_o,\Delta^i_e\}+ \{f^-_e,\Delta^i_o\}\right)
  +...,
\nonumber\\
     i\hbar\dot \kappa^i_o &=& -i[h^+_e\kappa^r_o]-i[h^-_o\kappa^r_e]-i[f^+_e\Delta^r_o]-i[f^-_o\Delta^r_e]
  +i\Delta^r_o  
\nonumber\\
&+&
  \frac{i\hbar}{2}\left(\{h^+_o,\kappa^i_o\}+\{h^-_e,\kappa^i_e\}
  +\{f^+_o,\Delta^i_o\}+ \{f^-_e,\Delta^i_e\}\right)
  +...,
 \label{WHFeo}
\end{eqnarray}
The following notation is introduced here:
 $h^{\pm}=h^{\uparrow\uparrow}\pm h^{\downarrow\downarrow}$,
 $[ab]=ab-\frac{\hbar^2}{8}\{\!\{a,b\}\!\}$.

These twelve equations will be solved by the method of moments in a small amplitude 
approximation. To this end 
all functions $f(\br,\bp,t)$ and $\kappa(\br,\bp,t)$ are divided into an equilibrium part 
and a deviation (variation): $f(\br,\bp,t)=f(\br,\bp)_{\rm eq}+\delta f(\br,\bp,t)$, 
$\kappa(\br,\bp,t)=\kappa(\br,\bp)_{\rm eq}+\delta \kappa(\br,\bp,t)$.
Then equations are linearized  neglecting quadratic in variations terms.

From general arguments one can expect that the phase of $\Delta$ (and
of $\kappa$, since both are linked, according to equation (\ref{DK}))
is much more flexible than its magnitude, since the former determines
the superfluid velocity. After linearization, the phase of $\Delta$
(and of $\kappa$) is expressed by $\delta\Delta^i$ (and $\delta\kappa^i$), while
$\delta\Delta^r$ (and $\delta\kappa^r$) describes oscillations of the
magnitude of $\Delta$ (and of $\kappa$). Let us therefore assume that
\begin{equation}
\delta\kappa^r(\br,\bp)\ll\delta\kappa^i(\br,\bp).
\label{approx1}
\end{equation}
This assumption was explicitly confirmed in~\cite{M.Urban} for the case
of superfluid trapped fermionic atoms, where it was shown that
$\delta\Delta^r$ is suppressed with respect to $\delta\Delta^i$ by one
order of $\Delta/E_{\rm F}$, where $E_{\rm F}$ denotes the Fermi energy.

The assumption~(\ref{approx1}) allows one to neglect all terms containing the variations
$\delta \kappa^r$ and $\delta \Delta^r$ in the equations~(\ref{WHFeo}) after their linearization. In this case the "small" variations 
$\delta \kappa^r$ and $\delta \Delta^r$  
will not affect the dynamics of the "big" variations 
$\delta \kappa^i$ and $\delta \Delta^i$ . This means that the
dynamical equations for the "big" variations can be considered
independently from that of the "small" variations, and we will finally
deal with a set of only ten equations.

\section{Model Hamiltonian}

 The microscopic Hamiltonian of the model, harmonic oscillator with 
spin orbit potential plus separable quadrupole-quadrupole and 
spin-spin interactions is given by
\begin{eqnarray}
\label{Ham}
 H=\sum\limits_{i=1}^A\left[\frac{\hat\bp_i^2}{2m}+\frac{1}{2}m\omega^2\br_i^2
-\eta\hat \bl_i\hat \bS_i\right]+H_{qq}+H_{ss}
\end{eqnarray}
with
\begin{eqnarray}
\label{Hqq}
&& H_{qq}=\!
\sum_{\mu=-2}^{2}(-1)^{\mu}
\left\{\bar{\kappa}
 \sum\limits_i^Z\!\sum\limits_j^N
+\frac{\kappa}{2}
\left[\sum\limits_{i,j(i\neq j)}^{Z}
+\sum\limits_{i,j(i\neq j)}^{N}
\right]
\right\}
q_{2-\mu}(\br_i)q_{2\mu}(\br_j)
,
\\
\label{Hss}
&&H_{ss}=\!
\sum_{\mu=-1}^{1}(-1)^{\mu}
\left\{\bar{\chi}
 \sum\limits_i^Z\!\sum\limits_j^N
+\frac{\chi}{2}
\left[
\sum\limits_{i,j(i\neq j)}^{Z}
+\sum\limits_{i,j(i\neq j)}^{N}
\right]
\right\}
\hat S_{-\mu}(i)\hat S_{\mu}(j)
\,\delta(\br_i-\br_j),
\end{eqnarray}
where $N$ and $Z$ are the numbers of neutrons and protons
and $\hat S_{\mu}$ are spin matrices~\cite{Var}:
\begin{equation}
\hat S_1=-\frac{\hbar}{\sqrt2}{0\quad 1\choose 0\quad 0},\quad
\hat S_0=\frac{\hbar}{2}{1\quad\, 0\choose 0\, -\!1},\quad
\hat S_{-1}=\frac{\hbar}{\sqrt2}{0\quad 0\choose 1\quad 0}.
\label{S}
\end{equation}

\subsection{Mean Field}

Let us analyze the mean field generated by this Hamiltonian.

\subsubsection{Spin-orbit Potential}

Written in cyclic coordinates, the spin orbit part of the
Hamiltonian reads
$$\hat h_{ls}=-\eta\sum_{\mu=-1}^1(-)^{\mu}\hat l_{\mu}\hat S_{-\mu}
=-\eta{\quad\hat l_0\frac{\hbar}{2}\quad\; \hat l_{-1}\frac{\hbar}{\sqrt2} \choose 
 -\hat l_{1}\frac{\hbar}{\sqrt2}\; -\hat l_0\frac{\hbar}{2}},
$$
where~\cite{Var}
\begin{equation}
\label{lqu}
\hat l_{\mu}=-\hbar\sqrt2\sum_{\nu,\alpha}C_{1\nu,1\alpha}^{1\mu}r_{\nu}\nabla_{\alpha},
\end{equation}
cyclic coordinates $r_{-1}, r_0, r_1$ 
are also defined in~\cite{Var},
$C_{1\sigma,1\nu}^{\lambda\mu}$ is a Clebsch-Gordan
coefficient, and
\begin{eqnarray}
&&\hat l_1=\hbar(r_0\nabla_1-r_1\nabla_0)=
-\frac{1}{\sqrt2}(\hat l_x+i\hat l_y),
\nonumber\\&&
\hat l_0=\hbar(r_{-1}\nabla_1-r_1\nabla_{-1})=\hat l_z,
\nonumber\\
&&\hat l_{-1}=\hbar(r_{-1}\nabla_0-r_0\nabla_{-1})=
\frac{1}{\sqrt2}(\hat l_x-i\hat l_y),
\nonumber\\
&&\hat l_x=-i\hbar(y\nabla_z-z\nabla_y),\quad
\hat l_y=-i\hbar(z\nabla_x-x\nabla_z),
\nonumber\\&&
\hat l_z=-i\hbar(x\nabla_y-y\nabla_x).
\label{lxyz}
\end{eqnarray}
 Matrix elements of $\hat h_{ls}$ in coordinate space can obviously be written~\cite{BaMo} as
\begin{eqnarray}
\langle \br_1,s_1|\hat h_{ls}|\br_2,s_2\rangle 
=-\frac{\hbar}{2}\eta\left[\hat l_{0}(\br_1)(\delta_{s_1\uparrow}\delta_{s_2\uparrow}
-\delta_{s_1\downarrow}\delta_{s_2\downarrow})\right.
\nonumber\\
+\left.\sqrt2\, \hat l_{-1}(\br_1)\delta_{s_1\uparrow}\delta_{s_2\downarrow}
-\sqrt2\, \hat l_{1}(\br_1)\delta_{s_1\downarrow}\delta_{s_2\uparrow}\right]\delta(\br_1-\br_2).\ 
\label{Hrr'}
\end{eqnarray}
Their Wigner transform reads~\cite{BaMo}:
\begin{eqnarray}
 h_{ls}^{s_1s_2}(\br,\bp)
=-\frac{\hbar}{2}\eta\left[l_{0}(\br,\bp)(\delta_{s_1\uparrow}\delta_{s_2\uparrow}
-\delta_{s_1\downarrow}\delta_{s_2\downarrow})\right.
\nonumber\\
+\left.\sqrt2 l_{-1}(\br,\bp)\delta_{s_1\uparrow}\delta_{s_2\downarrow}
-\sqrt2 l_{1}(\br,\bp)\delta_{s_1\downarrow}\delta_{s_2\uparrow}\right],\ 
\label{Hrp}
\end{eqnarray}
where
$l_{\mu}=-i\sqrt2\sum_{\nu,\alpha}C_{1\nu,1\alpha}^{1\mu}r_{\nu}p_{\alpha}$.

\subsubsection{Quadrupole-quadrupole interaction}

 The contribution of $H_{qq}$ to the mean field potential is easily
found by replacing one of the $q_{2\mu}$ operators by the average value.
We have
\begin{equation}
\label{potenirr}
V^{\tau}_{qq}=\sqrt6\sum_{\mu}(-1)^{\mu}Z_{2-\mu}^{\tau +}q_{2\mu}.
\end{equation}
 Here
\begin{eqnarray}
\label{Z2mu}
&&Z_{2\mu}^{n+}=\kappa R_{2\mu}^{n+}
+\bar{\kappa}R_{2\mu}^{p+},\quad
Z_{2\mu}^{p+}=\kappa R_{2\mu}^{p+}
+\bar{\kappa}R_{2\mu}^{n+},\nonumber\\&&  R_{2\mu}^{\tau+}(t)=
\frac{1}{\sqrt6}\int d(\bp,\br)
q_{2\mu}(\br)f^{\tau+}(\br,\bp,t)
\end{eqnarray}
 with
 $\int\! d(\bp,\br)\equiv
(2\pi\hbar)^{-3}\int\! d^3p\,\int\! d^3r$ and $\tau$ being the isospin index.

\subsubsection{Spin-spin interaction}

The analogous expression for $H_{ss}$ is found in the
standard way, with the Hartree-Fock contribution given~\cite{Ring} by:
\begin{equation}
\label{MeFi}
\Gamma_{kk'}(t)=\sum_{ll'}\bar v_{kl'k'l}\rho_{ll'}(t),
\end{equation}
where $\bar v_{kl'k'l}$ is the antisymmetrized matrix element of the two
body interaction $v(1,2)$. Identifying the indices $k,k',l,l'$
with the set of coordinates $(\br,s,\tau)$, i.e. (position, spin, isospin),
one rewrites (\ref{MeFi}) as
\begin{eqnarray}
&&V^{HF}(\br_1,s_1,\tau_1;\br_1',s_1',\tau_1';t)=
\nonumber\\
&&\int\!d\br_2\int\!d\br_2'\sum_{s_2,s_2'}\sum_{\tau_2,\tau_2'}
\langle\br_1,s_1,\tau_1;\br_2,s_2,\tau_2|\hat v|\br_1',s_1',\tau_1';\br_2',s_2',\tau_2'\rangle_{a.s.}
\rho(\br_2',s_2',\tau_2';\br_2,s_2,\tau_2;t).
\nonumber
\end{eqnarray}
Let us consider the neutron-proton part of the spin-spin interaction. 
In this case
$$\hat v=v(\hat\br_1-\hat\br_2)
\sum_{\mu=-1}^{1}(-1)^{\mu}
\hat S_{-\mu}(1)\hat S_{\mu}(2)
\delta_{\tau_1p}\delta_{\tau_2n},$$
where $\hat\br_1$ is the position operator: $\hat\br_1|\br_1\rangle=\br_1|\br_1\rangle, \quad
\langle\br_1|\hat\br_1|\br_1'\rangle=\langle\br_1|\br_1'\rangle\br_1'=\delta(\br_1-\br_1')\br_1'.$

For the Hartree term one finds:
\begin{eqnarray}
&&\langle\br_1,s_1,\tau_1;\br_2,s_2,\tau_2|\hat v|\br_1',s_1',\tau_1';\br_2',s_2',\tau_2'\rangle=
\nonumber\\
&&\delta(\br_1-\br_1')\delta(\br_2-\br_2')v(\br_1'-\br_2')
\sum_{\mu=-1}^{1}(-1)^{\mu}\langle s_1,\tau_1;s_2,\tau_2|\hat S_{-\mu}(1)\hat S_{\mu}(2)
\delta_{\tau_1p}\delta_{\tau_2n}|s_1',\tau_1';s_2',\tau_2'\rangle,
\nonumber
\end{eqnarray}
\begin{eqnarray}
&&V^{H}(\br_1,s_1,\tau_1;\br_1',s_1',\tau_1';t)=
\nonumber\\
&&\int\!d\br_2\int\!d\br_2'\sum_{s_2,s_2'}\sum_{\tau_2,\tau_2'}
\langle\br_1,s_1,\tau_1;\br_2,s_2,\tau_2|\hat v|\br_1',s_1',\tau_1';\br_2',s_2',\tau_2'\rangle
\rho(\br_2',s_2',\tau_2';\br_2,s_2,\tau_2;t)=
\nonumber\\
&&\delta_{\tau_1p}\delta_{\tau_1'p}\sum_{s_2,s_2'}\sum_{\mu=-1}^{1}(-1)^{\mu}
\langle s_1|\hat S_{-\mu}(1)|s_1'\rangle\langle s_2|\hat S_{\mu}(2)|s_2'\rangle
\nonumber\\ &&\times
\delta(\br_1-\br_1')\int\!d\br_2\,v(\br_1-\br_2)
\rho(\br_2,s_2',n;\br_2,s_2,n;t).
\nonumber
\end{eqnarray}
The Fock term reads:
\begin{eqnarray}
&&\langle\br_1,s_1,\tau_1;\br_2,s_2,\tau_2|\hat v|\br_2',s_2',\tau_2';\br_1',s_1',\tau_1'\rangle=
\nonumber\\
&&\delta(\br_1-\br_2')\delta(\br_2-\br_1')v(\br_2'-\br_1')
\sum_{\mu=-1}^{1}(-1)^{\mu}\langle s_1,\tau_1;s_2,\tau_2|
\hat S_{-\mu}(1)\hat S_{\mu}(2)
\delta_{\tau_1p}\delta_{\tau_2n}|s_2',\tau_2';s_1',\tau_1'\rangle,
\nonumber
\end{eqnarray}
\begin{eqnarray}
&&V^{F}(\br_1,s_1,\tau_1;\br_1',s_1',\tau_1';t)=
\nonumber\\
&&-\int\!d\br_2\int\!d\br_2'\sum_{s_2,s_2'}\sum_{\tau_2,\tau_2'}
\langle\br_1,s_1,\tau_1;\br_2,s_2,\tau_2|\hat v|\br_2',s_2',\tau_2';\br_1',s_1',\tau_1'\rangle
\rho(\br_2',s_2',\tau_2';\br_2,s_2,\tau_2;t)=
\nonumber\\
&&-\delta_{\tau_1p}\delta_{\tau_1'n}\sum_{s_2,s_2'}
\sum_{\mu=-1}^{1}(-1)^{\mu}\langle s_1|\hat S_{-\mu}(1)|s_2'\rangle\langle s_2|\hat S_{\mu}(2)|s_1'\rangle
v(\br_1-\br_1')\rho(\br_1,s_2',p;\br_1',s_2,n;t).
\nonumber
\end{eqnarray}
Taking into account the relations
$$
\langle s|\hat S_{-1}|s'\rangle=\frac{\hbar}{\sqrt2}\delta_{s\downarrow}\delta_{s'\uparrow},
\qquad 
\langle s|\hat S_{0}|s'\rangle=\frac{\hbar}{2}\delta_{s,s'}(\delta_{s\uparrow}-\delta_{s\downarrow}),
\qquad  
\langle s|\hat S_{1}|s'\rangle=-\frac{\hbar}{\sqrt2}\delta_{s\uparrow}\delta_{s'\downarrow}
$$
 and $v(\br-\br')=\bar\chi\delta(\br-\br')$
one finds for the mean field generated by the proton-neutron 
part of~$H_{ss}$:
\begin{eqnarray}
\label{Gam}
\Gamma_{pn}(\br,s,\tau;\br',s',\tau';t)&=&\bar\chi\frac{\hbar^2}{2}
\Bigg\{
\delta_{\tau p}\delta_{\tau'p}\Big[
\delta_{s\downarrow}\delta_{s'\uparrow}\rho(\br,\downarrow,n;\br',\uparrow,n;t)
+\delta_{s\uparrow}\delta_{s'\downarrow}\rho(\br,\uparrow,n;\br',\downarrow,n;t)\Big]
\nonumber\\
&-&\delta_{\tau p}\delta_{\tau'n}\Big[
\delta_{s\downarrow}\delta_{s'\downarrow}\rho(\br,\uparrow,p;\br',\uparrow,n;t)+
\delta_{s\uparrow}\delta_{s'\uparrow}\rho(\br,\downarrow,p;\br',\downarrow,n;t)\Big]
\nonumber\\
&+&\frac{1}{2}\delta_{\tau p}\delta_{\tau'p}\left(\delta_{s\uparrow}\delta_{s'\uparrow}-
\delta_{s\downarrow}\delta_{s'\downarrow}\right)\Big[\rho(\br,\uparrow,n;\br',\uparrow,n;t)
-\rho(\br,\downarrow,n;\br',\downarrow,n;t)\Big]
\nonumber\\
&+&\frac{1}{2}\delta_{\tau p}\delta_{\tau'n}\Big[
\delta_{s\uparrow}\delta_{s'\downarrow}\rho(\br,\uparrow,p;\br',\downarrow,n;t)
+\delta_{s\downarrow}\delta_{s'\uparrow}\rho(\br,\downarrow,p;\br',\uparrow,n;t)
\nonumber\\
&-&\delta_{s\uparrow}\delta_{s'\uparrow}\rho(\br,\uparrow,p;\br',\uparrow,n;t)
-\delta_{s\downarrow}\delta_{s'\downarrow}\rho(\br,\downarrow,p;\br',\downarrow,n;t)\Big]
\Bigg\}\delta(\br-\br').
\end{eqnarray}
The expression for the mean field $\Gamma_{pp}(\br,s,\tau;\br',s',\tau';t)$ 
generated by the proton-proton part of $H_{ss}$ can be obtained from 
(\ref{Gam}) by replacing index $n$ by $p$  and the strength constant~$\bar\chi$~by~$\chi$. 
The proton mean field is defined as the sum of these two terms 
$\Gamma_{pp}(\br,s,p;\br',s',p;t)+
\Gamma_{pn}(\br,s,p;\br',s',p;t)$. 
Its Wigner transform can be
written as
\begin{eqnarray}
\label{Vp}
V_{p}^{s s'}(\br,t)&=&
3\chi\frac{\hbar^2}{8}
\left\{
\delta_{s\downarrow}\delta_{s'\uparrow}n_p^{\downarrow\uparrow}+
\delta_{s\uparrow}\delta_{s'\downarrow}n_p^{\uparrow\downarrow}
-\delta_{s\downarrow}\delta_{s'\downarrow}n_p^{\uparrow\uparrow}
-\delta_{s\uparrow}\delta_{s'\uparrow}n_p^{\downarrow\downarrow}
\right\}
\nonumber\\
&+&\bar\chi\frac{\hbar^2}{8}
\left\{
2\delta_{s\downarrow}\delta_{s'\uparrow}n_n^{\downarrow\uparrow}+
2\delta_{s\uparrow}\delta_{s'\downarrow}n_n^{\uparrow\downarrow}
+(\delta_{s\uparrow}\delta_{s'\uparrow}-
\delta_{s\downarrow}\delta_{s'\downarrow})(n_n^{\uparrow\uparrow}-
n_n^{\downarrow\downarrow})
\right\},
\end{eqnarray}
where 
${\di n_{\tau}^{ss'}(\br,t)=\int\frac{d\bp}{(2\pi\hbar)^3}f^{ss'}_{\tau}(\br,\bp,t)}$.
The Wigner transform of the neutron mean field $V_n^{ss'}$ is 
obtained from (\ref{Vp}) by the obvious change of indices $p\leftrightarrow n$.
The Wigner function $f$ and density matrix 
$\rho$ are connected by the relation 
${\di f^{ss'}_{\tau\tau'}(\br,\bp,t)=\int\!d\bq\, e^{-i\bp\bq/\hbar}
\rho(\br_1,s,\tau;\br_2,s',\tau';t)}$, with $\bq=\br_1-\br_2$ and 
$\br=\frac{1}{2}(\br_1+\br_2)$. Integrating this relation over $\bp$
with $\tau'=\tau$ one finds:
$$n_{\tau}^{ss'}(\br,t)=\rho(\br,s,\tau;\br,s',\tau;t).$$
By definition the diagonal elements of the density matrix describe 
the proper densities. Therefore $n_{\tau}^{ss}(\br,t)$ is the density
of spin-up nucleons (if $s=\uparrow$) or spin-down nucleons 
(if $s=\downarrow$). Off diagonal in spin elements of the
density matrix $n_{\tau}^{ss'}(\br,t)$ are spin-flip characteristics
and can be called spin-flip densities.

\subsection{Pair potential}

The Wigner transform of the pair potential (pairing gap) $\Delta(\br,\bp)$ is related to 
the Wigner transform of the anomalous density by~\cite{Ring}
\begin{equation}
\Delta(\br,\bp)=-\int\! \frac{d\bp'}{(2\pi\hbar)^3}
v(|\bp-\bp'|)\kappa(\br,\bp'),
\label{DK}
\end{equation}
where $v(p)$ is a Fourier transform of the two-body interaction.
We take for the pairing interaction a simple Gaussian of strength $V_0$ 
and range $r_p$ ~\cite{Ring}
\begin{equation}
v(p)=\beta {\rm e}^{-\alpha p^2}\!,
\label{v_p}
\end{equation}
with $\beta=-|V_0|(r_p\sqrt{\pi})^3$ and $\alpha=r_p^2/4\hbar^2$. For the values of the parameters, see section~\ref{VA}.

\section{Equations of motion}

 Integrating the set of equations (\ref{WHFeo}) over phase space 
with the weights 
\begin{equation}
W =\{r\otimes p\}_{\lambda\mu},\,\{r\otimes r\}_{\lambda\mu},\,
\{p\otimes p\}_{\lambda\mu}, \mbox{ and } 1
\label{weightfunctions}
\end{equation}
one gets dynamic equations for 
the following collective variables:
\begin{eqnarray}
&&L^{\tau\varsigma}_{\lambda\mu}(t)=\int\! d(\bp,\br) \{r\otimes p\}_{\lambda\mu}
f^{\tau\varsigma}_o(\br,\bp,t),
\nonumber\\&&
R^{\tau\varsigma}_{\lambda\mu}(t)=\int\! d(\bp,\br) \{r\otimes r\}_{\lambda\mu}
f^{\tau\varsigma}_e(\br,\bp,t),
\nonumber\\
&&P^{\tau\varsigma}_{\lambda\mu}(t)=\int\! d(\bp,\br) \{p\otimes p\}_{\lambda\mu}
f^{\tau\varsigma}_e(\br,\bp,t),
\nonumber\\&&
F^{\tau\varsigma}(t)=\int\! d(\bp,\br)
f^{\tau\varsigma}_e(\br,\bp,t),
\nonumber\\
&&\tilde{L}^{\tau}_{\lambda\mu}(t)=\int\! d(\bp,\br) \{r\otimes p\}_{\lambda\mu}
\kappa^{\tau i}_o(\br,\bp,t),
\nonumber\\&&
\tilde{R}^{\tau}_{\lambda\mu}(t)=\int\! d(\bp,\br) \{r\otimes r\}_{\lambda\mu}
\kappa^{\tau i}_e(\br,\bp,t),
\nonumber\\
&&\tilde{P}^{\tau}_{\lambda\mu}(t)=\int\! d(\bp,\br) \{p\otimes p\}_{\lambda\mu}
\kappa^{\tau i}_e(\br,\bp,t),
\label{Varis}
\end{eqnarray}
 where 
$\varsigma\!=+,\,-,\,\uparrow\downarrow,\,\downarrow\uparrow,$
and $\displaystyle \{r\otimes r\}_{\lambda\mu}=\sum\limits_{\sigma,\nu}C_{1\sigma,1\nu}^{\lambda\mu}r_{\sigma}r_{\nu}.$
We already called the functions 
$f^+=f^{\uparrow\uparrow}+f^{\downarrow\downarrow}$ and
$f^-=f^{\uparrow\uparrow}-f^{\downarrow\downarrow}$
spin-scalar and spin-vector ones, respectively. It is, therefore, natural to call the
corresponding collective variables $X^{+}_{\lambda\mu}(t)$ and 
$X^{-}_{\lambda\mu}(t)$ spin-scalar and spin-vector variables.

The required expressions for 
$h^{\pm}$, $h^{\uparrow\downarrow}$ and $h^{\downarrow\uparrow}$ are
\begin{eqnarray}
h_{\tau}^{+}=\frac{p^2}{m}+m\,\omega^2r^2
+12\sum_{\nu}(-1)^{\nu}Z_{2\nu}^{\tau+}(t)\{r\otimes r\}_{2-\nu}
+V_{\tau}^+(\br,t)-\mu^{\tau},
\end{eqnarray}
$\mu^{\tau}$ being the chemical potential of protons ($\tau=p$) or neutrons ($\tau=n$),
\begin{eqnarray}
&&h_{\tau}^-=-\hbar\eta l_0+V_{\tau}^-(\br,t), \quad
h_{\tau}^{\uparrow\downarrow}=-\frac{\hbar}{\sqrt2}\eta l_{-1}+V_{\tau}^{\uparrow\downarrow}(\br,t),
\quad h_{\tau}^{\downarrow\uparrow}=\frac{\hbar}{\sqrt2}\eta l_{1}+V_{\tau}^{\downarrow\uparrow}(\br,t),\quad 
\end{eqnarray}
where according to (\ref{Vp})
\begin{eqnarray}
\label{Vss}
&&V_p^+(\br,t)=-3\frac{\hbar^2}{8}\chi n_p^+(\br,t),
\nonumber\\
&&V_p^-(\br,t)=3\frac{\hbar^2}{8}\chi n_p^-(\br,t)+\frac{\hbar^2}{4}\bar\chi n_n^-(\br,t),
\nonumber\\
&&V_p^{\uparrow\downarrow}(\br,t)=3\frac{\hbar^2}{8}\chi n_p^{\uparrow\downarrow}(\br,t)
+\frac{\hbar^2}{4}\bar\chi n_n^{\uparrow\downarrow}(\br,t),
\nonumber\\
&&V_p^{\downarrow\uparrow}(\br,t)=3\frac{\hbar^2}{8}\chi n_p^{\downarrow\uparrow}(\br,t)
+\frac{\hbar^2}{4}\bar\chi n_n^{\downarrow\uparrow}(\br,t)
\end{eqnarray}
and the neutron potentials $V_n^{\varsigma}$ are
obtained by the obvious change of indices $p\leftrightarrow n$.

The integration of equations (\ref{WHFeo})
with the weights (\ref{weightfunctions})
yields:
\begin{eqnarray}
\label{quadr}
     \dot L^{+}_{\lambda\mu}&=&
\frac{1}{m}P_{\lambda\mu}^{+}-
m\,\omega^2R^{+}_{\lambda \mu}
+12\sqrt5\sum_{j=0}^2\sqrt{2j+1}
\left\{_{2\lambda 1}^{11j}\right\}
\{Z_2^{+}\otimes R_j^{+}\}_{\lambda \mu}
\nonumber\\&&
-i\hbar\frac{\eta}{2}\left[\mu L_{\lambda\mu}^- 
+\sqrt{(\lambda-\mu)(\lambda+\mu+1)}L^{\uparrow\downarrow}_{\lambda\mu+1}+
\sqrt{(\lambda+\mu)(\lambda-\mu+1)}L^{\downarrow\uparrow}_{\lambda\mu-1}\right]
\nonumber\\&&
-\int\!d^3r\left[
\frac{1}{2}n^+\{r\otimes \nabla\}_{\lambda\mu}V^++
\frac{1}{2}n^-\{r\otimes \nabla\}_{\lambda\mu}V^-+
 n^{\downarrow\uparrow}\{r\otimes \nabla\}_{\lambda\mu}V^{\uparrow\downarrow}+
n^{\uparrow\downarrow}\{r\otimes \nabla\}_{\lambda\mu}V^{\downarrow\uparrow}
\right],
\nonumber
\\
     \dot L^{-}_{\lambda\mu}&=&
\frac{1}{m}P_{\lambda\mu}^{-}-
m\,\omega^2R^{-}_{\lambda \mu}
+12\sqrt5\sum_{j=0}^2\sqrt{2j+1}\left\{_{2\lambda 1}^{11j}\right\}
\{Z_2^{+}\otimes R_j^{-}\}_{\lambda \mu}-i\hbar\frac{\eta}{2}\mu L_{\lambda\mu}^+
\nonumber\\&&
-\frac{\hbar^2}{2}\eta\delta_{\lambda,1}
\left[\delta_{\mu,-1}F^{\uparrow\downarrow}+\delta_{\mu,1}F^{\downarrow\uparrow}\right]
-\frac{1}{2}\int\!d^3r\left[
n^-\{r\otimes \nabla\}_{\lambda\mu}V^++
n^+\{r\otimes \nabla\}_{\lambda\mu}V^-\right]
\nonumber\\&&
-2\frac{i}{\hbar}\int\! d(\bp,\br)\{r\otimes p\}_{\lambda\mu}
\left[h^{\uparrow\downarrow}f^{\downarrow\uparrow}-h^{\downarrow\uparrow}f^{\uparrow\downarrow}\right],
\nonumber
\end{eqnarray}\begin{eqnarray}
     \dot L^{\uparrow\downarrow}_{\lambda\mu+1}&=&
\frac{1}{m}P_{\lambda\mu+1}^{\uparrow\downarrow}-
m\,\omega^2R^{\uparrow\downarrow}_{\lambda \mu+1}
+12\sqrt5\sum_{j=0}^2\sqrt{2j+1}\left\{_{2\lambda 1}^{11j}\right\}
\{Z_2^{+}\otimes R_j^{\uparrow\downarrow}\}_{\lambda \mu+1}
\nonumber\\&&
-i\hbar\frac{\eta}{4} \sqrt{(\lambda-\mu)(\lambda+\mu+1)}L_{\lambda\mu}^+
+\frac{\hbar^2}{4}\eta\delta_{\lambda,1}
\left[\delta_{\mu,0}F^- +\sqrt2\delta_{\mu,-1}F^{\uparrow\downarrow}\right]
\nonumber\\&&
-\frac{1}{2}\int\!d^3r\left[
n^{\uparrow\downarrow}\{r\otimes \nabla\}_{\lambda\mu+1}V^++
n^+\{r\otimes \nabla\}_{\lambda\mu+1}V^{\uparrow\downarrow}\right]
\nonumber\\&&
-\frac{i}{\hbar}\int\! d(\bp,\br)\{r\otimes p\}_{\lambda\mu+1}
\left[h^{-}f^{\uparrow\downarrow}-h^{\uparrow\downarrow}f^-\right],
\nonumber
\\
     \dot L^{\downarrow\uparrow}_{\lambda\mu-1}&=&
\frac{1}{m}P_{\lambda\mu-1}^{\downarrow\uparrow}-
m\,\omega^2R^{\downarrow\uparrow}_{\lambda \mu-1}
+12\sqrt5\sum_{j=0}^2\sqrt{2j+1}\left\{_{2\lambda 1}^{11j}\right\}
\{Z_2^{+}\otimes R_j^{\downarrow\uparrow}\}_{\lambda \mu-1}
\nonumber\\&&
-i\hbar\frac{\eta}{4} \sqrt{(\lambda+\mu)(\lambda-\mu+1)}L_{\lambda\mu}^+
+\frac{\hbar^2}{4}\eta\delta_{\lambda,1}
\left[\delta_{\mu,0}F^- -\sqrt2\delta_{\mu,1}F^{\downarrow\uparrow}\right]
\nonumber\\&&
-\frac{1}{2}\int\!d^3r\left[
n^{\downarrow\uparrow}\{r\otimes \nabla\}_{\lambda\mu-1}V^++
n^+\{r\otimes \nabla\}_{\lambda\mu-1}V^{\downarrow\uparrow}\right]
\nonumber\\&&
-\frac{i}{\hbar}\int\! d(\bp,\br)\{r\otimes p\}_{\lambda\mu-1}
\left[h^{\downarrow\uparrow}f^- -h^{-}f^{\downarrow\uparrow}\right],
\nonumber
\\
     \dot F^{-}&=&
2\eta \left[L_{1-1}^{\downarrow\uparrow}+L_{11}^{\uparrow\downarrow}\right],
\nonumber\\
     \dot F^{\uparrow\downarrow}&=&
-\eta [L_{1-1}^- -\sqrt2L_{10}^{\uparrow\downarrow}],
\nonumber\\
     \dot F^{\downarrow\uparrow}&=&
-\eta \left[L_{11}^- +\sqrt2L_{10}^{\downarrow\uparrow}\right],
\nonumber\\
     \dot R^{+}_{\lambda\mu}&=&
\frac{2}{m}L^+_{\lambda\mu}
-i\hbar\frac{\eta}{2}\left[\mu R_{\lambda\mu}^- 
+\sqrt{(\lambda-\mu)(\lambda+\mu+1)}R^{\uparrow\downarrow}_{\lambda\mu+1}+
\sqrt{(\lambda+\mu)(\lambda-\mu+1)}R^{\downarrow\uparrow}_{\lambda\mu-1}\right],
\nonumber\\
     \dot R^{-}_{\lambda\mu}&=&
\frac{2}{m}L^-_{\lambda\mu}
-i\hbar\frac{\eta}{2}\mu R_{\lambda\mu}^+
-2\frac{i}{\hbar}\int\! d(\bp,\br)\{r\otimes r\}_{\lambda\mu}
\left[h^{\uparrow\downarrow}f^{\downarrow\uparrow}-h^{\downarrow\uparrow}f^{\uparrow\downarrow}\right],
\nonumber\\
     \dot R^{\uparrow\downarrow}_{\lambda\mu+1}&=&
\frac{2}{m}L^{\uparrow\downarrow}_{\lambda\mu+1}
-i\hbar\frac{\eta}{4} \sqrt{(\lambda-\mu)(\lambda+\mu+1)}R_{\lambda\mu}^+
-\frac{i}{\hbar}\int\! d(\bp,\br)\{r\otimes r\}_{\lambda\mu+1}
\left[h^{-}f^{\uparrow\downarrow}-h^{\uparrow\downarrow}f^-\right],
\nonumber\\
     \dot R^{\downarrow\uparrow}_{\lambda\mu-1}&=&
\frac{2}{m}L^{\downarrow\uparrow}_{\lambda\mu-1}
-i\hbar\frac{\eta}{4} \sqrt{(\lambda+\mu)(\lambda-\mu+1)}R_{\lambda\mu}^+
-\frac{i}{\hbar}\int\! d(\bp,\br)\{r\otimes r\}_{\lambda\mu-1}
\left[h^{\downarrow\uparrow}f^- -h^{-}f^{\downarrow\uparrow}\right],
\nonumber\\
     \dot P^{+}_{\lambda\mu}&=&
-2m\,\omega^2L^+_{\lambda \mu}
+24\sqrt5\sum_{j=0}^2\sqrt{2j+1}\left\{_{2\lambda 1}^{11j}\right\}
\{Z_2^+\otimes L^+_j\}_{\lambda \mu}
\nonumber\\
&&-i\hbar\frac{\eta}{2}\left[\mu P_{\lambda\mu}^- 
+\sqrt{(\lambda-\mu)(\lambda+\mu+1)}P^{\uparrow\downarrow}_{\lambda\mu+1}+
\sqrt{(\lambda+\mu)(\lambda-\mu+1)}P^{\downarrow\uparrow}_{\lambda\mu-1}\right]
\nonumber\\&&
-\int\!d^3r\left[
\{J^+\otimes \nabla\}_{\lambda\mu}V^++
\{J^-\otimes \nabla\}_{\lambda\mu}V^-+
2\{J^{\downarrow\uparrow}\otimes \nabla\}_{\lambda\mu}V^{\uparrow\downarrow}+
2\{J^{\uparrow\downarrow}\otimes \nabla\}_{\lambda\mu}V^{\downarrow\uparrow}
\right],
\nonumber\\
     \dot P^{-}_{\lambda\mu}&=&
-2m\,\omega^2L^-_{\lambda \mu}
+24\sqrt5\sum_{j=0}^2\sqrt{2j+1}\left\{_{2\lambda 1}^{11j}\right\}
\{Z_2^{+}\otimes L^-_j\}_{\lambda \mu}
-i\hbar\frac{\eta}{2}\mu P_{\lambda\mu}^+
\nonumber\\&&
-\int\!d^3r\left[
\{J^-\otimes \nabla\}_{\lambda\mu}V^++
\{J^+\otimes \nabla\}_{\lambda\mu}V^-\right]
-2\frac{i}{\hbar}\int\! d(\bp,\br)\{p\otimes p\}_{\lambda\mu}
\left[h^{\uparrow\downarrow}f^{\downarrow\uparrow}-h^{\downarrow\uparrow}f^{\uparrow\downarrow}\right],
\nonumber\\
     \dot P^{\uparrow\downarrow}_{\lambda\mu+1}&=&
-2m\,\omega^2L^{\uparrow\downarrow}_{\lambda \mu+1}
+24\sqrt5\sum_{j=0}^2\sqrt{2j+1}\left\{_{2\lambda 1}^{11j}\right\}
\{Z_2^{+}\otimes L^{\uparrow\downarrow}_j\}_{\lambda \mu+1}
-i\hbar\frac{\eta}{4} \sqrt{(\lambda-\mu)(\lambda+\mu+1)}P_{\lambda\mu}^+
\nonumber\\&&
-\int\!d^3r\left[
\{J^{\uparrow\downarrow}\otimes \nabla\}_{\lambda\mu+1}V^++
\{J^+\otimes \nabla\}_{\lambda\mu+1}V^{\uparrow\downarrow}\right]
\nonumber\\&&
-\frac{i}{\hbar}\int\! d(\bp,\br)\{p\otimes p\}_{\lambda\mu+1}
[h^{-}f^{\uparrow\downarrow}-h^{\uparrow\downarrow}f^-],
\nonumber\\
     \dot P^{\downarrow\uparrow}_{\lambda\mu-1}&=&
-2m\,\omega^2L^{\downarrow\uparrow}_{\lambda \mu-1}
+24\sqrt5\sum_{j=0}^2\sqrt{2j+1}\left\{_{2\lambda 1}^{11j}\right\}
\{Z_2^{+}\otimes L^{\downarrow\uparrow}_j\}_{\lambda \mu-1}
-i\hbar\frac{\eta}{4} \sqrt{(\lambda+\mu)(\lambda-\mu+1)}P_{\lambda\mu}^+
\nonumber\\&&
-\int\!d^3r\left[
\{J^{\downarrow\uparrow}\otimes \nabla\}_{\lambda\mu-1}V^++
\{J^+\otimes \nabla\}_{\lambda\mu-1}V^{\downarrow\uparrow}\right]
\nonumber\\&&
-\frac{i}{\hbar}\int\! d(\bp,\br)\{p\otimes p\}_{\lambda\mu-1}
\left[h^{\downarrow\uparrow}f^- -h^{-}f^{\downarrow\uparrow}\right],
\end{eqnarray}
 where $\left\{_{2\lambda 1}^{11j}\right\}$
is the Wigner $6j$-symbol and
${\di J_{\nu}^{\varsigma}(\br,t)=\int\frac{d\bp}{(2\pi\hbar)^3}\,
p_{\nu}f^{\varsigma}(\br,\bp,t)}$ is the current.
For the sake of simplicity the isospin and the time dependence of tensors is not
written out. It is easy to see that equations (\ref{quadr}) are
nonlinear due to quadrupole-quadrupole and spin-spin interactions.
We will solve them in the small amplitude approximation, by linearizing
the equations. This procedure helps also to solve another problem: to
represent the integral terms in (\ref{quadr}) as the linear combination
of collective variables (\ref{Varis}), that allows to close the whole 
set of  equations (\ref{quadr}). The detailed analysis of the integral
terms is given in the appendix~\ref{AppC}.

We are interested in the scissors mode with quantum number
$K^{\pi}=1^+$. Therefore, we only need the part of dynamic equations 
with $\mu=1$.

\subsection{Linearized equations  (\mbox{\boldmath${\mu=1}$}), isovector, isoscalar} 

 Writing all variables as a sum of their
equilibrium value plus a small deviation
\begin{eqnarray}
&&R_{\lambda\mu}(t)=R_{\lambda\mu}({\rm eq})+\R_{\lambda\mu}(t),\quad
  P_{\lambda\mu}(t)=P_{\lambda\mu}({\rm eq})+\P_{\lambda\mu}(t),\nonumber\\
&&L_{\lambda\mu}(t)=L_{\lambda\mu}({\rm eq})+\L_{\lambda\mu}(t),\quad
  F(t)=F({\rm eq})+\F(t),\nonumber
\end{eqnarray}
one gets dynamic equations for variations of the collective variables
(\ref{Varis}):
\begin{eqnarray}
&&\L^{\tau\varsigma}_{\lambda\mu}(t)=\int\! d(\bp,\br) \{r\otimes p\}_{\lambda\mu}
\delta f^{\tau\varsigma}_o(\br,\bp,t),
\nonumber\\&&
\R^{\tau\varsigma}_{\lambda\mu}(t)=\int\! d(\bp,\br) \{r\otimes r\}_{\lambda\mu}
\delta f^{\tau\varsigma}_e(\br,\bp,t),
\nonumber\\
&&\P^{\tau\varsigma}_{\lambda\mu}(t)=\int\! d(\bp,\br) \{p\otimes p\}_{\lambda\mu}
\delta f^{\tau\varsigma}_e(\br,\bp,t),
\nonumber\\&&
\F^{\tau\varsigma}(t)=\int\! d(\bp,\br)
\delta f^{\tau\varsigma}_e(\br,\bp,t),
\nonumber\\
&&\tilde{\L}^{\tau}_{\lambda\mu}(t)=\int\! d(\bp,\br) \{r\otimes p\}_{\lambda\mu}
\delta \kappa^{\tau i}_o(\br,\bp,t),
\nonumber\\&&
\tilde{\R}^{\tau}_{\lambda\mu}(t)=\int\! d(\bp,\br) \{r\otimes r\}_{\lambda\mu}
\delta \kappa^{\tau i}_e(\br,\bp,t),
\nonumber\\
&&\tilde{\P}^{\tau}_{\lambda\mu}(t)=\int\! d(\bp,\br) \{p\otimes p\}_{\lambda\mu}
\delta \kappa^{\tau i}_e(\br,\bp,t).
\label{VarisV}
\end{eqnarray}

Neglecting quadratic deviations, one obtains the linearized
equations. Naturally one needs to know the mean fields variations
and the equilibrium values of all variables.

Variations of mean fields read:

$$\delta h_{\tau}^{+}=12\sum_{\mu}(-1)^{\mu}\delta Z_{2\mu}^{\tau+}(t)\{r\otimes r\}_{2-\mu}
+\delta V_{\tau}^+(\br,t),$$
where
$\delta Z_{2\mu}^{p+}=\kappa \delta R_{2\mu}^{p+}
+\bar{\kappa}\delta R_{2\mu}^{n+}$,  $\delta R_{\lambda\mu}^{\tau+}(t)\equiv
\R_{\lambda\mu}^{\tau+}(t)$ and
\begin{eqnarray}
&&\delta V_p^+(\br,t)=-3\frac{\hbar^2}{8}\chi \delta n_p^+(\br,t),
\nonumber\\ \nonumber
&&\delta n_{p}^{+}(\br,t)=\int\frac{d^3p}{(2\pi\hbar)^3}\delta f^{+}_{p}(\br,\bp,t).
\end{eqnarray}
Variations of
$h^{-}$, $h^{\uparrow\downarrow}$ and $h^{\downarrow\uparrow}$ are obtained in a similar way.
Variation of the pair potential is
\begin{equation}
\delta \Delta(\br,\bp,t)=-\int\! \frac{d^3p'}{(2\pi\hbar)^3}
v(|\bp-\bp'|)\delta \kappa(\br,\bp',t).
\label{DKvar}
\end{equation}

Evident equilibrium conditions for an axially symmetric nucleus are:

\begin{equation}
R^{+}_{2\pm1}({\rm eq})=R^{+}_{2\pm2}({\rm eq})=0,\quad R^{+}_{20}({\rm eq})\neq0,\quad R^{+}_{00}({\rm eq})\neq0.
\label{equi1}
\end{equation}
It is obvious that all ground state properties of the system of spin
up nucleons are identical to the ones of the system of nucleons with spin
down. Therefore
\begin{equation}
R^{-}_{\lambda\mu}({\rm eq})=P^{-}_{\lambda\mu}({\rm eq})=L^{-}_{\lambda\mu}({\rm eq})=0.
\label{equi2}
\end{equation}
We also will suppose 
\begin{equation}
L^{+}_{\lambda\mu}({\rm eq})=L^{\uparrow\downarrow}_{\lambda\mu}({\rm eq})=L^{\downarrow\uparrow}_{\lambda\mu}({\rm eq})=0
\quad\mbox{ and }\quad
R^{\uparrow\downarrow}_{\lambda\mu}({\rm eq})=R^{\downarrow\uparrow}_{\lambda\mu}({\rm eq})=0.
\label{equi3}
\end{equation}
Let us recall
that all variables and equilibrium quantities $R^{+}_{\lambda 0}({\rm eq})$
and $Z^{+}_{20}({\rm eq})$ in (\ref{quadr}) have isospin indices 
$\tau=n,\,p$. All the difference 
between neutron and proton systems is contained in the mean field 
quantities $Z^{\tau+}_{20}({\rm eq})$ and $V_{\tau}^{\varsigma}$, which
are different for neutrons and protons (see eq.~(\ref{Z2mu}) and~(\ref{Vss})).

  It is convenient to rewrite the dynamical equations in terms
of isoscalar and isovector variables
\begin{eqnarray}
\label{Isovs}
&&\bar \R_{\lambda\mu}(t)=\R_{\lambda\mu}^{n}(t)+\R_{\lambda\mu}^{p}(t),\quad
\R_{\lambda\mu}(t)=\R_{\lambda\mu}^{n}(t)-\R_{\lambda\mu}^{p}(t),
\nonumber\\ 
&&\bar \P_{\lambda\mu}(t)=\P_{\lambda\mu}^{n}(t)+\P_{\lambda\mu}^{p}(t),\quad
\P_{\lambda\mu}(t)=\P_{\lambda\mu}^{n}(t)-\P_{\lambda\mu}^{p}(t),
\nonumber\\
&&\bar \L_{\lambda\mu}(t)=\L_{\lambda\mu}^{n}(t)+\L_{\lambda\mu}^{p}(t),\quad
\L_{\lambda\mu}(t)=\L_{\lambda\mu}^{n}(t)-\L_{\lambda\mu}^{p}(t).
\end{eqnarray}
It also is natural to define isovector and isoscalar strength constants
$\kappa_1=\frac{1}{2}(\kappa-\bar\kappa)$ and
$\kappa_0=\frac{1}{2}(\kappa+\bar\kappa)$ connected by the relation
$\kappa_1=\alpha\kappa_0$~\cite{BaSc}.
Then the equations for the neutron and proton systems are transformed
into isovector and isoscalar ones. Supposing that all equilibrium
characteristics of the proton system are equal to that of the neutron
system one decouples isovector and isoscalar equations. This 
approximations looks rather crude. In the paper~\cite{Urban} we have tried to
improve it by employing more accurate approximation which works very well in
the case of collective motion:
$$Q^n/N=\pm Q^p/Z,$$
where $Q$ is any of collective variables (\ref{VarisV}) and the sign +(-) is 
utilized for the isoscalar (isovector) motion. The corrections to the more simple
approximation turned out of the order $(\frac{N-Z}{A})^2$. For rare earth nuclei
this gives an error about 4$\%$, that is admissible for us, because the main goal
of this paper is to understand the influence of the simultaneous action of
pairing and spin degrees of freedom on the scissors mode. So, to keep final
formulae more transparent, we prefer to use the more simple approximations.

With the help of the above equilibrium relations one arrives at the 
following final set of equations for {\bf isovector} variables:
\begin{eqnarray}
     \dot {\L}^{+}_{21}&=&
\frac{1}{m}\P_{21}^{+}-
\left[m\,\omega^2
-4\sqrt3\alpha\kappa_0R_{00}^{\rm eq}
+\sqrt6(1+\alpha)\kappa_0 R_{20}^{\rm eq}\right]\R^{+}_{21}
-i\hbar\frac{\eta}{2}\left[\L_{21}^-
+2\L^{\uparrow\downarrow}_{22}+
\sqrt6\L^{\downarrow\uparrow}_{20}\right],
\nonumber\\
     \dot {\L}^{-}_{21}&=&
\frac{1}{m}\P_{21}^{-}
-
\left[m\,\omega^2+\sqrt6\kappa_0 R_{20}^{\rm eq}
-\frac{\sqrt{3}}{20}\hbar^2 
\left( \chi-\frac{\bar\chi}{3} \right)
\left(\frac{I_1}{a_0^2}+\frac{I_1}{a_1^2}\right)\left(\frac{a_1^2}{A_2}-\frac{a_0^2}{A_1}\right)
\right]\R^{-}_{21}
\nonumber\\&&
-i\hbar\frac{\eta}{2}\L_{21}^+ 
+\frac{4}{\hbar}|V_0| I_{rp}^{\kappa\Delta}(r') {\tilde\L}_{21},
\nonumber\\
     \dot {\L}^{\uparrow\downarrow}_{22}&=&
\frac{1}{m}\P_{22}^{\uparrow\downarrow}-
\left[m\,\omega^2-2\sqrt6\kappa_0R_{20}^{\rm eq}
-\frac{\sqrt{3}}{5}\hbar^2 
\left( \chi-\frac{\bar\chi}{3} \right)\frac{I_1}{A_2}
\right]\R^{\uparrow\downarrow}_{22}
-i\hbar\frac{\eta}{2}\L_{21}^+,
\nonumber\\
     \dot {\L}^{\downarrow\uparrow}_{20}&=&
\frac{1}{m}\P_{20}^{\downarrow\uparrow}-
\left[m\,\omega^2
+2\sqrt6\kappa_0 R_{20}^{\rm eq}\right]\R^{\downarrow\uparrow}_{20}
+\frac{2}{\sqrt3}\kappa_0 R_{20}^{\rm eq}\,\R^{\downarrow\uparrow}_{00}
-i\hbar\frac{\eta}{2}\sqrt{\frac{3}{2}}\L_{21}^+ 
\nonumber\\
&&+\frac{\sqrt{3}}{15}\hbar^2 
\left( \chi-\frac{\bar\chi}{3} \right)\frac{I_1}{A_1 A_2} 
\left[
\left(A_1-2A_2\right)
\R_{20}^{\downarrow\uparrow}+
\sqrt2
\left(A_1+A_2\right)
\R_{00}^{\downarrow\uparrow}
\right],
\nonumber\\
     \dot {\L}^{+}_{11}&=&
-3\sqrt6(1-\alpha)\kappa_0 R_{20}^{\rm eq}\,\R^{+}_{21}
-i\hbar\frac{\eta}{2}\left[\L_{11}^- 
+\sqrt2\L^{\downarrow\uparrow}_{10}\right],
\nonumber\\
     \dot {\L}^{-}_{11}&=&
-\left[3\sqrt6\kappa_0 R_{20}^{\rm eq}
-\frac{\sqrt{3}}{20}\hbar^2 
\left( \chi-\frac{\bar\chi}{3} \right)
\left(\frac{I_1}{a_0^2}-\frac{I_1}{a_1^2}\right)\left(\frac{a_1^2}{A_2}-\frac{a_0^2}{A_1}\right)
\right]\R^{-}_{21}
\nonumber\\&&
-\hbar\frac{\eta}{2}\left[i\L_{11}^+
+\hbar \F^{\downarrow\uparrow}\right]
+\frac{4}{\hbar}|V_0| I_{rp}^{\kappa\Delta}(r') {\tilde\L}_{11},
\nonumber\\
     \dot {\L}^{\downarrow\uparrow}_{10}&=&
-\hbar\frac{\eta}{2\sqrt2}\left[i\L_{11}^+
+\hbar \F^{\downarrow\uparrow}\right]
,
\nonumber\\
     \dot {\F}^{\downarrow\uparrow}&=&
-\eta\left[\L_{11}^- +\sqrt2\L^{\downarrow\uparrow}_{10}\right],
\nonumber\\
\dot {\R}^{+}_{21}&=&
\frac{2}{m}\L_{21}^{+}
-i\hbar\frac{\eta}{2}\left[\R_{21}^-
+2\R^{\uparrow\downarrow}_{22}+
\sqrt6\R^{\downarrow\uparrow}_{20}\right],
\nonumber\\
     \dot {\R}^{-}_{21}&=&
\frac{2}{m}\L_{21}^{-}
-i\hbar\frac{\eta}{2}\R_{21}^+,
\nonumber\\
     \dot {\R}^{\uparrow\downarrow}_{22}&=&
\frac{2}{m}\L_{22}^{\uparrow\downarrow}
-i\hbar\frac{\eta}{2}\R_{21}^+,
\nonumber\\
     \dot {\R}^{\downarrow\uparrow}_{20}&=&
\frac{2}{m}\L_{20}^{\downarrow\uparrow}
-i\hbar\frac{\eta}{2}\sqrt{\frac{3}{2}}\R_{21}^+,
\nonumber\\
     \dot {\P}^{+}_{21}&=&
-2\left[m\,\omega^2+\sqrt6\kappa_0 R_{20}^{\rm eq}\right]\L^{+}_{21}
+6\sqrt6\kappa_0 R_{20}^{\rm eq}\L^{+}_{11}
-i\hbar\frac{\eta}{2}\left[\P_{21}^- 
+2\P^{\uparrow\downarrow}_{22}+\sqrt6\P^{\downarrow\uparrow}_{20}\right]
\nonumber\\
&&+\frac{3\sqrt{3}}{4}\hbar^2 
\chi \frac{I_2}{A_1A_2}
\left[\left(A_1-A_2\right) \L_{21}^{+} +
\left(A_1+A_2\right) \L_{11}^{+}\right]
+\frac{4}{\hbar}|V_0| I_{pp}^{\kappa\Delta}(r') {\tilde\P}_{21},
\nonumber\\
     \dot {\P}^{-}_{21}&=&
-2\left[m\,\omega^2+\sqrt6\kappa_0 R_{20}^{\rm eq}\right]\L^{-}_{21}
+6\sqrt6\kappa_0 R_{20}^{\rm eq}\L^{-}_{11}
-6\sqrt2\alpha\kappa_0 L_{10}^-(\rm eq)\R^{+}_{21}
-i\hbar\frac{\eta}{2}\P_{21}^{+}
\nonumber\\
&&+\frac{3\sqrt{3}}{4}\hbar^2 
\chi \frac{I_2}{A_1A_2}
\left[\left(A_1-A_2\right)\L_{21}^{-} +
\left(A_1+A_2\right) \L_{11}^{-}\right],
\nonumber\\
     \dot {\P}^{\uparrow\downarrow}_{22}&=&
-\left[2m\,\omega^2-4\sqrt6\kappa_0 R_{20}^{\rm eq}
-\frac{3\sqrt{3}}{2}\hbar^2 
\chi \frac{I_2}{A_2}
\right]\L^{\uparrow\downarrow}_{22}
-i\hbar\frac{\eta}{2}\P_{21}^{+}
,
\nonumber\\
     \dot {\P}^{\downarrow\uparrow}_{20}&=&
-\left[2m\,\omega^2+4\sqrt6\kappa_0 R_{20}^{\rm eq}\right]\L^{\downarrow\uparrow}_{20}
+8\sqrt3\kappa_0 R_{20}^{\rm eq}\L^{\downarrow\uparrow}_{00}
-i\hbar\frac{\eta}{2}\sqrt{\frac{3}{2}}\P_{21}^{+}
\nonumber\\
&&+\frac{\sqrt{3}}{2}\hbar^2 
\chi \frac{I_2}{A_1A_2}
\left[\left(A_1-2A_2\right)\L_{20}^{\downarrow\uparrow}+
\sqrt2\left(A_1+A_2\right) \L_{00}^{\downarrow\uparrow}
\right],
\nonumber\\
     \dot {\L}^{\downarrow\uparrow}_{00}&=&
\frac{1}{m}\P_{00}^{\downarrow\uparrow}-m\,\omega^2\R^{\downarrow\uparrow}_{00}
+4\sqrt3\kappa_0 R_{20}^{\rm eq}\,\R^{\downarrow\uparrow}_{20}
\nonumber\\
&&+\frac{\hbar^2}{2\sqrt{3}A_1 A_2} 
\left[\left( \chi-\frac{\bar\chi}{3} \right)I_1-\frac{9}{4}\chi I_2\right]
\left[\left(2A_1-A_2\right)\R_{00}^{\downarrow\uparrow}+
\sqrt2\left(A_1+A_2\right)\R_{20}^{\downarrow\uparrow}
\right],
\nonumber\\
     \dot {\R}^{\downarrow\uparrow}_{00}&=&
\frac{2}{m}\L_{00}^{\downarrow\uparrow},
\nonumber\\
     \dot {\P}^{\downarrow\uparrow}_{00}&=&
-2m\,\omega^2\L^{\downarrow\uparrow}_{00}
+8\sqrt3\kappa_0 R_{20}^{\rm eq}\,\L^{\downarrow\uparrow}_{20}
+\frac{\sqrt{3}}{2}\hbar^2 
\chi \frac{I_2}{A_1 A_2}
\left[\left(2A_1-A_2\right)\L_{00}^{\downarrow\uparrow}+
\sqrt2\left(A_1+A_2\right)\L_{20}^{\downarrow\uparrow}
\right],
 \nonumber\\
     \dot{ {\tilde\R}}_{21} &=& -\frac{1}{\hbar}\left(\frac{16}{5} \alpha\kappa_0 K_4
     +\Delta_0(r') -\frac{3}{8}\hbar^2 \chi\kappa_0(r') \right)\R^+_{21}, 
 \nonumber\\
     \dot{ {\tilde\P}}_{21} &=& -\frac{1}{\hbar}\Delta_0(r') \P^+_{21} + 
 6 \hbar\alpha\kappa_0 K_0{\cal R}^+_{21}, 
\nonumber\\
     \dot {{\tilde\L}}_{21} &=& -\frac{1}{\hbar}\Delta_0(r') {\L}^-_{21},
\nonumber\\
     \dot {{\tilde\L}}_{11} &=& -\frac{1}{\hbar}\Delta_0(r') \L^-_{11},
\label{iv}
\end{eqnarray}
where 
\begin{eqnarray}
&&A_1=\sqrt2\, R_{20}^{\rm eq}-R_{00}^{\rm eq}=\frac{Q_{00}}{\sqrt3}\left(1+\frac{4}{3}\delta\right),
\nonumber\\
&&A_2= R_{20}^{\rm eq}/\sqrt2+R_{00}^{\rm eq}=-\frac{Q_{00}}{\sqrt3}\left(1-\frac{2}{3}\delta\right).
\nonumber
\end{eqnarray}
$a_{-1} = a_1 = R_0\left( \frac{1-(2/3)\delta}{1+(4/3)\delta} \right)^{1/6}$ and
$a_0 = R_0\left( \frac{1-(2/3)\delta}{1+(4/3)\delta} \right)^{-1/3}$ 
are semiaxes of ellipsoid by which the shape of nucleus is approximated, $\delta$ -- deformation parameter, 
$R_0=1.2A^{1/3}$~fm -- radius of nucleus, $R_{\lambda\mu}^{\rm eq}\equiv R_{\lambda\mu}({\rm eq})$.
$$I_1=\frac{\pi}{4}\int\limits_{0}^{\infty}dr\, r^4\left(\frac{\partial n(r)}{\partial r}\right)^2,
\
I_2=\frac{\pi}{4}\int\limits_{0}^{\infty}dr\, r^2 n(r)^2,$$
$n(r)= n_0\left(1+{\rm e}^{\frac{r-R_0}{a}}\right)^{-1}$ -- nuclear density,
$$K_0=\int d(\br,\bp) \kappa(\br,\bp), \, K_4=\int d(\br,\bp) r^4\kappa(\br,\bp).$$
The functions $\kappa_0(r')$, $\Delta_0(r')$, $I_{rp}^{\kappa\Delta}(r')$ and 
$I_{pp}^{\kappa\Delta}(r')$
are discussed in the next section and are outlined in Appendix~\ref{AppE}.
Deriving these equations we neglected double Poisson brackets containing $\kappa$~or~$\Delta$,
which are the quantum corrections to pair correlations.
The isoscalar set of equations is easily obtained from (\ref{iv}) by 
taking $\alpha=1$, replacing $\bar\chi \to -\bar\chi$
and putting the marks "bar" above all variables.

\subsection{Integrals of motion and the angular momentum conservation}
\label{4.3}

Imposing the time evolution via $\di{e^{iEt/\hbar}}$ for all variables
one transforms (\ref{iv}) into a set of algebraic equations. 
It contains 23 equations. 
To find the eigenvalues we construct the $23\times 23$
determinant and seek (numerically) for its zeros. 
We find seven roots with exactly $E=0$ and 16 roots which are non zero: 
eight positive ones (shown in the tables) and eight negative ones (not shown, situation exactly
same as with RPA; see~\cite{Ann} for connection of WFM and RPA). 

 The integrals of motion corresponding to Goldstone modes (zero roots)
can be found analytically.
In the isovector case we have 
\begin{eqnarray}
\label{intmot}
    && i\hbar\frac{\eta}{2} \left[\L^{+}_{11}-i\frac{\hbar}{2} \F^\d\right]
-3\sqrt6(1-\alpha)\kappa_0 R_{20}^{\rm eq}
\left[
\frac{2}{\sqrt3 c_2 m}\P^\d_{00}+\frac{c_1}{\sqrt3 c_2}\R^\d_{00}-\sqrt{\frac{2}{3}}\R^\d_{20}
\right]={\rm const},
\nonumber\\
    && \P^\u_{22} - \sqrt{\frac{2}{3}}\left(\P^\d_{20}+\sqrt2 \P^\d_{00}\right)+\frac{m}{2}(c_1-c_2)
    \left[\R^\u_{22} - \sqrt{\frac{2}{3}}\left(\R^\d_{20}+\sqrt2 \R^\d_{00}\right)\right] = {\rm const},
\nonumber\\
     &&i\hbar\frac{\eta}{2} {\L}^{+}_{21}
-\hbar^2\frac{\eta^2 m}{8}\left[\R^{-}_{21}+2\R^{\u}_{22}\right]
+\sqrt{\frac{2}{3}}\left(\frac{3}{8}\hbar^2\eta^2 m-c_3\right)\R^\d_{20}
+\sqrt{\frac{2}{3}}\frac{1}{m}\P^\d_{20}
\nonumber\\
&&+\frac{1}{2\sqrt3 c_2}\left((c_1-c_2)(c_1+2c_2)+2c_1c_3-\frac{3}{2}\hbar^2\eta^2 m\right)\R^\d_{00}
\nonumber\\
&&+\frac{1}{\sqrt3 c_2 m}\left(c_1+c_2+2c_3-\frac{3}{2}\hbar^2\eta^2 m\right)\P^\d_{00}
={\rm const},
\nonumber\\
 && i\hbar\frac{3}{4}\eta c_2\tilde\L_{11} + \frac{\Delta_0(r')}{\hbar}
 \left\{
 i\hbar\frac{\eta}{2}\left[\P^-_{21}+\frac{m}{4}(2c_1+c_2)\R^-_{21}-\sqrt{\frac{2}{3}}\P^\d_{20}\right]
\right.\nonumber\\
&& - \left(i\hbar\frac{\eta}{4\sqrt2} -\frac{4\sqrt6}{mc_2}\kappa_0\alpha L_{10}^{\rm eq}\right)\P^\d_{00}
   - \left(i\hbar\frac{\eta m}{2}\sqrt{\frac{2}{3}}(2c_1+c_2) +4\sqrt3\kappa_0\alpha L_{10}^{\rm eq}\right)\R^\d_{20}
\nonumber\\
&&\left.- \left(i\hbar\frac{\eta m}{8\sqrt3}(c_1-4c_2) - 2\sqrt2\frac{c_1}{c_2}\kappa_0\alpha L_{10}^{\rm eq}\right)\R^\d_{00}
\right\} = {\rm const},
\nonumber\\
&& i\hbar\frac{\eta}{2} \tilde\R_{21}
 -\left( \frac{16}{5\hbar} \kappa_0\alpha K_4+\frac{\Delta_0(r')}{\hbar}-\frac{3}{8}\hbar\chi\kappa_0(r')\! \right)\!
  \!\!\left[\sqrt{\frac{2}{3}}\R^\d_{20} -\frac{c_1}{\sqrt3 c_2} \R^\d_{00}- \frac{2}{\sqrt3 mc_2} \P^\d_{00}\right] = {\rm const},
\nonumber\\
&& i\hbar\frac{\eta}{2} \tilde\P_{21}
 -\frac{\Delta_0(r')}{\hbar}
\left[\sqrt{\frac{2}{3}}\P^\d_{20} + \frac{2(c_1+c_2)}{\sqrt3 c_2} \P^\d_{00}+\frac{m}{2}\frac{(c_1-c_2)(c_1+2c_2)}{\sqrt3 c_2} \R^\d_{00}
\right]
\nonumber\\
&& + \,6\hbar\kappa_0\alpha K_0
 \left[\sqrt{\frac{2}{3}}\R^\d_{20} -\frac{c_1}{\sqrt3 c_2} \R^\d_{00}- \frac{2}{\sqrt3 mc_2} \P^\d_{00}\right] = {\rm const},
\nonumber\\
&& \tilde\L_{21}
 +\frac{\Delta_0(r')}{\hbar}
\left[\frac{1}{\sqrt3 c_2}\P^\d_{00} + \frac{m}{2} \left( \R^-_{21}-\sqrt{\frac{2}{3}}\R^\d_{20}+\frac{c_1}{\sqrt3 c_2}\R^\d_{00}\right)\right]
= {\rm const},
\end{eqnarray}
where
\begin{eqnarray}
&&c_1= 2m\omega^2-
\frac{\sqrt3}{2}\hbar^2 \chi I_2 \frac{\left(2A_1-A_2\right)}{A_1A_2},
\qquad c_2= 4\sqrt6\kappa_0 R_{20}^{\rm eq}+
\frac{\sqrt3}{2}\hbar^2 \chi I_2 \frac{\left(A_1+A_2\right)}{A_1A_2},
\nonumber\\
&&c_3= m\,\omega^2-4\sqrt3\alpha\kappa_0R_{00}^{\rm eq}+\sqrt6(1+\alpha)\kappa_0 R_{20}^{\rm eq}.
\nonumber
\end{eqnarray}
Isoscalar integrals of motion are easily obtained from isovector ones by taking $\alpha=1$ and putting bars above all variables. 
In the case of harmonic oscillations all constants "const" are obviously equal to zero. 

The physical sense of variables entering into above integrals of motion
can be understood with the help of their definitions (\ref{VarisV}). The
variables (or matrix elements) $\R^{ss'}_{\lambda\mu}(t)$ describe the
quadrupole ($\lambda=2$) and monopole ($\lambda=0$) deformation of the 
density of nucleons with spin $s$, if $s=s'$, otherwise they describe
the simultaneous deformation and spin flip. The variables 
$\P^{ss'}_{\lambda\mu}(t)$ describe the analogous situation in the
momentum space, i.e. the Fermi surface deformation, if $s=s'$, or the
deformation accompanied by spin flip, if $s\neq s'$. 

The variables
$\L^{ss'}_{\lambda\mu}(t)$ with $\lambda=2, 0$ describe the similar
situation in the phase space ($\br,\bp$). 
Looking on the equations of motion for $\R^{ss'}_{\lambda\mu}(t)$
and $\P^{ss'}_{\lambda\mu}(t)$ in the case without spin orbital field and
spin-spin forces one can see that, roughly speaking, these variables ($\L^{ss'}_{\lambda\mu}$) determine the velocities 
$\frac{d}{dt}\R^{ss'}_{\lambda\mu}$ and
$\frac{d}{dt}\P^{ss'}_{\lambda\mu}$ of the density deformation 
and the Fermi surface deformation respectively.

The variables 
$\L^{ss'}_{1\mu}(t)$ describe the dynamics of the orbital angular
momentum of nucleons with spin $s$, if $s=s'$, otherwise they describe
the dynamics of the orbital angular momentum together with spin flip.
The variables $\F^{ss'}(t)$ describe the dynamics of the number of 
nucleons with spin $s$, if $s=s'$, or dynamics of spin, i.e. the spin
flip, if $s\neq s'$.

Having this information we can give the physical interpretation of 
some integrals of motion.
 The first isoscalar integral is the most simple one:
 $$
2i\bar{\L}^+_{11}(t)+\hbar \bar{\F}^{\downarrow\uparrow}(t)
={\rm const}
 $$
and has a clear physical interpretation -- the conservation of the total angular
momentum 
$\langle \hat J_1\rangle =\langle \hat l_1\rangle +\langle \hat S_1\rangle $. Really, by definition
\begin{eqnarray}
\label{l1}
\langle \hat l_1\rangle &=&Tr(\hat l_1 \hat\rho)=
\sum_{\tau,s}\int d^3r\int d^3r'\langle \br|\hat l_1|\br'\rangle \langle \br',s|\hat\rho^{\tau}|\br,s\rangle 
\nonumber\\
&=&\sum_{\tau,s}\int d^3r\int d^3r'\hat l_1(\br)\delta(\br-\br')\langle \br',s|\hat\rho^{\tau}|\br,s\rangle 
=\sum_{\tau}\int d^3r\hat l_1(\br)\left[\langle \br|\hat\rho^{\tau}|\br\rangle ^{\uparrow\uparrow}+
\langle \br|\hat\rho^{\tau}|\br\rangle ^{\downarrow\downarrow}\right]
\nonumber\\
&=&\int d(\bp,\br)l_1(\br,\bp)\bar f^+(\br,\bp,t)
=-i\sqrt2\int d(\bp,\br)\{r\otimes p\}_{11}\bar f^+(\br,\bp,t)=-i\sqrt2\bar L^+_{11}(t).
\end{eqnarray}
The average value of the spin operator $\hat S_1$ reads:
\begin{eqnarray}
\label{S1}
\langle \hat S_1\rangle &=&Tr(\hat S_1 \hat\rho)=
\sum_{\tau,s,s'}\int d^3r\langle s|\hat S_1|s'\rangle \langle \br,s'|\hat\rho^{\tau}|\br,s\rangle 
\nonumber\\
&=&\sum_{s,s'}\langle s|\hat S_1|s'\rangle \int d(\bp,\br)\bar f^{s's}(\br,\bp,t)
=-\frac{\hbar}{\sqrt2}\sum_{s,s'}\delta_{s\uparrow}\delta_{s'\downarrow}\bar F^{s's}(t)
=-\frac{\hbar}{\sqrt2}\bar F^{\downarrow\uparrow}(t).
\end{eqnarray}
As a result 
$\displaystyle\langle \hat J_1\rangle =-\frac{1}{\sqrt2}\left(2i\bar L^+_{11}+\hbar \bar F^{\downarrow\uparrow}\right)$. 
It is easy to see that such a combination of the respective equations of motion in 
(\ref{iv}) is equal to zero in the isoscalar case $(\alpha=1)$,
i.e. the total angular momentum is conserved.
 The isovector counterpart of this integral of motion implies that the 
relative (neutrons with respect of protons) total angular momentum 
oscillates in phase (we recall that $\kappa_0 < 0$) with the linear 
combination of three variables 
$ \R^{\downarrow\uparrow}_{20},
\R^{\downarrow\uparrow}_{00}$ and $\P^{\downarrow\uparrow}_{00}$.

 The second integral of motion can be interpreted saying that the 
definite combination of variables 
$\left(\sqrt{\frac{3}{2}}\P^{\uparrow\downarrow}_{22}
-\P^{\downarrow\uparrow}_{20}
-\sqrt{2}\P^{\downarrow\uparrow}_{00}\right)$, 
describing the quadrupole and monopole deformations of the Fermi surface
together with the spin flip, oscillates out of phase with the exactly 
the same combination of variables 
$\left(\sqrt{\frac{3}{2}}\R^{\uparrow\downarrow}_{22}
-\R^{\downarrow\uparrow}_{20}
-\sqrt{2}\R^{\downarrow\uparrow}_{00}\right)$, 
describing the quadrupole and monopole deformations of the density
distribution together with the spin flip. It is interesting to note
that in the analogous problem without spin~\cite{BaSc} there is the 
similar integral, saying that the nuclear density and the Fermi
surface oscillate out of phase. The physical interpretation of the
third integral and the integrals 4, 5, 6, 7 appearing due to pairing, seems not to be obvious.

Let us to prove that the conservation of the total angular momentum
follows from the set of equations (\ref{quadr}), which describe the
motion without any restrictions on the values (small or large) of
amplitudes. It is necessary to consider the first equation of 
(\ref{quadr}) in the isoscalar case for $\lambda=\mu=1$. Having in mind
that $R_{11}^+=P_{11}^+=0$ we find
\begin{eqnarray}
\label{L+11}
     \dot L^{\tau +}_{11}&=&
60\left\{_{211}^{112}\right\}
\{Z_2^{\tau +}\otimes R_2^{\tau +}\}_{11}
-i\hbar\frac{\eta}{2}\left[L_{11}^{\tau -}
+\sqrt2L^{\tau\downarrow\uparrow}_{10}\right]
\nonumber\\
&-&\int\!d^3r\left[
\frac{1}{2}n_\tau^{+}\{r\otimes \nabla\}_{11}V_{\tau}^{+}+
\frac{1}{2}n_\tau^{-}\{r\otimes \nabla\}_{11}V_{\tau}^{-}
\right.\nonumber\\&&\left.
+n_\tau^{\downarrow\uparrow}\{r\otimes \nabla\}_{11}V_{\tau}^{\uparrow\downarrow}+
n_\tau^{\uparrow\downarrow}\{r\otimes \nabla\}_{11}V_{\tau}^{\downarrow\uparrow}
\right]\!.\qquad
\end{eqnarray}
Let us analyze the first term. We have for protons:
\begin{eqnarray}
\label{Z+p}
\{Z_2^{p+}\otimes R_2^{p+}\}_{11}
&=&\sum_{\nu\sigma}C^{11}_{2\nu,2\sigma}Z^{p+}_{2\nu}R^{p+}_{2\sigma}
=\sum_{\nu\sigma}C^{11}_{2\nu,2\sigma}\left(\kappa R^{p+}_{2\nu}
+\bar\kappa R^{n+}_{2\nu}\right)R^{p+}_{2\sigma}
\nonumber\\
&=&\bar\kappa \sum_{\nu\sigma}C^{11}_{2\nu,2\sigma}
R^{n+}_{2\nu}R^{p+}_{2\sigma}.
\end{eqnarray}
We have used here the definition (\ref{Z2mu}) of $Z^{\tau +}_{2\mu}$ 
and the equality $C^{11}_{2\nu,2\sigma}=-C^{11}_{2\sigma,2\nu}$. 
Analogously one finds for neutrons:
\begin{eqnarray}
\label{Z+n}
\{Z_2^{n+}\otimes R_2^{n+}\}_{11}
=\sum_{\nu\sigma}C^{11}_{2\nu,2\sigma}\left(\kappa R^{n+}_{2\nu}
+\bar\kappa R^{p+}_{2\nu}\right)R^{n+}_{2\sigma}
=\bar\kappa \sum_{\nu\sigma}C^{11}_{2\nu,2\sigma}
R^{p+}_{2\nu}R^{n+}_{2\sigma}.
\end{eqnarray}
The sum of (\ref{Z+p}) and (\ref{Z+n}) is obviously equal to zero.

The integral in (\ref{L+11}) consists of four terms. The first one is
(see the definition of $V^+_{\tau}$ in~(\ref{Vss})):
\begin{eqnarray}
\label{Vpn}
-\frac{3}{16}\hbar^2\chi\int\!d^3r\,
n_{\tau}^+C^{11}_{11,10}\left[r_1\nabla_0-r_0\nabla_1\right]n_{\tau}^+
=-\frac{3}{32}\hbar^2\chi\int\!d^3r\,
C^{11}_{11,10}\left[r_1\nabla_0-r_0\nabla_1\right](n_{\tau}^+)^2.
\end{eqnarray}
Integrating by parts we find that this integral is equal to zero because
$\nabla_1r_0=\nabla_0r_1=0$. The second term of integral in (\ref{L+11})
can be written (for protons) as
\begin{eqnarray}
\label{Vpn2}
\frac{\hbar^2}{8}\int\!d^3r\,
n_p^{-}C^{11}_{11,10}\left[r_1\nabla_0-r_0\nabla_1\right]
\left(\frac{3}{2}\chi n_p^{-}+\bar\chi n_n^{-}\right)
=\frac{\hbar^2}{8}\bar\chi \int\!d^3r\,
C^{11}_{11,10}n_p^{-}\left[r_1\nabla_0-r_0\nabla_1\right]
n_n^{-}.\quad
\end{eqnarray}
Changing here the indices $p\leftrightarrow n$ we obtain the analogous
integral for neutrons. Their sum is obviously equal to zero.

The third and fourth terms of integral in (\ref{L+11}) must be analyzed
together. We have for protons:
\begin{eqnarray}
\label{Vpn3}
\frac{\hbar^2}{4}C^{11}_{11,10}\int\!d^3r
\left[n_p^{\downarrow\uparrow}\left(r_1\nabla_0-r_0\nabla_1\right)
\left(\frac{3}{2}\chi n_p^{\uparrow\downarrow}+\bar\chi n_n^{\uparrow\downarrow}\right)
+n_p^{\uparrow\downarrow}\left(r_1\nabla_0-r_0\nabla_1\right)
\left(\frac{3}{2}\chi n_p^{\downarrow\uparrow}+\bar\chi n_n^{\downarrow\uparrow}\right)\right]
\nonumber\\
=\frac{\hbar^2}{4}C^{11}_{11,10}\,\bar\chi\int\!d^3r
\left[n_p^{\downarrow\uparrow}\left(r_1\nabla_0-r_0\nabla_1\right) n_n^{\uparrow\downarrow}
+n_p^{\uparrow\downarrow}\left(r_1\nabla_0-r_0\nabla_1\right) n_n^{\downarrow\uparrow}\right].\quad
\end{eqnarray}
The sum of this integral with the analogous one for neutrons (which 
is obtained by changing indices $p\leftrightarrow n$) is obviously 
equal to zero.

So, finally we have found that the isoscalar variant of equation (\ref{L+11})
can be written (in variables defined in (\ref{Isovs})) as
\begin{eqnarray}
\label{L+11f}
\dot L^{p+}_{11}(t)+\dot L^{n+}_{11}(t)\equiv \dot {\bar L}^{+}_{11}(t)=
-i\hbar\frac{\eta}{2}\left[\bar L_{11}^{-}(t)
+\sqrt2\bar L^{\downarrow\uparrow}_{10}(t)\right].
\end{eqnarray}
It is easy to see that the proper combination of this equation with 
the seventh equation in (\ref{quadr}) gives the required result:
$$-\frac{1}{\sqrt2}\left(2i\dot {\bar L}^+_{11}+\hbar \dot {\bar F}^{\downarrow\uparrow}\right)=
\frac{d}{dt}\langle \hat J_1\rangle =0,$$
i.e. the total angular momentum is conserved for arbitrary amplitudes,
not only in a small amplitude approximation.
One must note that this result is not influenced by the approximate
treatment of integral terms in (\ref{quadr}).

\section{Results of calculations}

\subsection{Choice of parameters}\label{VA}
 
$\bullet$ Following our previous publications~\cite{BaSc,Ann} we take for the isoscalar strength 
constant of the quadrupole-quadrupole residual interaction $\kappa_0$ the self consistent 
value~\cite{BrMt} $\kappa_0=-m\bar\omega^2/(4Q_{00})$ with
$Q_{00}=\frac{3}{5}AR_0^2$, $\bar\omega^2=\omega_0^2/
[(1+\frac{4}{3}\delta)^{2/3}(1-\frac{2}{3}\delta)^{1/3}]$,
$\hbar\omega_0=41/A^{1/3}$~MeV. 

\begin{figure}
\centering\includegraphics[width=0.5\textwidth]{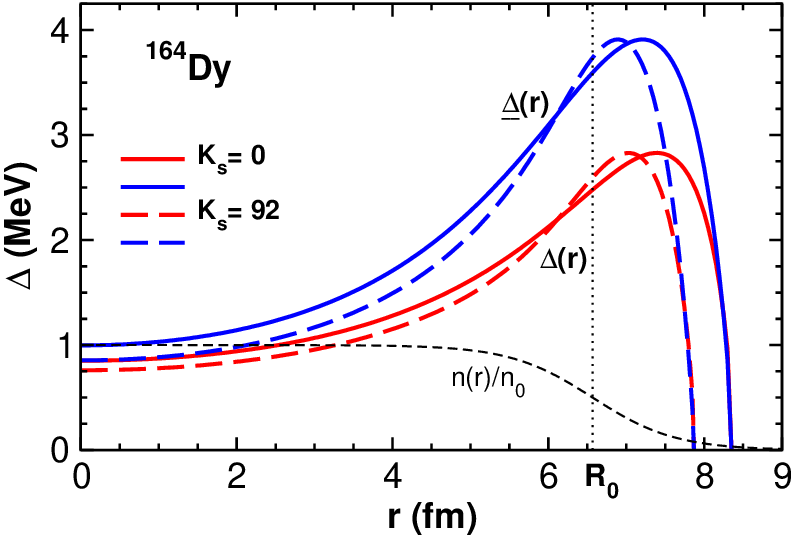}
\caption{The pair field (gap) $\Delta_0(r)$, the function 
$\underline{\Delta}(r)=|V_0|I_{pp}^{\kappa\Delta}(r)$ and the nuclear density $n(r)$ 
as the functions of radius $r$. The solid lines -- calculations without the spin-spin 
interaction $H_{ss}$, the dashed lines -- $H_{ss}$ is included.}
\label{fig1}\end{figure}

$\bullet$
The equations (\ref{iv}) contain the functions 
${\di\Delta_0(r')\equiv\Delta_{\rm eq}(r',p_F(r'))}$, 
$I_{rp}^{\kappa\Delta}(r')\equiv I_{rp}^{\kappa\Delta}(r',p_F(r'))$, 
$I_{pp}^{\kappa\Delta}(r')\equiv I_{pp}^{\kappa\Delta}(r',p_F(r'))$ 
and $\kappa_0(r')\equiv\kappa(r',r')$ 
depending on the radius $r'$  and the local Fermi momentum $p_F(r')$ 
(see Fig.~\ref{fig1} ).
The value of~$r'$ is not fixed by the theory and can be used 
as the fitting parameter. We have found in our previous paper~\cite{Urban} that the best agreement
of calculated results with experimental data is achieved at the point~$r'$ where the function
$I_{pp}^{\kappa\Delta}(r',p_F(r'))$ has its maximum. Nevertheless, to get rid off the fitting 
parameter, we use the averaged values of these functions:
$\bar\Delta_0=\int d\br\, n(\br)\Delta_0(r,p_F(r))/A$, etc.
The gap $\Delta(r,p_F(r))$, as well as the integrals $I^{\kappa\Delta}_{pp}(r,p_F(r))$, $K_4$
and $K_0$,
were calculated with the help of the semiclassical 
formulae for $\kappa(\br,\bp)$ and $\Delta(\br,\bp)$ (see
Appendix~\ref{AppE}), a Gaussian being
used for the pairing interaction with $r_p=1.9$ fm and $V_0=25$
MeV~\cite{Ring}. Those values reproduce usual nuclear pairing gaps.

$\bullet$
The used spin-spin interaction is repulsive, the values of its
strength constants being taken from the paper~\cite{Moya}, where the 
notation $\chi=K_s/A,\,\bar\chi=q\chi$ was introduced.
The constants were extracted by the authors of~\cite{Moya} 
from Skyrme forces following the standard procedure, the residual interaction 
being defined in terms of second derivatives of the Hamiltonian density 
$H(\rho)$ with respect to the one-body densities~$\rho$.
Different variants of Skyrme forces produce different strength constants of the
spin-spin interaction. The most consistent results are obtained with 
SG1, SG2~\cite{Giai} and Sk3~\cite{Floc} forces. 
To compare theoretical results with experiment the authors of~\cite{Moya} preferred 
to use the force SG2. Nevertheless they have noticed that "As is well known, the energy 
splitting of the HF states around the Fermi level is too large. This has an effect on the
spin $M1$ distributions that can be roughly compensated by reducing the $K_s$ value". According 
to this remark they changed the original self-consistent SG2 parameters from $K_s=88$~MeV, 
$q=-0.95$ to $K_s=50$~MeV, $q=-1$. It was  found that this modified set of parameters
gives better agreement with experiment for some nuclei in the description of spin-flip
resonance. So we will use \mbox{$K_s=50$~MeV and $q=-1$}.

$\bullet$
Our calculations without pairing~\cite{BaMoPRC} have shown that the results for scissors modes are strongly dependent on the values  of 
the strength constants of the spin-spin interaction. The natural question arises: how sensitive are they to the strength of the spin-orbital potential? 
\begin{figure}
\centering\includegraphics[width=\textwidth]{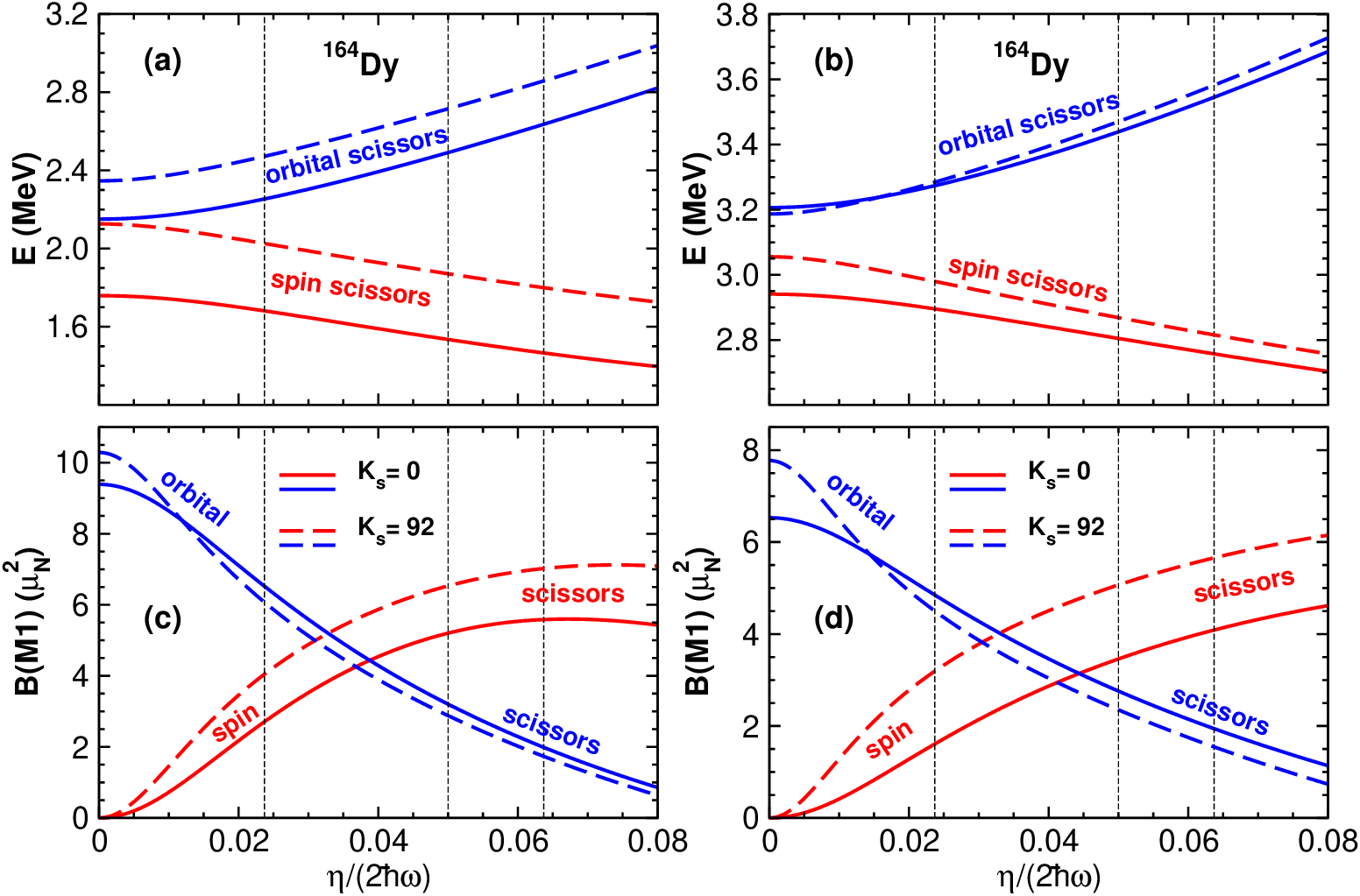}
\caption{The energies $E$ (a, b) and $B(M1)$-factors (c, d) as a functions of the spin-orbital strength 
constant~$\eta$. 
Left panel: solid lines -- without the spin-spin interaction $H_{ss}$, dashed lines -- $H_{ss}$ is included.
Right panel: The same as in left panel with pair correlations included.}
\label{fig2}\end{figure}

The results of the demonstrative calculations 
are shown in~Fig.~\ref{fig2}.
The $M1$ strengths were computed using effective spin giromagnetic factors 
$g_s^{\rm eff}=0.7g_s^{\rm free}$. One observes a rather strong dependence of the results
on the value of $\eta$: the splitting $\Delta E$ and the $M1$ strength of the spin scissors grow with 
increasing $\eta$, the $B(M1)$ of the orbital scissors being decreased.
At some critical point $\eta_c$ the $M1$ strength of the spin scissors becomes bigger than that of the
orbital scissors. 

The inclusion of the spin-spin interaction does not change the qualitative picture,
as well as the inclusion of pair correlations (see Fig.~\ref{fig2}).
Nevertheless, it is necessary to note, that the spin-spin interaction decreases remarkably the value of $\eta_c$, whereas pair correlations 
slightly increase it.

What value of $\eta$ to use?
 Accidentally, the choice of $\eta$ in our papers~\cite{BaMo,BaMoPRC} was not very realistic. 
The main purpose  of the first paper was the introduction 
of spin degrees of freedom into the WFM method, and the aim of the second paper
was to study the influence of spin-spin forces on both scissors -- we did not worry much about the comparison
with experiment. Now, both preliminary aims being achieved, one can think about the agreement with
experimental data, therefore the precise choice of the model parameters becomes important. Of course,
we could try to choose $\eta$ according to the standard requirement of the best agreement with experiment.
 However, in reality we are not absolutely free in our choice.
It turns out that we are already restricted by the other constraints. As a matter of fact
we work with the Nilsson potential, parameters of which are very well known. Really, the mean field of our
model (\ref{Ham}) is the deformed harmonic oscillator with the spin-orbit potential, the Nilsson 
$\ell^2$ term being neglected because it generates the fourth order moments and, anyway, they are probably not of great importance. 
In the original paper~\cite{Nils} Nilsson took the spin-orbit strength constant $\kappa_{\rm Nils}=0.05$ for rare earth nuclei. 
Later the best value of $\kappa_{\rm Nils}$
for rare earth nuclei was established~\cite{Ring} to be $0.0637$. For actinides there were established 
different values of $\kappa_{\rm Nils}$ for neutrons ($0.0635$) and protons ($0.0577$).
The numbers $\kappa_{\rm Nils}=0.0637$, $\kappa_{\rm Nils}=0.05$ 
and $\kappa_{\rm Nils}=0.024$ (corresponding to $\eta=0.36$ used in our calculations~\cite{BaMo,BaMoPRC}) 
are marked on~Figs~\ref{fig2},~\ref{figL} by 
the dotted vertical lines. Of course we will use the
conventional~\cite{Ring} parameters of the Nilsson potential and
from now on we will speak only about the Nilsson~\cite{Nils} spin-orbital strength parameter 
$\kappa_{\rm Nils}$, which is connected with~$\eta$ by the relation $\kappa_{\rm Nils}=\eta/2\hbar\omega$. 

\subsection{Discussion and interpretation of the results without pairing}

The dependence of energies and excitation probabilities on the spin-spin
interaction is demonstrated in Table~\ref{tab1} (isovector) and
in Table~\ref{tab2} (isoscalar). These results are obtained by the solution of the set of equations (\ref{iv}) without pairing.

\begin{table*}
\caption{\label{tab1}
 Isovector energies and excitation probabilities 
of $^{164}$Dy. Deformation parameter $\delta=0.26$, spin-orbit constant $\kappa_{\rm Nils}=0.0637/\hbar^2$. 
Spin-spin interaction constants are: I -- $K_s=0$~MeV/$\hbar^2$;
II -- $K_s=50$~MeV/$\hbar^2$, $q=-0.5$;
 III~--~$K_s=92$~MeV/$\hbar^2$, $q=-0.8$.
Quantum numbers (including indices $\varsigma = +,\,-,\,
\uparrow\downarrow,\,\downarrow\uparrow$) of variables responsible for 
the generation of the present level are shown in the first column. 
For example: $(1,1)^-$ -- spin scissors, $(1,1)^+$ -- conventional scissors, etc..
The numbers in the last line are imaginary, so they are marked by 
the letter i.}
\begin{tabular}{l|c|c|c|c|c|c|c|c|c}
\hline
 $(\lambda,\mu)^\varsigma$ &   
\multicolumn{3}{c|}{ $E_{\rm iv}$,~MeV}      &
\multicolumn{3}{c|}{ $B(M1)\!\!\uparrow,\ \mu_N^2$}      &
\multicolumn{3}{c}{ $B(E2)$, W.u.}        \\
\cline{2-10}
  & I & II & III & I & II & III & I & II & III \\ 
\hline
 (1,1)$^-$                    & ~1.47  & ~1.67  & ~1.80 & ~5.58 & ~6.45 & ~7.02 & ~0.14 & ~0.21 & ~0.28 \\   
 (1,1)$^+$                    & ~2.64  & ~2.76  & ~2.86 & ~1.99 & ~1.82 & ~1.73 & ~0.90 & ~1.01 & ~1.11 \\ 
 (0,0)$^{\downarrow\uparrow}$ & 12.66  & 14.60  & 15.87 & ~0.05 & ~0.07 & ~0.09 & ~0.27 & ~0.50 & ~0.72 \\  
 (2,1)$^-$                    & 14.45  & 16.34  & 17.66 & ~0.06 & ~0.08 & ~0.10 & ~0.22 & ~0.37 & ~0.58 \\  
 (2,0)$^{\downarrow\uparrow}$ & 16.10  & 18.22  & 19.62 & ~0.12 & ~0.18 & ~0.36 & ~1.06 & ~1.88 & ~4.29 \\  
 (2,2)$^{\uparrow\downarrow}$ & 16.25  & 19.32  & 21.21 & ~0    & ~0.09 & ~0.82 & ~0    & ~1.00 & 10.55 \\   
 (2,1)$^+$                    & 20.78  & 21.26  & 21.88 & ~2.83 & ~2.49 & ~1.40 & 34.33 & 31.49 & 18.58 \\
 (1,0)$^{\downarrow\uparrow}$ & i0.70  & i0.69  & i0.69 & -i9.8 & -i9.7 & -i9.7 & i0.04 & i0.04 & i0.04\\
\hline
\end{tabular}
\end{table*}
To avoid misunderstanding we want to recall here that quantum numbers 
of all levels are $K^{\pi}=1^+$ (the projection
of the total angular momentum and parity). The first columns of
tables~\ref{tab1} and~\ref{tab2} demonstrate the labels ($\lambda, \mu$ and spin 
projections $\uparrow,\downarrow$) of variables which
are responsible (approximately, because all equations
are coupled) for the generation of the corresponding eigenvalue. 

One can see from Table~\ref{tab1} that the spin-spin interaction does not change
the qualitative picture of the positions of excitations described 
in~\cite{BaMo}. It pushes all levels
up proportionally to its strength (2-20\% in the case II and 5-30\%
in the case III) without changing their order. The most interesting
result concerns the relative $B(M1)$ values of the two low lying scissors modes,
namely the spin scissors $(1,1)^-$ and the conventional (orbital) 
scissors $(1,1)^+$ mode. As can be noticed, the spin-spin interaction 
strongly redistributes M1 strength in the favour of the spin scissors 
mode. We tentatively want to link this fact to the recent experimental
finding in isotopes of Th and Pa~\cite{Siem}. The authors have studied 
deuteron and $^3$He-induced reactions on $^{232}$Th and found in the 
residual nuclei $^{231,232,233}$Th and $^{232,233}$Pa
"an unexpectedly strong integrated strength of $B(M1)=11-15~\mu_N^2$ in the 
$E_\gamma=1.0-3.5$~MeV region". The $B(M1)$ force in most nuclei shows 
evident splitting into two Lorentzians. "Typically, the experimental
splitting is $\Delta\omega_{M1}\sim 0.7$~MeV, and the ratio of the 
strengths between the lower and upper resonance components is
$B_L/B_U\sim 2$". (Note a misprint in that paper: it is written 
erroneously $B_2/B_1\sim 2$ whereas it should be $B_1/B_2\sim 2$. 
To avoid misunderstanding, we write here $B_L$ instead of $B_1$ 
and $B_U$ instead of $B_2$.) The authors have tried to explain the 
splitting by a $\gamma$-deformation. To describe the observed value of
$\Delta\omega_{M1}$ the deformation $\gamma\sim 15^\circ$ is required, 
that leads to the ratio $B_L/B_U\sim 0.7$ in an obvious contradiction 
with experiment. The authors conclude that "the splitting may be 
due to other mechanisms". In this sense, we tentatively may argue as
follows. On one side, theory~\cite{Gogny} and experiment~\cite{Korten}
give zero value of $\gamma$-deformation for $^{232}$Th. On the other 
side, it is easy to see that our theory suggests the required mechanism. 
The calculations performed for $^{232}$Th give 
$\Delta\omega_{M1}\sim 0.32$~MeV and $B_L/B_U\sim 1.6$ for the first 
variant of the spin-spin interaction and 
$\Delta\omega_{M1}\sim 0.28$~MeV and $B_L/B_U\sim 4.1$ for second one 
in reasonable agreement with experimental values. 

The general picture of the influence of the spin-spin interaction on isoscalar energies and excitation probabilities (Table~\ref{tab2})
 is quite close to that observed in the isovector
case. The only difference is the low lying mode marked by $(1,1)^+$
which is practically insensitive to the spin-spin interaction.
In ref~\cite{Siem} the assignment of the resonances to be of isovector type 
is only tentative based on the assumption that at such low energies there 
is no  collective mode other than the isovector scissors mode. However, 
from~\cite{Siem} one cannot exclude that also an isoscalar spin scissors mode 
is mixed in. From our analysis we see that the isoscalar spin scissors 
(the level $(1,1)^-$) where all nucleons with spin up counter-rotate with respect the ones of spin down
comes more or less at the same energy as the isovector scissors. So it 
would be very important for the future to pin down precisely the quantum 
numbers of the resonances.

\begin{table*}
\caption{\label{tab2} The same as in Table~\ref{tab1}, but for isoscalar excitations.}
\begin{tabular}{l|c|c|c|c|c|c|c|c|c}
\hline
 $(\lambda,\mu)^\varsigma$ &   
\multicolumn{3}{c|}{ $E_{\rm is}$,~MeV}      &
\multicolumn{3}{c|}{ $B(M1)\!\!\uparrow,\ \mu_N^2$}      &
\multicolumn{3}{c}{ $B(E2)$, W.u.}        \\
\cline{2-10}
  & I & II & III & I & II & III & I & II & III \\ 
\hline
 (1,1)$^+$                    & ~0.90  & ~0.91  & ~0.91  & -0.22 & -0.22 & -0.22 & 37.51 & 39.76 & 39.76 \\
 (1,1)$^-$                    & ~1.87  & ~2.00  & ~2.00  & ~0.13 & ~0.12 & ~0.12 & ~9.21 & ~7.11 & ~7.11 \\
 (2,1)$^+$                    & ~9.99  & 10.75  & 10.75  & ~0    & ~0    & ~0    & 62.09 & 58.96 & 58.96 \\
 (0,0)$^{\downarrow\uparrow}$ & 12.89  & 14.04  & 14.04  & ~0    & ~0    & ~0    & ~3.40 & ~2.61 & ~2.61 \\
 (2,1)$^-$                    & 14.54  & 15.77  & 15.77  & ~0    & ~0    & ~0    & ~0.77 & ~0.60 & ~0.60 \\
 (2,2)$^{\uparrow\downarrow}$ & 16.25  & 17.69  & 17.69  & ~0    & ~0    & ~0    & ~0    & ~0.44 & ~0.44 \\
 (2,0)$^{\downarrow\uparrow}$ & 16.37  & 17.90  & 17.90  & ~0    & ~0    & ~0    & ~1.31 & ~0.49 & ~0.49 \\
 (1,0)$^{\downarrow\uparrow}$ & i0.54  & i0.54  & i0.54  & i0.10 & i0.10 & i0.10 & i14.3 & i14.0 & i14.0 \\
\hline
\end{tabular}
\end{table*}

Let us discuss in more detail the nature of the predicted excitations. 
As one sees, the generalization of
the WFM method by including spin dynamics allowed one to
reveal a variety of new types of nuclear collective motion involving 
spin degrees of freedom. Two isovector and two isoscalar low
lying eigenfrequencies and five isovector and five isoscalar high 
lying eigenfrequencies have been found. 

Three low lying levels correspond to the excitation of new types of 
modes. For example the isovector level marked by $(1,1)^-$ describes
rotational oscillations of nucleons with the spin projection "up"
with respect of nucleons with the spin projection "down", i.e. one
can talk of a nuclear spin scissors mode. Having in mind that this 
excitation is an isovector one, we can see that the resulting motion 
looks rather complex -- proton spin scissors counter-rotates with 
respect to the neutron spin scissors (see Fig.~\ref{fig0}). Thus the experimentally 
observed group of 1$^+$ peaks in the interval 2-4~MeV, associated 
usually with the nuclear scissors mode, in reality consists of the 
excitations of the "spin" scissors mode together with the conventional
\cite{Heyd} scissors mode (the level $(1,1)^+$ in our case).
The isoscalar level $(1,1)^-$ describes the real spin 
scissors mode: all spin up nucleons (protons together with neutrons)
oscillate rotationally out of phase with all spin down 
nucleons.

Such excitations were, undoubtedly, produced implicitly
by other methods (e.g. RPA~\cite{Heyd,Pena,Moya,Oster}), but they never 
were analyzed in such terms. It is interesting to note, for example,
that in~\cite{Pena} the scissors mode was analyzed in 
so-called spin and orbital components. Roughly speaking there are two groups of 
states corresponding to these two types of components, not completely 
dissimilar to our finding. Whereas the nature of the orbital, i.e. conventional 
scissors is quite clear, the authors did not analyze the character of their 
states which consist of the spin component. It can be speculated that those 
spin components just correspond to the isovector spin scissors mode discussed 
in our work here. It would be interesting to study whether our suggestion is 
correct or not. This could for example be done in analyzing the current 
patterns.

One more new low lying mode (isoscalar at 0.90~MeV, marked by 
$(1,1)^+$) is generated by 
the relative motion of the orbital angular momentum and spin of the nucleus. 
They can change their absolute values and directions keeping the total spin 
unchanged. If there was not the spin-orbit coupling, orbital angular momentum 
and spin would be constants of motion separately, see dynamical equations
for ${\L}^{+}_{11}$ (the orbital angular momentum variable) and 
${\F}^{\downarrow\uparrow}$ (the spin variable) in the isoscalar variant of
the set of equations (\ref{iv}). Apparently spin-orbit
force is too weak to lift up this zero mode strongly. Physically it is 
quite understandable that such a mode can exist. We want to call it 
'collective spin-orbit mode'. 
Another question is whether such a collective spin-orbit mode can be 
excited experimentally. In any case, to our knowledge a low lying 
mode of this type with a 
strong $B(E2)$ has so far not been identified experimentally. On the 
other hand the negligibly small negative $B(M1)$ value probably has to 
do with the 
approximate treatment of integrals in the equations of 
motion (\ref{quadr}) (especially the neglect by the terms generating
fourth order moments, see appendix~\ref{AppA}).

In order to complete the picture of the low-lying states, it is important to 
discuss the state which is slightly imaginary. Let us first state that the 
nature of this state has nothing to do with neither spin scissors nor with 
conventional scissors. It can namely be seen from the structure of our 
equations that this state corresponds to a spin flip induced by the spin-orbit 
potential. Such a state is of purely quantal character and it cannot be hoped 
that we can accurately describe it with our WFM approach restricting 
the consideration by second order moments only.
 For its correct treatment, we certainly should consider higher moments
like fourth order moments, for instance. The spin-orbit 
potential is the only term in our theory which couples  the second order 
moments to the fourth order ones. As mentioned, we decoupled the system in 
neglecting the fourth order moments. Therefore, it is no surprise that this 
particular spin flip mode is not well described. Nevertheless, one may try to 
better understand the origin of this mode almost at zero energy. For this, we 
make the following approximation of our diagonalisation procedure to get the 
eight eigenvalues listed in Table~\ref{tab1}. We neglect in (\ref{iv}) all 
couplings between the set of variables 
$X^+_{\lambda\mu}, X^-_{\lambda\mu}$ and the set of variables
$X^{\uparrow\downarrow}_{\lambda\mu}, X^{\downarrow\uparrow}_{\lambda\mu}$ ($X\equiv\L,\R,\P,\F$).
To this end in the dynamical equations for 
$X^+_{\lambda\mu}, X^-_{\lambda\mu}$ we omit all terms containing
$X^{\uparrow\downarrow}_{\lambda\mu}, X^{\downarrow\uparrow}_{\lambda\mu}$
and in the dynamical equations for
$X^{\uparrow\downarrow}_{\lambda\mu}, X^{\downarrow\uparrow}_{\lambda\mu}$
we omit all terms containing $X^+_{\lambda\mu}, X^-_{\lambda\mu}$.
In such a way we get two independent sets of dynamical equations. The
first one (for $X^+_{\lambda\mu}, X^-_{\lambda\mu}$) was already 
studied in~\cite{BaMo}, where we have found that such approximation
gives satisfactory (in comparison with the exact solution) results but
must be used cautiously because of the problems with the angular 
momentum conservation. The second set of equations (for 
$X^{\uparrow\downarrow}_{\lambda\mu}, X^{\downarrow\uparrow}_{\lambda\mu}$)
splits into three independent subsets. Two of them were already analyzed
in~\cite{BaMo} (it turns out that these subsets can be obtained also
in the limit $\eta\to 0$, which was studied there), where it
was shown that the results of approximate calculations are very close
to that of exact calculations, i.e. the coupling between the respective
variables 
$X^{\uparrow\downarrow}_{\lambda\mu}, X^{\downarrow\uparrow}_{\lambda\mu}$
and $X^+_{\lambda\mu}, X^-_{\lambda\mu}$ is very weak. The only new
subset of equations reads:
\begin{eqnarray}
\label{LF}
     \dot {\L}^{\downarrow\uparrow}_{10}&=&
-\hbar^2\frac{\eta}{2\sqrt2}
\F^{\downarrow\uparrow},
\nonumber\\
     \dot {\F}^{\downarrow\uparrow}&=&
-\eta\sqrt2\L^{\downarrow\uparrow}_{10}.
\end{eqnarray}
The solution of these equations is  ${\di E =i\frac{\hbar}{\sqrt2}\eta = i\,0.676}$
what practically coincides with 
the number of the full diagonalisation. So the non-zero (purely imaginary) 
value of this root only comes from the fact that z-component of orbital angular 
momentum is not conserved (only total spin J is conserved). However, the 
violation of the conservation of orbital angular momentum is very small as can 
be seen from the numbers. In any case, we see that this spin flip state has 
nothing to do with neither the spin scissors nor with the conventional 
scissors.

Two high lying excitations of a new nature are found. They are
marked by $(2,1)^-$ and following the paper~\cite{Oster} can be 
called spin-vector giant quadrupole
resonances. The isovector one corresponds to the following quadrupole
motion: the proton system oscillates out of phase with the neutron 
system, whereas inside of each system spin up nucleons oscillate out 
of phase with spin down nucleons. The respective isoscalar resonance 
describes out of phase oscillations of all spin up nucleons
(protons together with neutrons) with respect of all spin down 
nucleons.

Six high lying modes can be interpreted 
as spin-flip giant monopole (marked by $(0,0)^{\downarrow\uparrow}$)
and quadrupole (marked by $(2,0)^{\downarrow\uparrow}$ and 
$(2,2)^{\uparrow\downarrow}$) resonances.

\subsection{Discussion of results with pairing}

The results of calculations with pairing taken into account are compared
with that of without pairing in Tables~\ref{tab1x} (isovector) and~\ref{tab2x} (isoscalar).
\begin{table*}
\caption{\label{tab1x}
 Isovector energies and excitation probabilities 
of $^{164}$Dy. Deformation parameter \mbox{$\delta=0.26$}, spin-orbit constant $\kappa_{\rm Nils}=0.0637$, spin-spin interaction constants are
 $K_s=50$~MeV, $q=-0.5$. Results: a -- without pair correlations, b -- with pair correlations. The notation of the first column is the same as
 in Table~\ref{tab1}.}
\begin{tabular}{l|c|c|c|c|c|c}
\hline
 $(\lambda,\mu)^\varsigma$ &   
\multicolumn{2}{c|}{ $E_{\rm iv}$,~MeV}      &
\multicolumn{2}{c|}{ $B(M1)\!\!\uparrow,\ \mu_N^2$} &
\multicolumn{2}{c }{ $B(E2)$, W.u.}        \\
\cline{2-7}
  & a  &  b  &  a  &  b  & a  &  b  \\ 
\hline
 (1,1)$^-$                    & ~1.67  & ~2.80  & ~6.45 & ~4.98 & ~0.21 & ~0.52  \\   
 (1,1)$^+$                    & ~2.76  & ~3.57  & ~1.82 & ~1.71 & ~1.01 & ~1.94  \\ 
 (0,0)$^{\downarrow\uparrow}$ & 14.60  & 14.60  & ~0.07 & ~0.07 & ~0.50 & ~0.47  \\  
 (2,1)$^-$                    & 16.34  & 16.50  & ~0.08 & ~0.08 & ~0.37 & ~0.38  \\  
 (2,0)$^{\downarrow\uparrow}$ & 18.22  & 18.23  & ~0.18 & ~0.17 & ~1.88 & ~1.76  \\  
 (2,2)$^{\uparrow\downarrow}$ & 19.32  & 19.32  & ~0.09 & ~0.08 & ~1.00 & ~0.93  \\   
 (2,1)$^+$                    & 21.26  & 21.33  & ~2.49 & ~2.47 & 31.49 & 31.32  \\
 (1,0)$^{\downarrow\uparrow}$ & i0.69  & i0.59  & -i9.7 & -i5.4 & i0.04 & -i0.02 \\
 \hline
\end{tabular}
\end{table*}
\begin{table*}
\caption{\label{tab2x} The same as in Table~\ref{tab1x}, but for isoscalar excitations.}
\begin{tabular}{l|c|c|c|c|c|c}
\hline
 $(\lambda,\mu)^\varsigma$ &   
\multicolumn{2}{c|}{ $E_{\rm is}$,~MeV}      &
\multicolumn{2}{c|}{ $B(M1)\!\!\uparrow,\ \mu_N^2$} &
\multicolumn{2}{c }{ $B(E2)$, W.u.}        \\
\cline{2-7}
  & a  &  b  &  a  &  b  & a  &  b  \\ 
\hline
 (1,1)$^+$                    & ~0.91  & ~1.55  & -0.22 & -0.08 & 39.76 & 50.97 \\
 (1,1)$^-$                    & ~2.00  & ~2.97  & ~0.04 & ~0.04 & ~7.11 & ~2.14 \\
 (2,1)$^+$                    & 10.75  & 11.05  & ~0    & ~0    & 58.96 & 53.47 \\
 (0,0)$^{\downarrow\uparrow}$ & 14.04  & 14.05  & ~0    & ~0    & ~2.61 & ~2.90 \\
 (2,1)$^-$                    & 15.77  & 15.93  & ~0    & ~0    & ~0.60 & ~0.58 \\
 (2,2)$^{\uparrow\downarrow}$ & 17.69  & 17.69  & ~0    & ~0    & ~0.44 & ~0.45 \\
 (2,0)$^{\downarrow\uparrow}$ & 17.90  & 17.90  & ~0    & ~0    & ~0.49 & ~0.52 \\
 (1,0)$^{\downarrow\uparrow}$ & i0.54  & i0.49  & i0.10 & i0.02 & i14.0 & i6.93 \\ 
\hline
\end{tabular}
\end{table*}
As it was expected the energies of all four low lying modes increased approximately
by 1 Mev after inclusion of pairing. The behaviour of transition probabilities turned out less 
predictable. The $B(M1)$ value of the spin scissors decreased approximately by 1.5 $\mu_N^2$, whereas
$B(M1)$ value of the orbital scissors turned out practically insensitive to the inclusion of pair correlations, $B(E2)$ values of both isovector
scissors increased two times. 

The $B(M1)$ value of the isoscalar scissors (excitation $(1,1)^-$) does not feel pairing, whereas its $B(E2)$ value decreased three times. 
Rather interesting is the situation with $B(M1)$ value of the isoscalar excitation $(1,1)^+$.
It has small negative value because the appropriate linear response is
calculated not enough accurately, that in its turn is explained by the
neglect of the fourth order moments. It is seen that taking into account
pairing makes the role of this error less important, reducing this
(ridiculous) negative value three times. It is necessary to note also the
remarkable ($\sim 30\%$) increase of $B(E2)$ value of this mode.

The influence of pairing on energies and excitation probabilities of all
high lying modes is negligible. The more or less remarkable change happens
with isoscalar GQR $(2,1)^+$ only -- its energy increased by 0.3~MeV 
($\sim 3\%$), whereas its $B(E2)$ value decreased about $9\%$.

We are interested mostly in the scissors modes. Let us
compare the summed \mbox{$B(M1)_{\Sigma}= B(M1)_{\rm or}+B(M1)_{\rm sp}$} values and the centroid
of both scissors energies
$$E_{\rm cen}=[E_{\rm or}B(M1)_{\rm or}+E_{\rm sp}B(M1)_{\rm sp}]/B(M1)_{\Sigma}$$
with the results of the paper~\cite{Urban} where no spin degrees of freedom had been considered and with the experimental data. 
The respective results are shown in the Table~\ref{tab12}. 
\begin{table*}[h!]
\caption{\label{tab12} Scissors modes energy centroid $E_{\rm cen}$ and summarized
transition probabilities $B(M1)_{\Sigma}$. 
The experimental values
of $E_{\rm cen}$ and $B(M1)_{\Sigma}$ are from~\cite{Pietr95,Pietr98}.}
\begin{tabular}{c|c|c|c|c|c|c}\hline
 $^{164}$Dy  & \multicolumn{3}{c|}{ $E_{\rm cen}$,~MeV} & \multicolumn{3}{c}{ $B(M1)_{\Sigma},\ \mu_N^2$} \\
\cline{2-7}
    &
 $K_{\rm s}=50$&  Ref.~\cite{Urban} & Exp. &
 $K_{\rm s}=50$&  Ref.~\cite{Urban} & Exp.  \\
\hline
$\bar\Delta_0=0$     & 1.91 & 2.17 &      & 8.27 & 9.59 &       \\
                     &      &      & 3.14 &      &      & 3.18  \\
$\bar\Delta_0\neq 0$ & 2.99 & 3.60 &      & 6.69 & 5.95 &       \\
\hline
\end{tabular}
\end{table*}
It is seen that the inclusion of spin degrees of freedom in the WFM
method does not change markedly our results (in comparison with previous ones
\cite{Urban}). Of course, the energy changed in the desired direction and now practically 
coincides with the experimental value. However,
the situation with the $B(M1)$ values did not change (and even becomes worse). Our hope, that spin degrees of freedom can improve
the situation with the  $B(M1)$ values, did not become true: the theory so far gives two-times-bigger values of 
$B(M1)$ than the experimental ones, exactly as it was the case in the paper~\cite{Urban}.

The result look discouraging. However,
a phenomenon, which was missed in our previous papers and  described in the next
section will save the situation.

\section{Counter-rotating angular momenta of spins up/down
(hidden angular momenta)}

The equilibrium (ground state) orbital angular momentum  of any
nucleus is composed of two equal parts: half of nucleons (protons + neutrons) having spin projection up and other half having 
spin projection down. It is known that 
the huge majority of nuclei have zero angular momentum in the ground state. We will show 
below that as a rule this zero is just the sum of two rather big counter directed angular 
momenta (hidden angular momenta, because they are not manifest in the ground state)
of the above mentioned two parts of any nucleus.  Being connected with the spins of 
nucleons this phenomenon naturally has great influence 
on all nuclear properties connected with the spin, in particular, the spin scissors mode.

Let us analyze the procedure of linearization of the equations of motion for collective variables
(\ref{Varis}). We consider small deviations of the system from  equilibrium, so all variables 
are written as a sum of their equilibrium value plus a small deviation:
$$L(t)=L({\rm eq})+\L(t),\quad \mbox{et al.}$$
Neglecting quadratic deviations one obtains the set of linearized equations for deviations depending on the equilibrium 
values $R^{\tau\varsigma}_{\lambda\mu}({\rm eq})$ and $L^{\tau\varsigma}_{\lambda\mu}({\rm eq})$, which are the input data
of the problem. In the paper~\cite{BaMoPRC} we made the choice shown in
equations (\ref{equi1})-(\ref{equi3}). For the sake of convenience we
write it again:
\begin{eqnarray}
&&R^{+}_{2\pm1}({\rm eq})=R^{+}_{2\pm2}({\rm eq})=0,\nonumber\\ 
&&R^{+}_{20}({\rm eq})\neq0,\quad R^{+}_{00}({\rm eq})\neq0,
\label{eq1}
\\
&&R^{\uparrow\downarrow}_{\lambda\mu}({\rm eq})=R^{\downarrow\uparrow}_{\lambda\mu}({\rm eq})=0,
\label{eq3}
\\
&&L^{\tau\varsigma}_{\lambda\mu}({\rm eq})=0,\quad
R^{-}_{\lambda\mu}({\rm eq})=0.
\label{eq4}
\end{eqnarray}

At first glance, this choice looks quite natural. Really, relations (\ref{eq1}) 
follow from the axial symmetry of nucleus. Relations (\ref{eq3}) are justified by the fact that these quantities should be diagonal in spin at equilibrium. The variables
$L^{\tau\varsigma}_{\lambda\mu}(t)$ contain the momentum $\bp$ in their definition which 
incited us to suppose  zero equilibrium values as well (we will show below that it is not 
true for $L^-_{10}$ because of quantum effects connected with spin). 

The relation $R^{-}_{\lambda\mu}({\rm eq})=0$ follows from the shell model considerations: the 
nucleons with spin projection "up" and "down" are sitting in pairs on the same levels, therefore all
average properties of the "spin up" part of the nucleus must be identical to that of the "spin down" part.
However, a careful analysis shows that being undoubtedly true for variables
$R^{\uparrow\uparrow}_{\lambda\mu},\,R^{\downarrow\downarrow}_{\lambda\mu}$ this statement turns out
erroneous for variables $L^{\uparrow\uparrow}_{10},\,L^{\downarrow\downarrow}_{10}$. Let us demonstrate it. 
By definition
\begin{eqnarray}
L^{ss'}_{\lambda\mu}(t)=
\int\! d^3r\,\int\! \frac{d^3p}{(2\pi\hbar)^{3}} \{r\otimes p\}_{\lambda\mu}
f^{ss'}(\br,\bp,t)
=\int\! d^3r \{r\otimes J^{ss'}\}_{\lambda\mu},
\label{LJ}
\end{eqnarray}
where 
\begin{eqnarray}
J^{ss'}_i(\br,t)=\int\! \frac{d^3p}{(2\pi\hbar)^3} p_i f^{ss'}(\br,\bp,t)
=\int\! \frac{d^3p p_i}{(2\pi\hbar)^{3}}\int\!d^3q{\rm e}^{-\frac{i}{\hbar}\bp\cdot\bq}
\rho\left(\br+\frac{\bq}{2},s;\br-\frac{\bq}{2},s';t\right)\qquad 
\label{J}
\end{eqnarray}
is the i-th component of the nuclear current. In the last relation the definition (\ref{B.1}) of Wigner function is used. 
Performing the integration over $\bp$ one finds:
\begin{eqnarray}
&&J^{ss'}_i(\br,t)=i\hbar\int\!d^3q[\frac{\partial}{\partial q_i}\delta(\bq)]
\rho(\br+\frac{\bq}{2},s;\br-\frac{\bq}{2},s';t)
\nonumber\\
&&=-i\hbar\int\!d^3q\delta(\bq)\frac{\partial}{\partial q_i}
\rho(\br+\frac{\bq}{2},s;\br-\frac{\bq}{2},s';t)
=-\frac{i\hbar}{2}[(\nabla_{1i}-\nabla_{2i})\rho(\br_1,s;\br_2,s';t)]_{\br_1=\br_2=\br},\qquad
\label{J1}
\end{eqnarray}%
where $\br_1=\br+\frac{\bq}{2},\, \br_2=\br-\frac{\bq}{2}$. The density matrix of the ground state
nucleus is defined~\cite{Ring} as
\begin{equation}
\rho(\br_1,s;\br_2,s';t)=\sum_{\nu}v^2_{\nu}\psi_{\nu}(\br_1s)\psi^*_{\nu}(\br_2s'),
\label{rho}
\end{equation}
where $v^2_{\nu}$ are occupation numbers and $\psi_{\nu}$ are single particle wave functions. For the sake
of simplicity we will consider the case of spherical symmetry. Then $\nu=nljm$ and
\begin{eqnarray}
\psi_{nljm}(\br,s)=\R_{nlj}(r)\sum_{\Lambda,\sigma}C^{jm}_{l\Lambda,\frac{1}{2}\sigma}
Y_{l\Lambda}(\theta,\phi)\chi_{\frac{1}{2}\sigma}(s),
\label{phi}
\end{eqnarray}
\begin{eqnarray}
&&J^{ss'}_i(\br)=-\frac{i\hbar}{2}\sum_{\nu}v^2_{\nu}
[\nabla_i\psi_{\nu}(\br,s)\cdot\psi^*_{\nu}(\br,s')
-\psi_{\nu}(\br,s)\cdot\nabla_i\psi^*_{\nu}(\br,s')]
\label{Jdef}
\\
&&=-\frac{i\hbar}{2}\sum_{nljm}v^2_{nljm}\R^2_{nlj}\sum_{\Lambda,\sigma,\Lambda',\sigma'}
C^{jm}_{l\Lambda,\frac{1}{2}\sigma}C^{jm}_{l\Lambda',\frac{1}{2}\sigma'}
[Y^*_{l\Lambda'}\nabla_iY_{l\Lambda}-Y_{l\Lambda}\nabla_iY^*_{l\Lambda'}]
\chi_{\frac{1}{2}\sigma}(s)\chi_{\frac{1}{2}\sigma'}(s').
\label{Jphi}
\end{eqnarray}
Inserting this expression into (\ref{LJ}) one finds:
\begin{eqnarray}
&&L^{ss'}_{10}({\rm eq})=
\nonumber\\&&
-\frac{i\hbar}{2}\!\!\sum_{nljm}\!\!v^2_{nljm}\!\!\!\!\sum_{\Lambda\sigma,\Lambda'\sigma'}\!\!\!\!
C^{jm}_{l\Lambda,\frac{1}{2}\sigma}C^{jm}_{l\Lambda',\frac{1}{2}\sigma'}
\chi_{\frac{1}{2}\sigma}(s)\chi_{\frac{1}{2}\sigma'}(s')
\!\!\int\!\! d^3r\,\R^2_{nlj}[Y^*_{l\Lambda'}\{r\otimes\nabla\}_{10}Y_{l\Lambda}-Y_{l\Lambda}\{r\otimes\nabla\}_{10}Y^*_{l\Lambda'}]
\nonumber\\
&&=\frac{i}{2\sqrt2}\sum_{nljm}v^2_{nljm}\sum_{\Lambda\sigma,\Lambda'\sigma'}
C^{jm}_{l\Lambda,\frac{1}{2}\sigma}C^{jm}_{l\Lambda',\frac{1}{2}\sigma'}
\chi_{\frac{1}{2}\sigma}(s)\chi_{\frac{1}{2}\sigma'}(s')
\int\! d^3r\,\R^2_{nlj}[Y^*_{l\Lambda'}\hat l_0Y_{l\Lambda}-Y_{l\Lambda}\hat l_0Y^*_{l\Lambda'}]
\nonumber\\
&&=\frac{i}{2\sqrt2}\sum_{nljm}v^2_{nljm}\sum_{\Lambda\sigma,\Lambda'\sigma'}
C^{jm}_{l\Lambda,\frac{1}{2}\sigma}C^{jm}_{l\Lambda',\frac{1}{2}\sigma'}
\chi_{\frac{1}{2}\sigma}(s)\chi_{\frac{1}{2}\sigma'}(s')(\Lambda+\Lambda')\delta_{\Lambda,\Lambda'}
\nonumber\\
&&=\frac{i}{\sqrt2}\sum_{nljm}v^2_{nljm}\sum_{\Lambda\sigma}\Lambda\,
\left(C^{jm}_{l\Lambda,\frac{1}{2}\sigma}\right)^2
\chi_{\frac{1}{2}\sigma}(s)\chi_{\frac{1}{2}\sigma}(s').
\label{L10}
\end{eqnarray}
Here the definition $\hat l_{\mu}=-\hbar\sqrt{2}\{r\otimes\nabla\}_{1\mu}$, formula 
$\hat l_0Y_{l\Lambda}=\Lambda Y_{l\Lambda}$ and normalization of functions $\R_{nlj}$ were used.
Remembering the definition of the spin function $\chi_{\frac{1}{2}\sigma}(s)=\delta_{\sigma s}$
we get finally:
\begin{equation}
L^{ss'}_{10}({\rm eq})=
\frac{i}{\sqrt2}\sum_{nljm}v^2_{nljm}\sum_{\Lambda}\Lambda\,
\left(C^{jm}_{l\Lambda,\frac{1}{2}s}\right)^2\delta_{ss'}=
\delta_{ss'}\frac{i}{\sqrt2}\sum_{nljm}v^2_{nljm}
\left(C^{jm}_{lm-s,\frac{1}{2}s}\right)^2(m-s).
\label{L10f}
\end{equation}
Now, with the help of analytic expressions for Clebsh-Gordan coefficients one obtains the final  
expressions
\begin{eqnarray}
&&L^{\uparrow\uparrow}_{10}({\rm eq})
=\frac{i}{\sqrt2}\sum_{nl}\left[
\sum_{m=-\left(l+\frac{1}{2}\right)}^{l+\frac{1}{2}}v^2_{nlj^+m}\frac{l+\frac{1}{2}+m}{2l+1} +
\sum_{m=-\left(l-\frac{1}{2}\right)}^{l-\frac{1}{2}}v^2_{nlj^-m}\frac{l+\frac{1}{2}-m}{2l+1}
\right]\left(m-\frac{1}{2}\right),\qquad
\label{L10up}
\\
&&L^{\downarrow\downarrow}_{10}({\rm eq})
=\frac{i}{\sqrt2}\sum_{nl}\left[
\sum_{m=-\left(l+\frac{1}{2}\right)}^{l+\frac{1}{2}}v^2_{nlj^+m}\frac{l+\frac{1}{2}-m}{2l+1} +
\sum_{m=-\left(l-\frac{1}{2}\right)}^{l-\frac{1}{2}}v^2_{nlj^-m}\frac{l+\frac{1}{2}+m}{2l+1}
\right]\left(m+\frac{1}{2}\right),
\label{L10d}
\end{eqnarray}
where the notation $j^{\pm}=l\pm\frac{1}{2}$ is introduced. Replacing in (\ref{L10up}) $m$
by $-m$ we find that
\begin{eqnarray}
L^{\uparrow\uparrow}_{10}({\rm eq})=-L^{\downarrow\downarrow}_{10}({\rm eq}).
\label{L10upd}
\end{eqnarray}
By definition (\ref{Varis}) 
$L^{\pm}_{10}({\rm eq})=L^{\uparrow\uparrow}_{10}({\rm eq}) \pm L^{\downarrow\downarrow}_{10}({\rm eq})$.
Combining linearly (\ref{L10up}) and (\ref{L10d}) one finds:
\begin{eqnarray}
&&L^+_{10}({\rm eq})
=\frac{i}{\sqrt2}\sum_{nl}\left[
\sum_{m=-\left(l+\frac{1}{2}\right)}^{l+\frac{1}{2}}v^2_{nlj^+m}\frac{2l}{2l+1}m+
\sum_{m=-\left(l-\frac{1}{2}\right)}^{l-\frac{1}{2}}v^2_{nlj^-m}\frac{2l+2}{2l+1}m
\right],
\label{L10+}
\\
&&L^-_{10}({\rm eq})=\frac{i}{\sqrt2}\sum_{nl}\left[
\sum_{m=-\left(l+\frac{1}{2}\right)}^{l+\frac{1}{2}}v^2_{nlj^+m}\frac{2m^2-l-\frac{1}{2}}{2l+1}-
\sum_{m=-\left(l-\frac{1}{2}\right)}^{l-\frac{1}{2}}v^2_{nlj^-m}\frac{2m^2+l+\frac{1}{2}}{2l+1}
\right].
\label{L10-}
\end{eqnarray}
These formulas are valid for spherical nuclei. However, with the scissors and spin-scissors modes, we are considering deformed nuclei. 
For the sake of the discussion,
let us consider the case of infinitesimally small deformation, when one can continue to use formulae~(\ref{L10+},~\ref{L10-}). 
Now only levels with quantum numbers $\pm m$ are degenerate.
According to, for example, the Nilsson scheme~\cite{Nils} nucleons will occupy pairwise 
precisely those levels which leads to the zero 
value of~$L^+_{10}({\rm eq})$.

What about~$L^-_{10}({\rm eq})$? It only enters (\ref{iv}) in the equation for $\dot {\P}^-_{21}$. Let us analyze the structure of formula (\ref{L10-}) 
considering for the sake of simplicity the case without pairing. Two sums over $m$ (let us note them $\Sigma_1$
and $\Sigma_2$) represent the two spin-orbital partners: in the first sum the summation goes over the
levels of the lower partner ($j=l+\frac{1}{2}$) and in the second sum -- over the levels of the
higher partner ($j=l-\frac{1}{2}$). The values of both sums depend naturally on the values of
occupation numbers $n_{nljm}= 0,1$. There are three possibilities. The first one is trivial: if all
levels of both spin-orbital partners are disposed above the Fermi surface, then the respective
occupation numbers $n_{nljm}=0$ and both sums are equal to zero identically. The second 
possibility: all levels of both spin-orbital partners are disposed below the Fermi surface.
Then all respective occupation numbers $n_{nlj^+m}=n_{nlj^-m}=1$. The elementary analytical
calculation (for arbitrary $l$) shows that in this case the two sums in~(\ref{L10-}) exactly compensate 
each other,
i.e. $\Sigma_1+\Sigma_2=0$. The most interesting is the third possibility, when one part of
levels of two spin-orbital partners is disposed below the Fermi surface and another part is
disposed above it. In this case the compensation does not happen and one gets 
$\Sigma_1+\Sigma_2\neq 0$ what leads to $L^-_{10}({\rm eq})\neq 0$. In the case of pairing, things 
are not so sharply separated and $L_{10}^-({\rm eq})$ has always a finite value. However, the modifications with respect to mean field are very small.

Let us illustrate the above analysis by the example of $^{164}$Dy (protons). Its deformation is
$\delta=0.26$ $(\epsilon=0.28)$ and Z=66. Looking on the Nilsson scheme (for example, Fig. 1.5 of~\cite{Solov}
or Fig. 2.21c of~\cite{Ring}) one easily finds, that only three pairs of spin-orbital partners 
give a nonzero contribution to $L^-_{10}({\rm eq})$. They are: $N=4, d_{5/2}-d_{3/2}$ (two 
levels of $d_{5/2}$ are below the Fermi surface, all the rest -- above); 
$N=4, g_{9/2}-g_{7/2}$ (one level of $g_{7/2}$ is above the Fermi surface, all the rest -- below);
$N=5, h_{11/2}-h_{9/2}$ (four levels of $h_{11/2}$ are below the Fermi surface, all the rest 
-- above).
It is possible to make the crude evaluation of $L^-_{10}({\rm eq})$ using the quantum numbers
indicated on Fig. 1.5 of~\cite{Solov} or Fig. 2.21c of~\cite{Ring}. The result turns out rather
close to the exact one,
computed with the help of formulas (\ref{LJ},\ref{Jdef}) and Nilsson wave functions.
 The influence of pair correlations is very small.

\begin{figure}
\centering\includegraphics[width=0.5\textwidth]{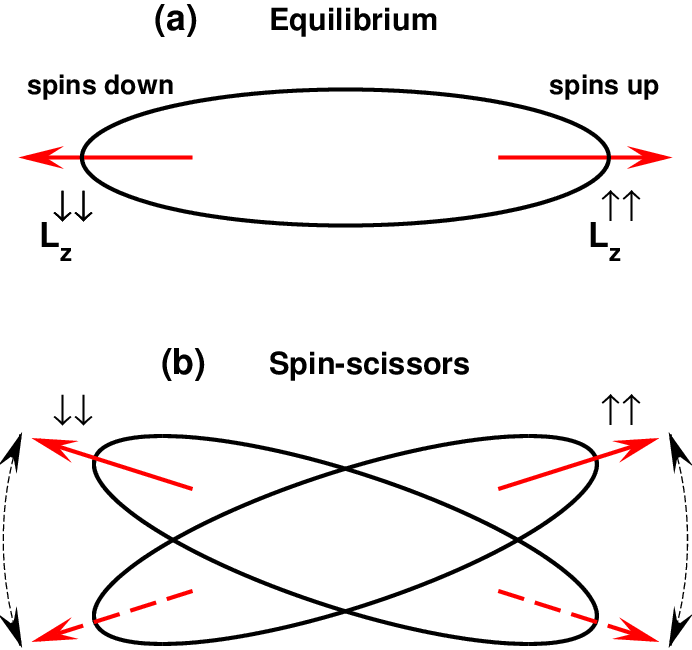}
\caption{(a) Protons with spins $\uparrow$ (up) and $\downarrow$ (down) having nonzero orbital
angular momenta at equilibrium. (b) Protons from Fig.(a) vibrating against one-another.}
\label{figSch}\end{figure}

Indeed, from the definitions (\ref{LJ}) and (\ref{L10}) one can see that $L^{ss}_{10}({\rm eq})$ is just 
the average value of the z-component of the orbital angular momentum of nucleons with the 
spin projection $s$~($\frac{1}{2}$~or~$-\frac{1}{2}$). So,  
the ground state nucleus consists 
of two equal parts having nonzero angular momenta with opposite directions, which compensate 
each other resulting in the zero total angular momentum. 
This is graphically depicted in Fig.~\ref{figSch}(a). 

On the other hand, when the opposite angular momenta become tilted, one excites the system and the opposite 
angular momenta are vibrating with a tilting angle, see Fig.~\ref{figSch}(b). 
Actually the two opposite angular momenta are oscillating, one in the opposite sense of the other. 
It is rather obvious from Fig.~\ref{fig0} that these tilted vibrations 
happen separately in each of the neutron and proton lobes. 
These spin-up against spin-down motions certainly influence the 
excitation of the spin scissors mode. 
So, classically speaking
the proton and neutron parts of the ground state nucleus 
consist each of two identical gyroscopes rotating in opposite directions. One knows that it 
is very difficult to deviate gyroscope from an equilibrium. So one can expect, 
that the probability to force two gyroscopes to oscillate as scissors (spin scissors) 
should be small. This picture is confirmed in the next section.

\section{Results of calculations continued}

\begin{figure}
\centering\includegraphics[width=0.5\textwidth]{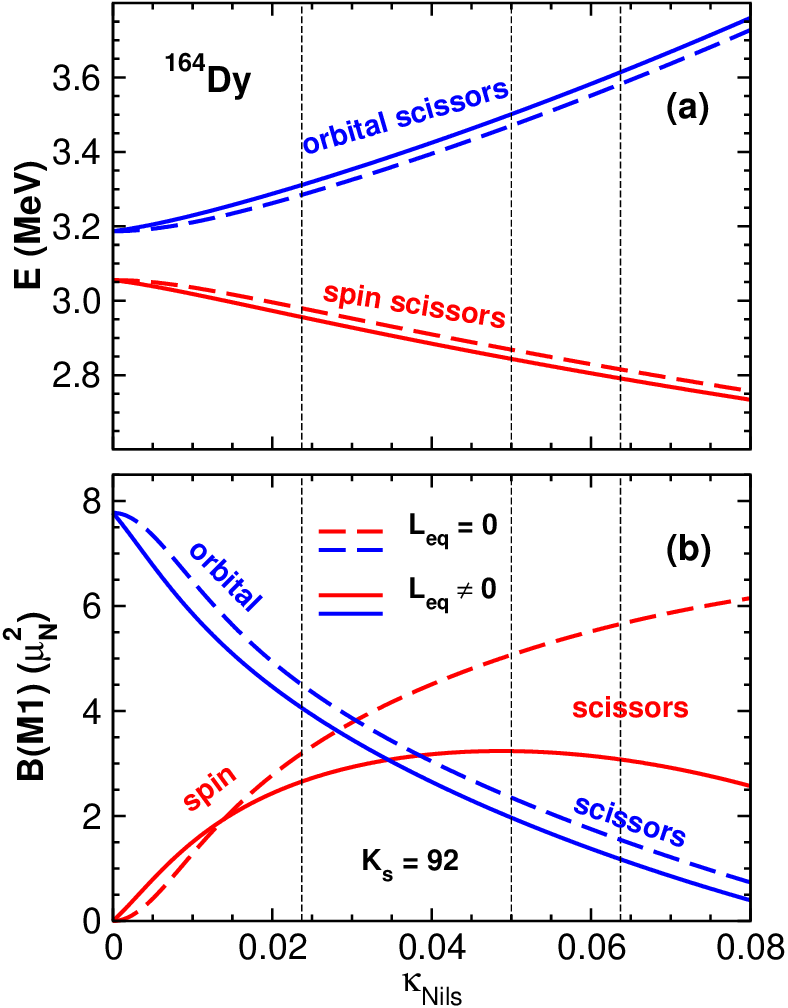}
\caption{The energies $E$ and $B(M1)$-factors as a functions of the spin-orbital strength constant 
$\kappa_{\rm Nils}$. The dashed lines -- calculations without 
$L_{10}^-({\rm eq})$, the solid lines -- $L_{10}^-({\rm eq})$ are taken into account. $H_{ss}$ and pairing 
are included.}
\label{figL}\end{figure}

We made the calculations taking into account the non zero 
value of $L_{10}^-({\rm eq})$ (which was computed according to formulas
(\ref{LJ},\ref{Jdef}) and Nilsson wave functions). 
The results are presented in~Fig.~\ref{figL} and Table~\ref{tab1y}.

\begin{table*}
\caption{\label{tab1y}
 Isovector energies and excitation probabilities 
of $^{164}$Dy with pair correlations taken into account. Deformation parameter \mbox{$\delta=0.26$}, spin-orbit constant $\kappa_{\rm Nils}=0.0637$, 
spin-spin interaction constants are
$K_s=50$~MeV, $q=-0.5$. Results: a -- without $L_{10}^-({\rm eq})$, 
b -- with $L_{10}^-({\rm eq})$.}
\begin{tabular}{l|c|c|c|c|c|c}
\hline
 $(\lambda,\mu)^\varsigma$ &   
\multicolumn{2}{c|}{ $E_{\rm iv}$,~MeV}      &
\multicolumn{2}{c|}{ $B(M1)\!\!\uparrow,\ \mu_N^2$}      &
\multicolumn{2}{c }{ $B(E2)$, W.u.}        \\
\cline{2-7}
  & a &  b  &  a &  b  &  a &  b  \\ 
\hline
 (1,1)$^-$                    & ~2.80  & ~2.77  & ~4.98 & ~2.44 & ~0.52 & ~0.49  \\   
 (1,1)$^+$                    & ~3.57  & ~3.60  & ~1.71 & ~1.36 & ~1.94 & ~2.00  \\ 
 (0,0)$^{\downarrow\uparrow}$ & 14.60  & 14.60  & ~0.07 & ~0.07 & ~0.47 & ~0.47  \\  
 (2,1)$^-$                    & 16.50  & 16.51  & ~0.08 & ~0.07 & ~0.38 & ~0.36  \\  
 (2,0)$^{\downarrow\uparrow}$ & 18.23  & 18.22  & ~0.17 & ~0.17 & ~1.76 & ~1.80  \\  
 (2,2)$^{\uparrow\downarrow}$ & 19.32  & 19.32  & ~0.08 & ~0.07 & ~0.93 & ~0.94  \\   
 (2,1)$^+$                    & 21.33  & 21.32  & ~2.47 & ~2.48 & 31.32 & 31.30  \\
 (1,0)$^{\downarrow\uparrow}$ & i0.59  & i0.58  & -i5.4 & -i65  & -i0.02& i0.0   \\
 \hline
\end{tabular}
\end{table*}
Figure~\ref{figL} demonstrates the dependence of the scissors modes energies and $B(M1)$ values on the spin orbital strength constant 
$\kappa_{\rm Nils}$. One can observe the strong
influence of the hidden angular momenta on the spin scissors mode, whose $B(M1)$ value is strongly 
decreasing with increasing $\kappa_{\rm Nils}$. The $B(M1)$ value of the orbital scissors also
is reduced, but not so much, the value of the reduction being practically independent on
$\kappa_{\rm Nils}$. The influence of $L_{10}^-({\rm eq})$ on the energies of both scissors is negligible,
leading to a small increase of their splitting. Now the energy centroid of both scissors and
their summed $B(M1)$ value at $\kappa_{\rm Nils}=0.0637$ are $E_{\rm cen}=3.07$~MeV and 
$B(M1)_{\Sigma}=3.78\ \mu_N^2$. The general agreement with experiment becomes considerably 
better (compare with Table~\ref{tab12}).

Table~\ref{tab1y} demonstrates the energies and transition probabilities
of all (low and high lying) isovector modes of $^{164}$Dy obtained by the solution of equations (\ref{iv}) with hidden angular momenta taken into account. The results for both scissors coincide with that of Fig. 
\ref{figL} for $\kappa_{\rm Nils}=0.0637$. It is seen also that all high
lying modes are completely insensitive to the value of 
$L_{10}^-({\rm eq})$. The same is also true for all isoscalar modes. It is
worth noting, nevertheless , that the small negative $B(M1)$ value of the
isoscalar excitation $(1,1)^+$ (see Tab.~\ref{tab2x}) goes at last to zero!

\begin{figure}
\centering\includegraphics[width=0.5\textwidth]{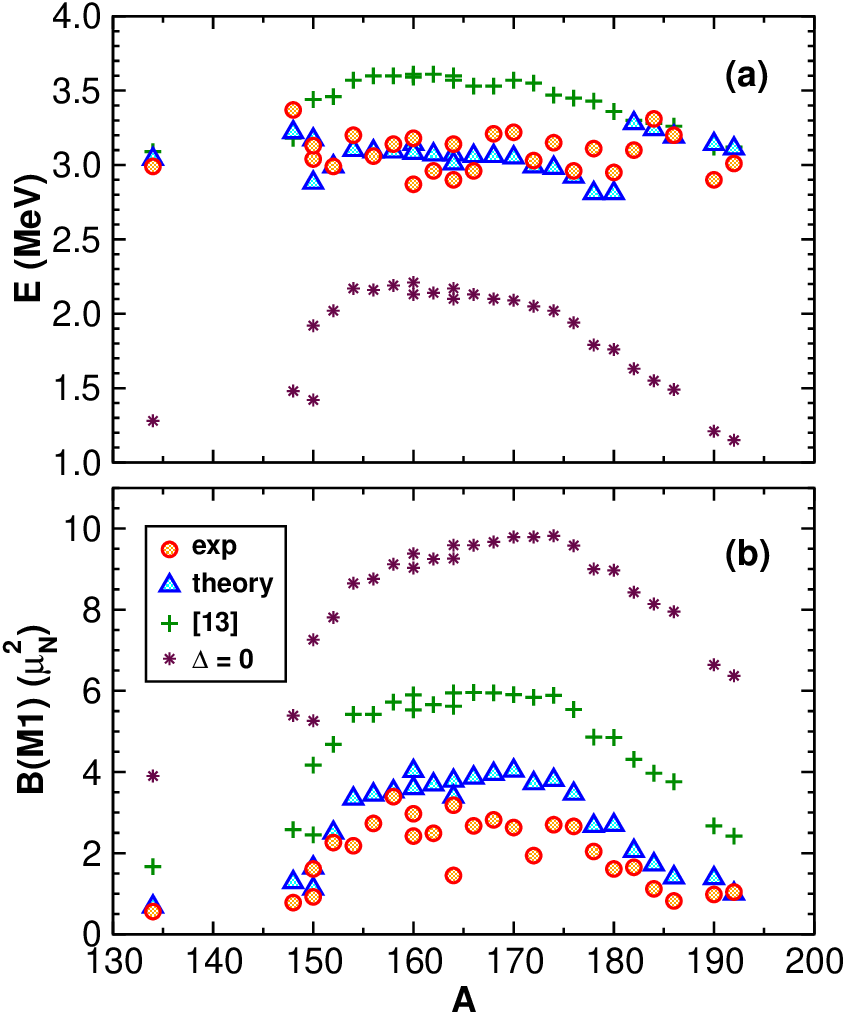}
\caption{The energies $E$ and $B(M1)$-factors as a function of the mass number A
for nuclei listed in the Tables~\ref{tab3},~\ref{tab4}.}
\label{figMalov}\end{figure}

The results of systematic calculations for the rare-earth nuclei are presented in Tables~\ref{tab3} and
~\ref{tab4} and desplayed in Fig.~\ref{figMalov}. Table~\ref{tab3} contains the results for well deformed nuclei with
$\delta \geq 0.18.$
It is easy to see that the overall (general) agreement of theoretical
results with experimental data is substantially improved (in comparison with our previous
calculations~\cite{Urban}). 

\begin{table*}
\caption{\label{tab3}Scissors modes energy centroids $E_{\rm cen}$ and summarized
transition probabilities $B(M1)_{\Sigma}$. Parameters: 
$\kappa_{\rm Nils}=0.0637$, pairing strength constant $V_0=25$
($V_0=27$ for $^{182, 184, 186}$W). The experimental values
of $E_{\rm cen}$, $\delta$, and $B(M1)_{\Sigma}$ are from~\cite{Pietr95,Pietr98} and
references therein.}
\begin{tabular}{c|c|c|c|c|c|c|c|c|c}
\hline
 Nuclei & $\delta$ &
\multicolumn{4}{c|}{ $E_{\rm cen}$,~MeV}    &
\multicolumn{4}{c }{ $B(M1)_{\Sigma},\ \mu_N^2$}    \\
\cline{3-10}
 &  & Exp. & WFM & Ref.~\cite{Urban} & $\Delta=0$ 
    & Exp. & WFM & Ref.~\cite{Urban} & $\Delta=0$ \\
\hline
 $^{150}$Nd & 0.22 & 3.04 & 2.88 & 3.44 & 1.92 & 1.61 & 1.64 & 4.17 & 7.26 \\
 $^{152}$Sm & 0.24 & 2.99 & 2.99 & 3.46 & 2.02 & 2.26 & 2.50 & 4.68 & 7.81 \\
 $^{154}$Sm & 0.26 & 3.20 & 3.10 & 3.57 & 2.17 & 2.18 & 3.34 & 5.42 & 8.65 \\
 $^{156}$Gd & 0.26 & 3.06 & 3.09 & 3.60 & 2.16 & 2.73 & 3.44 & 5.42 & 8.76 \\
 $^{158}$Gd & 0.26 & 3.14 & 3.09 & 3.60 & 2.19 & 3.39 & 3.52 & 5.72 & 9.12 \\
 $^{160}$Gd & 0.27 & 3.18 & 3.14 & 3.61 & 2.21 & 2.97 & 4.02 & 5.90 & 9.38 \\
 $^{160}$Dy & 0.26 & 2.87 & 3.08 & 3.59 & 2.13 & 2.42 & 3.60 & 5.53 & 9.03 \\
 $^{162}$Dy & 0.26 & 2.96 & 3.07 & 3.61 & 2.14 & 2.49 & 3.69 & 5.66 & 9.25 \\
 $^{164}$Dy & 0.26 & 3.14 & 3.07 & 3.60 & 2.17 & 3.18 & 3.78 & 5.95 & 9.59 \\
 $^{164}$Er & 0.25 & 2.90 & 3.01 & 3.57 & 2.10 & 1.45 & 3.39 & 5.62 & 9.26 \\
 $^{166}$Er & 0.26 & 2.96 & 3.06 & 3.53 & 2.13 & 2.67 & 3.86 & 5.96 & 9.59 \\
 $^{168}$Er & 0.26 & 3.21 & 3.06 & 3.53 & 2.10 & 2.82 & 3.95 & 5.95 & 9.67 \\
 $^{170}$Er & 0.26 & 3.22 & 3.05 & 3.57 & 2.09 & 2.63 & 4.03 & 5.91 & 9.79 \\
 $^{172}$Yb & 0.25 & 3.03 & 2.99 & 3.55 & 2.05 & 1.94 & 3.72 & 5.84 & 9.79 \\
 $^{174}$Yb & 0.25 & 3.15 & 2.98 & 3.47 & 2.02 & 2.70 & 3.80 & 5.89 & 9.82 \\
 $^{176}$Yb & 0.24 & 2.96 & 2.92 & 3.45 & 1.94 & 2.66 & 3.46 & 5.54 & 9.58 \\
 $^{178}$Hf & 0.22 & 3.11 & 2.81 & 3.43 & 1.79 & 2.04 & 2.67 & 4.86 & 9.00 \\
 $^{180}$Hf & 0.22 & 2.95 & 2.81 & 3.36 & 1.76 & 1.61 & 2.69 & 4.85 & 8.97 \\
 $^{182}$W  & 0.20 & 3.10 & 3.28 & 3.30 & 1.63 & 1.65 & 2.05 & 4.31 & 8.43 \\
 $^{184}$W  & 0.19 & 3.31 & 3.24 & 3.28 & 1.55 & 1.12 & 1.72 & 3.97 & 8.14 \\
 $^{186}$W  & 0.18 & 3.20 & 3.19 & 3.26 & 1.49 & 0.82 & 1.40 & 3.76 & 7.95 \\
 \hline
\end{tabular}
\end{table*}

\begin{table*}
\caption{\label{tab4}Scissors modes energy centroids $E_{\rm cen}$ and summarized
transition probabilities $B(M1)_{\Sigma}$. Parameters: $\kappa_{\rm Nils}=0.05$ 
($\kappa_{\rm Nils}=0.0637$ for $^{182, 184, 186}$W), pairing strength constant~$V_0=27$.}
\begin{tabular}{c|c|c|c|c|c|c|c|c|c}
\hline
 Nuclei & $\delta$ &
\multicolumn{4}{c|}{ $E_{\rm cen}$,~MeV}    &
\multicolumn{4}{c }{ $B(M1)_{\Sigma},\ \mu_N^2$}    \\
\cline{3-10}
 &  & Exp. & WFM & Ref.~\cite{Urban} & $\Delta=0$ 
    & Exp. & WFM & Ref.~\cite{Urban} & $\Delta=0$ \\
\hline
 $^{134}$Ba & 0.14 & 2.99 & 3.04 & 3.09 & 1.28 & 0.56 & 0.68 & 1.67 & 3.90 \\
 $^{148}$Nd & 0.17 & 3.37 & 3.22 & 3.18 & 1.48 & 0.78 & 1.28 & 2.58 & 5.39 \\
 $^{150}$Sm & 0.16 & 3.13 & 3.17 & 3.13 & 1.42 & 0.92 & 1.12 & 2.45 & 5.26 \\
 $^{182}$W  & 0.20 & 3.10 & 3.28 & 3.30 & 1.63 & 1.65 & 2.05 & 4.31 & 8.43 \\
 $^{184}$W  & 0.19 & 3.31 & 3.24 & 3.28 & 1.55 & 1.12 & 1.72 & 3.97 & 8.14 \\
 $^{186}$W  & 0.18 & 3.20 & 3.19 & 3.26 & 1.49 & 0.82 & 1.40 & 3.76 & 7.95 \\
 $^{190}$Os & 0.15 & 2.90 & 3.14 & 3.12 & 1.21 & 0.98 & 1.38 & 2.67 & 6.64 \\
 $^{192}$Os & 0.14 & 3.01 & 3.11 & 3.12 & 1.15 & 1.04 & 1.00 & 2.42 & 6.37 \\
 \hline
\end{tabular}
\end{table*}

The results of calculations for two groups ("light" and "heavy") of weakly deformed nuclei
with deformations $0.14\leq \delta \leq 0.17$ are shown in the Table~\ref{tab4}.
They require some discussion, because of the 
self-consistency problem. These two groups of nuclei are transitional between well deformed
and spherical nuclei. Systematic calculations of equilibrium deformations~\cite{Solov} 
predict $\delta_{\rm eq}^{th}=0.0$ for $^{134}$Ba, $\pm 0.1$ for $^{148}$Nd,
 0.15 or -0.12 for $^{150}$Sm, 0.1 or -0.14 for $^{190}$Os and -0.1 for $^{192}$Os,
whereas their experimental values are $\delta_{\rm eq}=0.14,\, 0.17,\, 0.16,\, 0.15$ and 0.14
respectively. As one sees, the discrepancy between theoretical and experimental 
$\delta_{\rm eq}$ is large. Uncertain signs of theoretical equilibrium deformations are 
connected with very small ($\sim $0.1-0.2~MeV) difference between the values of deformation 
energies $\E_{\rm def}$ at positive and negative $\delta_{\rm eq}$. 
Even more so, the values of deformation
energies of these nuclei are very small: $\E_{\rm def}=0.20,~0.50,~0.80$ and 0.70~MeV for
$^{148}$Nd, $^{150}$Sm, $^{190}$Os and $^{192}$Os respectively. This means that these
nuclei are very "soft" with respect of $\beta$- or $\gamma$-vibrations and probably they have
more complicated equilibrium shapes, for example,  hexadecapole or octupole deformations in
addition to the quadrupole one. This means that for the correct description of their dynamical
and equilibrium properties it is necessary to include higher order Wigner function moments
(at least fourth order) in addition to the second order ones. In this case it would be natural
also to use more complicate mean field potentials (for example, the Woods-Saxon one or the potential extracted 
from some of the numerous variants of Skyrme forces) instead of the too simple Nilsson potential.
Naturally, this will be the subject of further investigations. However, to be sure that the
situation with these nuclei is not absolutely hopeless, one can try
to imitate the properties of the more perfect potential by fitting
parameters of the Nilsson potential. As a matter of fact this potential contains one single but
essential parameter -- the spin-orbital strength $\kappa_{\rm Nils}$.
It turns out that changing its value from 0.0637 to 0.05 (the value used by Nilsson in his 
original paper~\cite{Nils}) is enough to obtain a reasonable description of $B(M1)$ factors
(see Table~\ref{tab4}).
To obtain a reasonable description of the scissors energies we use the "freedom" of
choosing the value of the pairing interaction constant $V_0$ in (\ref{v_p}). It turns out
that changing its value from 25~MeV to 27~MeV is enough to obtain satisfactory agreement
between the theoretical and experimental values of~$E_{sc}$~(Table~\ref{tab4}).

 The isotopes
$^{182-186}$W turn out intermediate between weakly deformed and well deformed nuclei: 
reasonable results  are obtained with $\kappa_{\rm Nils}=0.0637$ (as for well deformed)
and $V_0=27$~MeV (as for weakly deformed). That is why they appear in both Tables.
 
Returning to the group of well deformed nuclei with $\delta \geq 0.18$ (Table~\ref{tab3}) 
it is necessary to emphasize that all presented results for these nuclei were obtained without any 
fitting. In spite of it the agreement between the theory and experiment looks more
or less satisfactory
for all nuclei of this group except two: $^{164}$Er and $^{172}$Yb, where the 
theory overestimates $B(M1)$ values approximately two times. However, these two nuclei fall out of the systematics and one can suspect, that
there the experimental $B(M1)$ values are underestimated. Therefore one
can hope, that new experiments will correct the situation with these
nuclei, as it happened, for example, with $^{232}$Th~\cite{Adekola}. 

It is interesting to compare our results with that of RPA calculations. The only
systematic calculations for rare earth nuclei was done in the frame of the extended
RPA formalism (Quasiparticle-Phonon Nuclear Model (QPNM))~\cite{Sushkov2}. We
took the Table~\ref{tab5} from this paper adding there, for the sake of comparison, the
column with our results (WFM). It is easy to see that QPNM results practically coincide
with experimental ones, whereas deviations of our results from experimental data
reach sometimes 50$\%$. However, it is necessary to emphasize here, that such 
naive comparison is not fully legitimate, because the objects of comparison are
slightly different.
The numbers presented in third column of Table~\ref{tab5} are just the sums of all M1
strength found experimentally in the energy interval shown in second column.
Theorists, working in RPA, represent their results exactly in the same manner --
the sum of $B(M1)$ values of all pikes in the respective energy interval.
\begin{table}
\caption{\label{tab5}Scissors modes summarized
transition probabilities $\sum B(M1)$. The experimental values
$\sum B(M1)$ are from~\cite{Pietr95}.}
\begin{tabular}{c|c|c|c|c}
\hline
 Nuclei & $E$,~MeV & \multicolumn{3}{c}{ $\sum B(M1),\ \mu_N^2$ }  \\ \cline{3-5} 
        &          & Exp.~\cite{Pietr95}     & QPNM~\cite{Sushkov2}  & WFM \\
\hline
 $^{156}$Gd & $2.7-3.7$ & 2.73 & 2.95 & 3.44 \\
 $^{158}$Gd & $2.7-3.7$ & 3.39 & 3.41 & 3.52 \\
 $^{160}$Gd & $2.7-3.7$ & 2.97 & 2.86 & 4.02 \\
 $^{160}$Dy & $2.7-3.7$ & 2.42 & 2.46 & 3.60 \\
 $^{162}$Dy & $2.7-3.7$ & 2.49 & 2.60 & 3.69 \\
 $^{164}$Dy & $2.7-3.7$ & 3.18 & 2.92 & 3.78 \\
 $^{166}$Er & $2.4-3.7$ & 2.67 & 2.51 & 3.86 \\
 $^{168}$Er & $2.4-3.7$ & 2.82 & 2.87 & 3.95 \\
 $^{172}$Yb & $2.4-3.7$ & 1.94 & 2.27 & 3.72 \\
 $^{174}$Yb & $2.4-3.7$ & 2.70 & 2.84 & 3.80 \\
 $^{178}$Hf & $2.4-3.7$ & 2.04 & 2.30 & 2.67 \\
 \hline
\end{tabular}
\end{table}

In principle, RPA calculations~\cite{Zaw,Sushkov2} predict some $M1$ strength at
energies higher than 3.7~MeV (up to 10~MeV). "Because of the dominance of 
spin-flip and the high level density in this region there is little hope that
reliable measurements of this srength will ever be possible"~\cite{Zaw}.
This just the point: the WFM approach implicitly takes in account the whole configuration space. 
Then, the two scissors modes (spin and orbital), 
found by the WFM method, include this 
part of the $M1$ strength which is inaccessible, even for the modern experiments.

In the light of the aforesaid it becomes clear that the summarized $M1$ strength
of spin and orbital scissors is to become somewhat bigger than the number 
presented as the experimental $B(M1)$ value of the scissors mode. So, in evaluating
the quality of agreement between theoretical and experimental results, one has 
to have in mind this element of uncertainty. 

\section{Concluding remarks}

In this work, we continued the investigation of spin modes~\cite{BaMo} with the 
Wigner Function Method in studying the influence of 
spin-spin forces. 
This method, when pushed to high order moments, is 
equivalent to the exact solution of TDHFB equation. For lower rank moments, it yields a coarse 
grained spectrum. It has the advantage that the moments allow for a direct 
physical interpretation and, thus, the spin or orbital structure of the 
found states comes directly at hand.
 
The inclusion of spin-spin interaction does not change qualitatively
the picture concerning the spectrum of the spin modes found in~\cite{BaMo}. It 
pushes all levels up without changing their order. However, it strongly 
redistributes M1 strength 
between the conventional and spin scissors mode in the favour of the last one.

We mentioned the recent experimental
work~\cite{Siem}, where for the two low lying magnetic states a stronger 
B(M1) transition for the lower state with respect to the higher one was 
found. A tentative explanation in terms of a slight triaxial deformation 
in~\cite{Siem} failed. However, our theory can naturally predict such a scenario with
a non vanishing spin-spin force.
It would indeed be very exciting, if the results of~\cite{Siem} had 
already discovered the isovector spin scissors mode. However, much 
deeper experimental and theoretical results must be obtained before a 
firm conclusion on this point is possible.

The method of Wigner function moments is generalized to take into account spin degrees of 
freedom and pair correlations simultaneously~\cite{BaMoPRC15}. 
The inclusion of the spin into the theory
allows one to discover several new phenomena. One of them, the nuclear spin scissors, was
described and studied in~\cite{BaMo,BaMoPRC}.
Another phenomenon, the 
opposite rotation of spin up/down nucleons, or in other words, the phenomenon of hidden 
angular momenta, is described in paper~\cite{BaMoPRC15}. Being determined by the spin degrees 
of freedom this phenomenon has  great influence on the excitation probability of the
spin scissors mode. On the other hand the spin scissors $B(M1)$ values and the energies
of both, spin and orbital, scissors are very sensitive to the action of pair correlations.
As a result, these two factors, the spin up/down counter-rotation and pairing, working together, 
improve substantially the agreement between the theory and experiment in the description 
of the energy centroid of two nuclear scissors and their summed excitation probability.
More precisely, a satisfactory agreement is achieved for well deformed 
nuclei of the rare earth region with standard values of all possible parameters.   
The accuracy of the description of the scissors mode by the WFM method is comparable 
with that of RPA, if to take into account the principal difference in definitions
of scissors in WFM method and RPA and experiment.
A satisfactory agreement is also achieved for weakly deformed (transitional)
nuclei of the same region by a very modest re-fit of the spin-orbit and pairing strength.
We suppose that fourth order moments and more realistic interactions are required for the 
adequate description of transitional nuclei. This shall be the object of future work.

\begin{acknowledgments}
The work was supported by the IN2P3/CNRS-JINR Collaboration agreement.
Valuable discussions with  A. V. Sushkov are gratefully acknowledged.
\end{acknowledgments}

\appendix

\section{ }
\label{AppA}

{\bf Abnormal density}

According to formula (D.47) of~\cite{Ring} the abnormal density in 
coordinate representation $\kappa(\br,s;\br',s')$ is connected with 
the abnormal density in the representation of the
harmonic oscillator quantum numbers
$\kappa_{\nu,\nu'}=\langle\Phi|a_{\nu}a_{\nu'}|\Phi\rangle$ by the relation
\begin{equation}
\kappa(\br,s;\br',s')=\langle\Phi|a(\br,s)a(\br',s')|\Phi\rangle=\sum_{\nu,\nu'}
\psi_{\nu}(\br,s)\psi_{\nu'}(\br',s')\langle\Phi|a_{\nu}a_{\nu'}|\Phi\rangle,
\label{kap1}
\end{equation}
where
$\nu\equiv k,\varsigma$, (with $k\equiv n,l,j,|m|$ and  $\varsigma\equiv{\rm sign}(m)=\pm$), 
($k,+\equiv \nu$;  $k,-\equiv \bar \nu$), 
$\psi_{\bar \nu}(\br,s)=T\psi_{\nu}(\br,s).$
$T$ -- time reversal operator defined by formula (XV.85) of~\cite{Mess}:
$T=-i\sigma_yK_0,$
where $\sigma_y$ is the Pauli matrix and $K_0$ is the complex-conjugation
operator.

According to formula (7.12) of~\cite{Ring}
$$a_{k,\varsigma}=u_k\alpha_{k,\varsigma}-\varsigma v_k\alpha^{\dagger}_{k,-\varsigma},\quad
\alpha_{\nu}|\Phi\rangle=0,
$$
\begin{equation}
\langle\Phi|a_{\nu}a_{\nu'}|\Phi\rangle\equiv\kappa_{\nu\nu'}=-\varsigma'u_kv_{k'}
\langle\Phi|\alpha_{k,\varsigma}\alpha^{\dagger}_{k',-\varsigma'}|\Phi\rangle
=-\varsigma'u_kv_{k'}\delta_{k,k'}\delta_{-\varsigma,\varsigma'}.
\label{kapnn}
\end{equation}
This result means that in accordance with the theorem of Bloch and 
Messiah we have found the basis $|\nu\rangle$ in which the abnormal 
density $\kappa_{\nu,\nu'}$ has the canonical form. Therefore the spin
structure of $\kappa_{\nu,\nu'}$ is
\begin{equation}
\kappa_{\nu,\nu'}={\qquad 0\;\qquad u_{k}v_{k} 
\choose -u_{k}v_{k}\;\qquad 0 \qquad},
\label{anom}
 \end{equation}
or $\kappa_{\bar\nu,\nu}=-\kappa_{\nu,\bar\nu}$ and $\kappa_{\nu,\nu}=\kappa_{\bar\nu,\bar\nu}=0.$

With the help of (\ref{kapnn}) formula (\ref{kap1}) can be transformed into
\begin{eqnarray}
\kappa(\br,s;\br',s')
&=&\sum_{k,\varsigma}\varsigma u_kv_k
\psi_{k,\varsigma}(\br,s)\psi_{k,-\varsigma}(\br',s')
\nonumber\\
&=&\sum_{\nu>0}u_{\nu}v_{\nu}
[\psi_{\nu}(\br,s)\psi_{\bar \nu}(\br',s')
-\psi_{\bar \nu}(\br,s)\psi_{\nu}(\br',s')].
\label{kap2}
\end{eqnarray}
that reproduces formula (D.48) of~\cite{Ring}.

{\bf What is the spin structure of $\kappa(\br,s;\br',s')$?}

Let us consider the spherical case:
\begin{equation}
\psi_{\nu}(\br,s)=\R_{nlj}(r)
\sum_{\Lambda,\sigma}C^{j m}_{l\Lambda,\frac{1}{2}\sigma}
Y_{l\Lambda}(\theta,\phi)\chi_{\frac{1}{2}\sigma}(s)
\equiv \R_{nlj}(r)\phi_{ljm}(\Omega,s),
\label{psi}
 \end{equation}
where $\phi_{ljm}(\Omega,s)=
\sum_{\Lambda,\sigma}C^{j m}_{l\Lambda,\frac{1}{2}\sigma}
Y_{l\Lambda}(\theta,\phi)\chi_{\frac{1}{2}\sigma}(s)$,
 spin function $\chi_{\frac{1}{2}\sigma}(s)=\delta_{\sigma s}$
 and angular variables are denoted by $\Omega$.

Time reversal:
\begin{eqnarray}
&&TY_{l\Lambda}=Y^*_{l\Lambda}=(-1)^{\Lambda}Y_{l-\Lambda},
\nonumber\\
&&T\chi_{\frac{1}{2}\frac{1}{2}}=\chi_{\frac{1}{2}-\frac{1}{2}},\quad
T\chi_{\frac{1}{2}-\frac{1}{2}}=-\chi_{\frac{1}{2}\frac{1}{2}}\quad
\to\quad T\chi_{\frac{1}{2}\sigma}
=(-1)^{\sigma-\frac{1}{2}}\chi_{\frac{1}{2}-\sigma},
\nonumber\\
&&T\sum_{\Lambda,\sigma}C^{jm}_{l\Lambda,\frac{1}{2}\sigma}Y_{l\Lambda}
\chi_{\frac{1}{2}\sigma}=
\sum_{\Lambda,\sigma}C^{jm}_{l\Lambda,\frac{1}{2}\sigma}Y_{l-\Lambda}
\chi_{\frac{1}{2}-\sigma}(-1)^{\Lambda+\sigma-\frac{1}{2}}=
\sum_{\Lambda,\sigma}C^{jm}_{l-\Lambda,\frac{1}{2}-\sigma}Y_{l\Lambda}
\chi_{\frac{1}{2}\sigma}(-1)^{-\Lambda-\sigma-\frac{1}{2}}
\nonumber\\
&&=\sum_{\Lambda,\sigma}C^{j-m}_{l\Lambda,\frac{1}{2}\sigma}Y_{l\Lambda}
\chi_{\frac{1}{2}\sigma}(-1)^{l+\frac{1}{2}-j-\Lambda-\sigma-\frac{1}{2}}=
\sum_{\Lambda,\sigma}C^{j-m}_{l\Lambda,\frac{1}{2}\sigma}Y_{l\Lambda}
\chi_{\frac{1}{2}\sigma}(-1)^{l-j+m}.\nonumber
\end{eqnarray}
As a result
\begin{equation}
\psi_{\bar \nu}(\br,s)
=(-1)^{l-j+m}\R_{nlj}(r)\sum_{\Lambda,\sigma}
C^{j -m}_{l\Lambda,\frac{1}{2}\sigma}
Y_{l\Lambda}(\theta,\phi)\chi_{\frac{1}{2}\sigma}(s)
=(-1)^{l-j+m}\R_{nlj}(r)\phi_{lj-m}(\Omega,s),
\label{bark}
\end{equation}
that coincides with formula (2.45) of~\cite{Ring}. Formula (\ref{kap2})
can be rewritten now as \begin{eqnarray}
&&\kappa(\br_1,s_1;\br_2,s_2)=
\nonumber\\
&&\!\!\!\!\sum_{nljm>0}\!\!\!\!(uv)_{nljm}
\R_{nlj}(r_1)\R_{nlj}(r_2)(-1)^{l-j+m}
\!\left[
\phi_{ljm}(\Omega_1,s_1)\phi_{lj-m}(\Omega_2,s_2)-
\phi_{ljm}(\Omega_2,s_2)\phi_{lj-m}(\Omega_1,s_1)
\right]
\nonumber\\
&&=\!\!\!\!\sum_{nljm>0}\!\!\!\!(uv)_{nljm}
\R_{nlj}(r_1)\R_{nlj}(r_2)(-1)^{l-j+m}
\nonumber\\
&&\times\sum_{\Lambda,\Lambda'}\left[
C^{j m}_{l\Lambda,\frac{1}{2}s_1}C^{j -m}_{l\Lambda',\frac{1}{2}s_2}
Y_{l\Lambda}(\Omega_1)Y_{l\Lambda'}(\Omega_2)
-C^{j m}_{l\Lambda,\frac{1}{2}s_2}C^{j -m}_{l\Lambda',\frac{1}{2}s_1}
Y_{l\Lambda}(\Omega_2)Y_{l\Lambda'}(\Omega_1)
\right]
\nonumber\\
&&=\!\!\!\!\sum_{nljm>0}\!\!\!\!(uv)_{nljm}
\R_{nlj}(r_1)\R_{nlj}(r_2)(-1)^{l-j+m}
\nonumber\\ &&\times
\sum_{\Lambda,\Lambda'}
Y_{l\Lambda}(\Omega_1)Y_{l\Lambda'}(\Omega_2)
\left[
C^{j m}_{l\Lambda,\frac{1}{2}s_1}C^{j -m}_{l\Lambda',\frac{1}{2}s_2}
-C^{j m}_{l\Lambda',\frac{1}{2}s_2}C^{j -m}_{l\Lambda,\frac{1}{2}s_1}
\right].\qquad
\label{Rkap}
\end{eqnarray}
It is obvious that
$\kappa(\br,\uparrow;\br',\downarrow)\neq -\kappa(\br,\downarrow;\br',\uparrow)$, 
i.e. in the coordinate
representation the spin structure of $\kappa$ has nothing common with
(\ref{anom}).

The anomalous density defined by (\ref{Rkap}) has not definite angular
momentum~$J$ and spin~$S$. It can be represented as the sum of several 
terms with definite $J, S$. We have:
\begin{eqnarray}
\phi_{ljm}(1)\phi_{lj-m}(2)&=&
\sum_{0\le J \le 2j}C^{J0}_{jm,j-m}
\{\phi_j(1)\otimes\phi_j(2)\}_{J0}
\nonumber\\
&=&C^{00}_{jm,j-m}\{\phi_j(1)\otimes\phi_j(2)\}_{00}+
\sum_{1\le J \le 2j}C^{J0}_{jm,j-m}
\{\phi_j(1)\otimes\phi_j(2)\}_{J0}.
\label{phi12}
\end{eqnarray}
We are interested in the monopole pairing only, so we omit all terms 
except the first one:
\begin{eqnarray}
&&\left[\phi_{ljm}(1)\phi_{lj-m}(2)\right]_{J=0}=
C^{00}_{jm,j-m}\{\phi_j(1)\otimes\phi_j(2)\}_{00}
\nonumber\\
&&=(-1)^{j-m}\frac{1}{\sqrt{2j+1}}
\sum_{\nu,\sigma}C^{00}_{j\nu,j\sigma}
\phi_{j\nu}(1)\phi_{j\sigma}(2)
=\frac{1}{2j+1}\sum_{\nu}(-1)^{\nu-m}
\phi_{j\nu}(1)\phi_{j-\nu}(2).
\label{phiJ0}
\end{eqnarray}
Remembering the definition of $\phi$ function we find
\begin{eqnarray}
&&(-1)^m\left[\phi_{ljm}(\Omega_1,s_1)\phi_{lj-m}(\Omega_2,s_2)\right]_{J=0}=
\nonumber\\
&&=\frac{1}{2j+1}\sum_{\nu}(-1)^{\nu}
\sum_{\Lambda,\sigma}
\sum_{\Lambda',\sigma'}
C^{j \nu}_{l\Lambda,\frac{1}{2}\sigma}
C^{j-\nu}_{l\Lambda',\frac{1}{2}\sigma'}
Y_{l\Lambda}(\Omega_1)
Y_{l\Lambda'}(\Omega_2)
\chi_{\frac{1}{2}\sigma}(s_1)
\chi_{\frac{1}{2}\sigma'}(s_2).
\label{phiJ0'}
\end{eqnarray}
The direct product of spin functions in this formula can be written as
\begin{eqnarray}
&&\chi_{\frac{1}{2}\sigma}(s_1)\chi_{\frac{1}{2}\sigma'}(s_2)=
\sum_{S,\Sigma}C^{S\Sigma}_{\frac{1}{2}\sigma,\frac{1}{2}\sigma'}
\{\chi_{\frac{1}{2}}(s_1)\otimes\chi_{\frac{1}{2}}(s_2)\}_{S\Sigma}
\nonumber\\
&&=C^{00}_{\frac{1}{2}\sigma,\frac{1}{2}\sigma'}
\{\chi_{\frac{1}{2}}(s_1)\otimes\chi_{\frac{1}{2}}(s_2)\}_{00}
+\sum_{\Sigma}C^{1\Sigma}_{\frac{1}{2}\sigma,\frac{1}{2}\sigma'}
\{\chi_{\frac{1}{2}}(s_1)\otimes\chi_{\frac{1}{2}}(s_2)\}_{1\Sigma}.
\label{chi12}
\end{eqnarray}
According to this result the formula for $\kappa$ consists of two
terms: the one with $S=0$ and another one with $S=1$.
It was shown in the paper~\cite{Sandu} that the term 
with $S=1$ is an order of magnitude less than the term with $S=0$, so
we can neglect by it. Then
\begin{eqnarray}
\chi_{\frac{1}{2}\sigma}(s_1)\chi_{\frac{1}{2}\sigma'}(s_2)&&=
(-1)^{\frac{1}{2}-\sigma}\frac{1}{\sqrt2}\delta_{\sigma,-\sigma'}
\{\chi_{\frac{1}{2}}(s_1)\otimes\chi_{\frac{1}{2}}(s_2)\}_{00}
\nonumber\\
&&=(-1)^{\frac{1}{2}-\sigma}\frac{1}{\sqrt2}\delta_{\sigma,-\sigma'}
\sum_{\nu,\nu'}C^{00}_{\frac{1}{2}\nu,\frac{1}{2}\nu'}
\chi_{\frac{1}{2}\nu}(s_1)\chi_{\frac{1}{2}\nu'}(s_2)
\nonumber\\
&&=(-1)^{\frac{1}{2}-\sigma}\frac{1}{\sqrt2}\delta_{\sigma,-\sigma'}
\sum_{\nu=-1/2}^{1/2}(-1)^{\frac{1}{2}-\nu}\frac{1}{\sqrt2}
\chi_{\frac{1}{2}\nu}(s_1)\chi_{\frac{1}{2}-\nu}(s_2)
\nonumber\\
&&=(-1)^{\frac{1}{2}-\sigma}\frac{1}{2}\delta_{\sigma,-\sigma'}
\left[
\chi_{\frac{1}{2}\frac{1}{2}}(s_1)\chi_{\frac{1}{2}-\frac{1}{2}}(s_2)
-\chi_{\frac{1}{2}-\frac{1}{2}}(s_1)\chi_{\frac{1}{2}\frac{1}{2}}(s_2)
\right]
\nonumber\\
&&=\frac{1}{2}\delta_{\sigma,-\sigma'}(-1)^{\frac{1}{2}-\sigma}
\left[
\delta_{s_1 \frac{1}{2}}\delta_{s_2 -\frac{1}{2}}
-\delta_{s_1 -\frac{1}{2}}\delta_{s_2 \frac{1}{2}}
\right].
\label{chiS0}
\end{eqnarray}
Inserting this result into (\ref{phiJ0'}) we find
\begin{eqnarray}
&&(-1)^m\left[\phi_{ljm}(\Omega_1,s_1)\phi_{lj-m}(\Omega_2,s_2)\right]_{J=0}^{S=0}=
\nonumber\\
&&=\frac{1}{2}
\left[
\delta_{s_1 \frac{1}{2}}\delta_{s_2 -\frac{1}{2}}
-\delta_{s_1 -\frac{1}{2}}\delta_{s_2 \frac{1}{2}}
\right]
\frac{1}{2j+1}
\sum_{\Lambda,\Lambda'}
Y_{l\Lambda}(\Omega_1)Y_{l\Lambda'}(\Omega_2)
\sum_{\nu,\sigma}(-1)^{\nu+\frac{1}{2}-\sigma}
C^{j \nu}_{l\Lambda,\frac{1}{2}\sigma}
C^{j-\nu}_{l\Lambda',\frac{1}{2}-\sigma}
\nonumber\\
&&=\frac{1}{2}
\left[
\delta_{s_1 \frac{1}{2}}\delta_{s_2 -\frac{1}{2}}
-\delta_{s_1 -\frac{1}{2}}\delta_{s_2 \frac{1}{2}}
\right]
\frac{1}{2j+1}
\sum_{\Lambda,\Lambda'}
Y_{l\Lambda}(\Omega_1)Y_{l\Lambda'}(\Omega_2)
\nonumber \\ && \times
\sum_{\nu,\sigma}
(-1)^{\frac{1}{2}+\Lambda}
\frac{2j+1}{2l+1}(-1)^{1+j+\frac{1}{2}-l}
C^{l\Lambda}_{j\nu,\frac{1}{2}-\sigma}
C^{l-\Lambda'}_{j\nu,\frac{1}{2}-\sigma}
\nonumber\\
&&=\frac{1}{2}
\left[
\delta_{s_1 \frac{1}{2}}\delta_{s_2 -\frac{1}{2}}
-\delta_{s_1 -\frac{1}{2}}\delta_{s_2 \frac{1}{2}}
\right]
\frac{1}{2l+1}
(-1)^{j-l}\sum_{\Lambda,\Lambda'}
Y_{l\Lambda}(\Omega_1)Y_{l\Lambda'}(\Omega_2)
(-1)^{\Lambda}\delta_{\Lambda,-\Lambda'}
\nonumber\\
&&=\frac{1}{2}
\left[
\delta_{s_1 \frac{1}{2}}\delta_{s_2 -\frac{1}{2}}
-\delta_{s_1 -\frac{1}{2}}\delta_{s_2 \frac{1}{2}}
\right]
(-1)^{j-l}\frac{1}{4\pi}
P_{l}(\cos\Omega_{12}),
\label{phiJS0}
\end{eqnarray}
where $P_{l}(\cos\Omega_{12})$ is Legendre polynomial and $\Omega_{12}$ 
is the angle between vectors $\br_1$ and $\br_2$.
With the help of this result formula (\ref{Rkap}) is transformed into
\begin{eqnarray}
\kappa(\br_1,s_1;\br_2,s_2)_{J=0}^{S=0}=
\left[
\delta_{s_1 \frac{1}{2}}\delta_{s_2 -\frac{1}{2}}
-\delta_{s_1 -\frac{1}{2}}\delta_{s_2 \frac{1}{2}}
\right]
\frac{1}{4\pi}\sum_{nljm>0}(uv)_{nljm}
\R_{nlj}(r_1)\R_{nlj}(r_2)P_{l}(\cos\Omega_{12}).\qquad
\label{kapJS0}
\end{eqnarray}
Now it is obvious that in the coordinate representation $\kappa$ with
$J=0, S=0$ has the spin structure similar to the one demonstrated by
formula (\ref{anom}):
\begin{equation}
\kappa(\br_1,s_1;\br_2,s_2)_{J=0}^{S=0}=
{\qquad 0\;\qquad \kappa(\br_1,\br_2)
\choose -\kappa(\br_1,\br_2)\;\qquad 0 \qquad}
\label{Ranom}
\end{equation}
with
\begin{eqnarray}
\kappa(\br_1,\br_2)=
\frac{1}{4\pi}\sum_{nljm>0}(uv)_{nljm}
\R_{nlj}(r_1)\R_{nlj}(r_2)P_{l}(\cos\Omega_{12}).
\label{kap0}
\end{eqnarray}

\section{ }
\label{AppB}

{\bf Wigner transformation}

The Wigner Transform (WT) of the single-particle operator matrix
$\hat F_{\br_1,\sigma;\br_2,\sigma'}$ is defined as
\begin{eqnarray}\label{B.1}
[\hat F_{\br_1,\sigma;\br_2,\sigma'}]_{\rm WT}\equiv
F_{\sigma,\sigma'}(\br,\bp)
=\int d\bs\, {\rm e}^{-i\bp\cdot\bs/\hbar}
\hat F_{\br+\bs/2,\sigma;\br-\bs/2,\sigma'}
\end{eqnarray}
with $\br=(\br_1+\br_2)/2$ and $\bs=\br_1-\br_2.$
It is easy to derive a pair of useful relations. The first one is
\begin{eqnarray}\label{B.2}
F_{\sigma,\sigma'}^*(\br,\bp)\!\!\!\!\!
&&=\int d\bs\, {\rm e}^{i\bp\cdot\bs/\hbar}
\hat F^*_{\br+\bs/2,\sigma;\br-\bs/2,\sigma'}
=\int d\bs\, {\rm e}^{-i\bp\cdot\bs/\hbar}
\hat F^*_{\br-\bs/2,\sigma;\br+\bs/2,\sigma'}
\nonumber\\
&&=\int d\bs\, {\rm e}^{-i\bp\cdot\bs/\hbar}
\hat F^{\dagger}_{\br+\bs/2,\sigma';\br-\bs/2,\sigma}=
[\hat F^{\dagger}_{\br_1,\sigma';\br_2,\sigma}]_{\rm WT},
\end{eqnarray}
i.e., $[\hat F^{\dagger}_{\br_1,\sigma;\br_2,\sigma'}]_{\rm WT}
=[\hat F_{\br_1,\sigma';\br_2,\sigma}]_{\rm WT}^*
=F_{\sigma'\sigma}^*(\br,\bp).$
The second relation is
\begin{eqnarray}\label{B.3}
\bar F_{\sigma\sigma'}(\br,\bp)\!\!\!\!\!&&\equiv F_{\sigma\sigma'}(\br,-\bp)
=\int d\bs\, {\rm e}^{i\bp\cdot\bs/\hbar}
\hat F_{\br+\bs/2,\sigma;\br-\bs/2,\sigma'}
\nonumber\\
&&=\int d\bs\, {\rm e}^{-i\bp\cdot\bs/\hbar}
\hat F_{\br-\frac{\bs}{2},\sigma;\br+\frac{\bs}{2},\sigma'}
=\int d\bs\, {\rm e}^{-i\bp\cdot\bs/\hbar}
[\hat F^{\dagger}_{\br+\bs/2,\sigma';\br-\bs/2,\sigma}]^*.
\end{eqnarray}
For the hermitian operators $\hat \rho$ and $\hat h$ this latter relation gives
$$[\hat\rho^*_{\br_1,\sigma;\br_2,\sigma}]_{\rm WT}
=\rho_{\sigma\sigma}(\br,-\bp)\ \mbox{ and }\
[\hat h^*_{\br_1,\sigma;\br_2,\sigma}]_{\rm WT}
=h_{\sigma\sigma}(\br,-\bp).$$

The Wigner transform of the product of two matrices $F$ and $G$ is
\begin{equation}\label{B.4}
[\hat F\hat G]_{\rm WT}=F(\br,\bp)\exp\left(\frac{i\hbar}{2}
\stackrel{\leftrightarrow}{\Lambda}\right)G(\br,\bp),
\end{equation}
where the symbol
 $\stackrel{\leftrightarrow}{\Lambda}$
 stands for the Poisson bracket operator
\begin{equation}
\nonumber \displaystyle
\stackrel{\leftrightarrow}{\Lambda}
=\sum_{i=1}^3\left(
\frac{\stackrel{\gets}{\partial} }{\partial r_i}\frac{
\stackrel{\to}{\partial} }{\partial p_i}
-\frac{\stackrel{\gets}{\partial} }{\partial p_i}\frac{
\stackrel{\to}{\partial} }{\partial r_i}\right).
\end{equation}

\section{ }
\label{AppC}

All derivations of this section will be done in the approximation of
spherical symmetry. The inclusion of deformation makes the calculations
more cumbersome without changing the final conclusions.
Let us consider, as an example, the integral 
$$I_h=\int\! d(\bp,\br)\{r\otimes p\}_{\lambda\mu}
[h^{\uparrow\downarrow}f^{\downarrow\uparrow}-h^{\downarrow\uparrow}f^{\uparrow\downarrow}].$$
It can be divided in two parts corresponding to the contributions
of spin-orbital and spin-spin potentials: $I_h=I_{so}+I_{ss},$
where
$$I_{so}=-\frac{\hbar}{\sqrt2}\eta\int\! d(\bp,\br)\{r\otimes p\}_{\lambda\mu}
[l_{-1}f^{\downarrow\uparrow} + l_{1}f^{\uparrow\downarrow}],$$
$$I_{ss}=\int\! d(\bp,\br)\{r\otimes p\}_{\lambda\mu}
[V_{\tau}^{\uparrow\downarrow}f^{\downarrow\uparrow}-V_{\tau}^{\downarrow\uparrow}f^{\uparrow\downarrow}],$$
$V_{\tau}^{ss'}$ being defined in (\ref{Vss}).
It is easy to see that the integral $I_{so}$ generate moments of 
fourth order. 
According to the rules of the WFM method~\cite{Bal} this integral 
is neglected.

Let us analyze the integral $I_{ss}$ (to be definite, for protons).
In this case
$$V_p^{\uparrow\downarrow}(\br)=3\frac{\hbar^2}{8}\chi n_p^{\uparrow\downarrow}(\br)
+\frac{\hbar^2}{4}\bar\chi n_n^{\uparrow\downarrow}(\br),$$
$$V_p^{\downarrow\uparrow}(\br)=3\frac{\hbar^2}{8}\chi n_p^{\downarrow\uparrow}(\br)
+\frac{\hbar^2}{4}\bar\chi n_n^{\downarrow\uparrow}(\br).$$
It is seen that $I_{ss}$ is split into four terms of 
identical structure, so it will be sufficient to analyze in detail 
only one part. For example
\begin{eqnarray}
\label{Iss4}
I_{ss4}
=\int\! d(\bp,\br)\{r\otimes p\}_{\lambda\mu}n^{\downarrow\uparrow}f^{\uparrow\downarrow}
=\int\! d^3r\{r\otimes J^{\uparrow\downarrow}\}_{\lambda\mu}n^{\downarrow\uparrow}
=\sum_{\nu,\alpha}C^{\lambda\mu}_{1\nu,1\alpha}\int\! d^3r 
r_{\nu}J^{\uparrow\downarrow}_{\alpha}n^{\downarrow\uparrow},
\end{eqnarray}
where ${ J_{\alpha}^{\uparrow\downarrow}(\br,t)=\int\frac{d^3p}{(2\pi\hbar)^3}
p_{\alpha}f^{\uparrow\downarrow}(\br,\bp,t)}$.
The variation of this integral reads
\begin{eqnarray}
\label{Var4}
\delta I_{ss4}
=\sum_{\nu,\alpha}C^{\lambda\mu}_{1\nu,1\alpha}\int\! d\,^3r r_{\nu}
\left[n^{\downarrow\uparrow}({\rm eq})\delta J^{\uparrow\downarrow}_{\alpha}+
J^{\uparrow\downarrow}_{\alpha}({\rm eq})\delta n^{\downarrow\uparrow}\right].
\end{eqnarray}
It is necessary to represent this integral in terms of the collective
variables (\ref{VarisV}). This problem can not be solved exactly, so 
we will use the approximation suggested in~\cite{Bal} and expand the 
density and current variations as a series (see appendix~\ref{AppD}).

Let us consider the second part of integral (\ref{Var4}). With the 
help of formula (\ref{Varn1}) we find
\begin{eqnarray}
\label{Var42}
&&I_{2}
\equiv\sum_{\nu,\alpha}C^{\lambda\mu}_{1\nu,1\alpha}\int\! d^3r\, r_{\nu}
J^{\uparrow\downarrow}_{\alpha}({\rm eq})\delta n^{\downarrow\uparrow}
\nonumber\\
&&=-\sum_{\nu,\alpha}C^{\lambda\mu}_{1\nu,1\alpha}\int\! d^3r\, r_{\nu}
J^{\uparrow\downarrow}_{\alpha}({\rm eq})
\sum_{\beta}(-1)^{\beta}\left\{
N^{\downarrow\uparrow}_{\beta,-\beta}(t)n^+
+\sum_{\gamma}(-1)^{\gamma}N^{\downarrow\uparrow}_{\beta,\gamma}(t)\frac{1}{r}
\frac{\partial n^+}{\partial r}r_{-\beta}r_{-\gamma}
\right\}.\qquad
\end{eqnarray}
Let us analyze at first the more simple part of this expression:
\begin{eqnarray}
\label{Var422}
I_{2,1}\equiv -\sum_{\beta}(-1)^{\beta}N^{\downarrow\uparrow}_{\beta,-\beta}(t)
\int\! d^3r\, 
\sum_{\nu,\alpha}C^{\lambda\mu}_{1\nu,1\alpha}
r_{\nu}J^{\uparrow\downarrow}_{\alpha}({\rm eq})n^+=
-\sum_{\beta}(-1)^{\beta}N^{\downarrow\uparrow}_{\beta,-\beta}X_{\lambda\mu}.
\end{eqnarray}
We are interested in the value of $\mu=1$, therefore it is necessary 
to analyze two possibilities: $\lambda=1$ and $\lambda=2$.

In the case $\lambda=1,\,\mu=1$ we have
\begin{eqnarray}
\label{Lam11}
X_{11}\equiv \int\! d^3r\, n^+\sum_{\nu,\alpha}C^{11}_{1\nu,1\alpha}
r_{\nu}J^{\uparrow\downarrow}_{\alpha}({\rm eq})
=\int\! d^3r\, n^+\frac{1}{\sqrt2}\left[r_1J^{\uparrow\downarrow}_0({\rm eq})-r_0J^{\uparrow\downarrow}_1({\rm eq})\right].
\end{eqnarray}

Inserting the definition (\ref{Jphi}) into (\ref{Lam11}) one finds
\begin{eqnarray}
\label{L11}
X_{11}=\frac{i\hbar}{2}\frac{1}{\sqrt2}\sum_{nljm} v_{nljm}^2
\int\! d^3r\, n^+(r)\R_{nlj}^2(r)C^{jm}_{l\Lambda,\frac12 \frac12}
C^{jm}_{l\Lambda',\frac12 -\frac12}\left[Y_{l\Lambda}(r_1\nabla_0-r_0\nabla_1)Y_{l\Lambda'}^*\right.
\nonumber\\
\left.-Y_{l\Lambda'}^*(r_1\nabla_0-r_0\nabla_1)Y_{l\Lambda}\right]
\end{eqnarray}
with $\Lambda=m-\frac12$ and $\Lambda'=m+\frac12$. Remembering the definition
(\ref{lxyz}) of the angular momentum $\hat l_1=\hbar(r_0\nabla_1-r_1\nabla_0)$
and using the relation~\cite{Var} 
$\hat l_{\pm1}Y_{l\Lambda}=\mp\frac1{\sqrt2}\sqrt{(l\mp\Lambda)(l\pm\Lambda+1)}Y_{l\Lambda\pm1}$
one transforms (\ref{L11}) into
\begin{eqnarray}
\label{L11a}
X_{11}
=-\frac{i\hbar}{2}\frac{1}{\sqrt2}\sum_{nljm} v_{nljm}^2
\int\!dr n^+(r)r^2\R_{nlj}^2(r)C^{jm}_{l\Lambda,\frac12 \frac12}
C^{jm}_{l\Lambda',\frac12 -\frac12}\frac2{\sqrt2}\sqrt{(l-\Lambda)(l+\Lambda+1)}
\nonumber\\
=-i\hbar\sum_{nl}\sum_{m=\frac12}^{|l-\frac12|}
\frac{[(l+\frac12)^2-m^2]}{2l+1}
\int\!dr n^+(r)r^2\left[v_{nll+\frac12m}^2\R_{nll+\frac12}^2(r)
-v_{nl|l-\frac12|m}^2\R_{nl|l-\frac12|}^2(r)\right].
\end{eqnarray}
As it is seen, the value of this integral is determined by the difference
of the wave functions of spin-orbital partners $(v\R)_{nll+\frac12m}^2
-(v\R)_{nl|l-\frac12|m}^2$, which is usually very small, so we will
neglect it. The only remarkable contribution can appear in the vicinity
of the Fermi surface, where some spin-orbital partners with 
$j=l+\frac12$ and $j=|l-\frac12|$ can be disposed on different sides of
the Fermi surface. In reality such situation happens very frequently,
nevertheless we will not take into account this effect, because the 
values of the corresponding integrals are considerably smaller than
$R_{20}({\rm eq})$, the typical input parameter of our model.

Let us consider now the integral $I_{2,1}$ (formula (\ref{Var422}))
for the case $\lambda=2,\,\mu=1$. We have
\begin{eqnarray}
\label{Lam21}
X_{21}\equiv \int\! d^3r n^+\sum_{\nu,\alpha}C^{21}_{1\nu,1\alpha}
r_{\nu}J^{\uparrow\downarrow}_{\alpha}({\rm eq})
=\int\! d^3r n^+C^{21}_{11,10}\left[r_1J^{\uparrow\downarrow}_0({\rm eq})+r_0J^{\uparrow\downarrow}_1({\rm eq})\right].
\end{eqnarray}
With the help of formulae (\ref{Jphi}) one can show by
simple algebraic transformations that 
\begin{eqnarray}
\label{Lam21'}
\int\! d\Omega \,r_1J^{\uparrow\downarrow}_0({\rm eq})=-\int\! d\Omega \,r_0J^{\uparrow\downarrow}_1({\rm eq}),
\end{eqnarray}
where $\int\! d\Omega$ means the integration over angles. As a result
$X_{21}=0$.

Let us consider the second, more complicated, part of 
integral $I_2$:
\begin{eqnarray}
\label{Var421}
I_{2,2}
&=&-\sum_{\beta,\gamma}(-1)^{\beta+\gamma}N^{\downarrow\uparrow}_{-\beta,-\gamma}(t)
\sum_{\nu,\alpha}C^{\lambda\mu}_{1\nu,1\alpha}\int\! d\,^3r r_{\nu}
J^{\uparrow\downarrow}_{\alpha}({\rm eq})
\frac{1}{r}\frac{\partial n^+}{\partial r}r_{\beta}r_{\gamma}
\nonumber\\
&=&-\sum_{\beta,\gamma}(-1)^{\beta+\gamma}
N^{\downarrow\uparrow}_{-\beta,-\gamma}(t)X_{\lambda\mu}'(\beta,\gamma).
\end{eqnarray}
The case $\lambda=1,\,\mu=1$:
\begin{eqnarray}
\label{Lam11'}
&&X_{11}'(\beta,\gamma)
=\frac{1}{\sqrt2}\int\! d^3r\, \frac{1}{r}\frac{\partial n^+}{\partial r}
\left[r_1J^{\uparrow\downarrow}_0({\rm eq})-r_0J^{\uparrow\downarrow}_1({\rm eq})\right]r_{\beta}r_{\gamma}
\nonumber\\
&&=-\frac{i\hbar}{4}\sum_{nljm} v_{nljm}^2
\int\! d^3r\, \frac{1}{r}\frac{\partial n^+}{\partial r}
\R_{nlj}^2(r)C^{jm}_{l\Lambda,\frac12 \frac12}
C^{jm}_{l\Lambda',\frac12 -\frac12}
\nonumber\\&&\times
\sqrt{(l-\Lambda)(l+\Lambda+1)}
\left[Y_{l\Lambda}Y_{l\Lambda}^*
+Y_{l\Lambda'}^*Y_{l\Lambda'}\right]
r_{\beta}r_{\gamma}.\qquad
\end{eqnarray}
The angular part of this integral is
\begin{eqnarray}
\label{Ome}
\int\!\! d\Omega
\left[Y_{l\Lambda}Y_{l\Lambda}^*
+Y_{l\Lambda'}^*Y_{l\Lambda'}\right]r_{\beta}r_{\gamma}
=\sum_{L,M}C^{LM}_{1\beta,1\gamma}
\int\!\! d\Omega
\left[Y_{l\Lambda}Y_{l\Lambda}^*
+Y_{l\Lambda'}^*Y_{l\Lambda'}\right]\{r\otimes r\}_{LM}
\nonumber\\
=-\frac{2}{\sqrt3}\,r^2 C^{00}_{1\beta,1\gamma}+\sqrt{\frac{8\pi}{15}}r^2
\sum_{M}C^{2M}_{1\beta,1\gamma}
\int\!\! d\Omega 
\left[Y_{l\Lambda}Y_{l\Lambda}^*+Y_{l\Lambda'}^*Y_{l\Lambda'}\right]Y_{2M}
\nonumber\\
=\frac{2}{3}\,r^2\delta_{\gamma,-\beta}\left\{1-\sqrt{\frac{5}{2}}\, C^{l0}_{l0,20}
C^{1\beta}_{1\beta,20}\left[C^{l\Lambda}_{l\Lambda,2M}+C^{l\Lambda'}_{l\Lambda',2M}\right]\right\}.
\end{eqnarray}
Therefore
\begin{eqnarray}
\label{I21}
X_{11}'(\beta,\gamma)
=
-\frac{i\hbar}{6}
\delta_{\gamma,-\beta}
\int\! dr\, \frac{\partial n^+(r)}{\partial r}\,r^3
\sum_{nljm}
\left\{1-\sqrt{\frac{5}{2}}\,C^{1\beta}_{1\beta,20}
C^{l0}_{l0,20}\left[C^{l\Lambda}_{l\Lambda,20}+C^{l\Lambda'}_{l\Lambda',20}\right]
\right\}
\nonumber\\
 \times v_{nljm}^2\R_{nlj}^2(r)
C^{jm}_{l\Lambda,\frac12 \frac12}
C^{jm}_{l\Lambda',\frac12 -\frac12}
\sqrt{(l-\Lambda)(l+\Lambda+1)}
\nonumber\\
=
-\frac{i\hbar}{3}
\delta_{\gamma,-\beta}
\sum_{nl}
\left\{1-\sqrt{\frac{5}{2}}\,C^{1\beta}_{1\beta,20}
C^{l0}_{l0,20}\left[C^{l\Lambda}_{l\Lambda,20}+C^{l\Lambda'}_{l\Lambda',20}\right]
\right\}
\nonumber\\
\times\sum_{m=\frac12}^{|l-\frac12|}
\frac{[(l+\frac12)^2-m^2]}{2l+1}
\int\! dr \frac{\partial n^+(r)}{\partial r}r^3
\left[v_{nll+\frac12m}^2\R_{nll+\frac12}^2(r)
-v_{nl|l-\frac12|m}^2\R_{nl|l-\frac12|}^2(r)\right].
\end{eqnarray}
One sees that, exactly as in formula (\ref{L11a}), the value of this
integral is determined by the difference of the wave functions of 
spin-orbital partners $(v\R)_{nll+\frac12m}^2-(v\R)_{nl|l-\frac12|m}^2$ near the Fermi surface, 
so it can be omitted together with $X_{11}$ following the same arguments.

The case $\lambda=2, \mu=1$ can be analyzed in full analogy with
formulae (\ref{Lam21},\ref{Lam21'}) that allows us to take $X_{21}'=0$.

So, we have shown that the integral $I_2$ can be approximated by zero.
Let us consider now the first part of the integral (\ref{Var4}):
\begin{eqnarray}
\label{Var41}
I_1
&=&\sum_{\nu,\alpha}C^{\lambda\mu}_{1\nu,1\alpha}\int\! d^3r r_{\nu}
n^{\downarrow\uparrow}({\rm eq})\delta J^{\uparrow\downarrow}_{\alpha}
=\sum_{\nu,\alpha}C^{\lambda\mu}_{1\nu,1\alpha}\int\! d^3r r_{\nu}
n^{\downarrow\uparrow}({\rm eq})n^+(r)
\sum_{\gamma}(-1)^{\gamma}K^{\uparrow\downarrow}_{\alpha,-\gamma}(t)r_{\gamma}
\nonumber\\
&=&\sum_{\nu,\alpha}C^{\lambda\mu}_{1\nu,1\alpha}
\sum_{\gamma}(-1)^{\gamma}K^{\uparrow\downarrow}_{\alpha,-\gamma}(t)
\int\! d^3r 
n^{\downarrow\uparrow}({\rm eq})n^+(r)
\sum_{L,M}C^{LM}_{1\nu,1\gamma}
\{r\otimes r\}_{LM}.
\end{eqnarray}
This integral can be estimated in the approximation of constant density
$n^+(r)=n_0$. Then
\begin{eqnarray}
\label{Var41a}
I_1
=n_0\sum_{\nu,\alpha}C^{\lambda\mu}_{1\nu,1\alpha}
\sum_{\gamma}(-1)^{\gamma}K^{\uparrow\downarrow}_{\alpha,-\gamma}(t)
\sum_{L,M}C^{LM}_{1\nu,1\gamma}
R^{\downarrow\uparrow}_{LM}({\rm eq})=0.
\end{eqnarray}
It is easy to show, that $R^{\downarrow\uparrow}_{LM}({\rm eq})=0.$ Let us consider, for 
example, the case with $L=2$:
\begin{eqnarray}
\label{Rd}
R^{\downarrow\uparrow}_{2M}
=\int\! d(\bp,\br) \{r\otimes r\}_{2M}f^{\downarrow\uparrow}(\br,\bp)
=\int\! d^3r \{r\otimes r\}_{2M}n^{\downarrow\uparrow}(\br)
=\sqrt{\frac{8\pi}{15}}\int\! d^3r r^2Y_{2M}n^{\downarrow\uparrow}(\br).\qquad
\end{eqnarray}
By definition
\begin{eqnarray}
\label{neq}
n^{ss'}(\br)=
\int\frac{d^3p}{(2\pi\hbar)^3}f^{ss'}(\br,\bp)
=\sum_k v_k^2
\psi_k(\br,s)\psi_k^*(\br,s')
\end{eqnarray}
with $\psi_k$ defined in (\ref{phi}). Therefore
\begin{eqnarray}
\label{Rd'}
R^{\downarrow\uparrow}_{2M}
=\sqrt{\frac{8\pi}{15}}
\int\!d^3r r^2Y_{2M}\sum_{nljm} v_{nljm}^2\R_{nlj}^2(r)
C^{jm}_{l\Lambda',\frac12 -\frac12}
C^{jm}_{l\Lambda,\frac12 \frac12}
Y_{l\Lambda'}Y_{l\Lambda}^*
\nonumber\\
=\sqrt{\frac{2}{3}}\sum_{nljm} v_{nljm}^2
\int\!dr r^4\R_{nlj}^2(r)
C^{jm}_{l\Lambda,\frac12 \frac12}
C^{jm}_{l\Lambda',\frac12 -\frac12}
C^{l0}_{20,l0}
C^{l\Lambda}_{2M,l\Lambda'}=0,
\end{eqnarray}
where $\Lambda=m-\frac12$ and $\Lambda'=m+\frac12$.
The zero is obtained due to summation over $m$. Really, the
product $C^{jm}_{l\Lambda,\frac12 \frac12}
C^{jm}_{l\Lambda',\frac12 -\frac12}=\pm\frac{\sqrt{(l+\frac12)^2-m^2}}{2l+1}$
(for $j=l\pm\frac12$) does not depend on the sign of $m$, whereas the
Clebsh-Gordan coefficient $C^{l\Lambda}_{2M,l\Lambda'}
=C^{lm-\frac{1}{2}}_{2-1,lm+\frac12}$ changes its
sign together with~$m$.

Summarizing, we have demonstrated that $I_1+I_2\simeq 0$, hence one can 
neglect the contribution of the integrals $I_h$ in the equations of motion.

$\bullet$ It is necessary to analyze also the integrals with the weight
$\{p\otimes p\}_{\lambda\mu}$:
$$I'_h=\int\! d(\bp,\br)\{p\otimes p\}_{\lambda\mu}
\left[h^{\uparrow\downarrow}f^{\downarrow\uparrow}-h^{\downarrow\uparrow}f^{\uparrow\downarrow}\right]=I'_{so}
+I'_{ss}.$$
Again we neglect the contribution of the spin-orbital part $I'_{so}$,
which generates fourth order moments. For the spin-spin contribution, we have
\begin{eqnarray}
\label{I'ss4}
I'_{ss4}
=\int\! d(\bp,\br)\{p\otimes p\}_{\lambda\mu}n^{\downarrow\uparrow}(\br,t)f^{\uparrow\downarrow}(\br,\bp,t)
=\int\! d^3r\Pi^{\uparrow\downarrow}_{\lambda\mu}(\br,t)n^{\downarrow\uparrow}(\br,t),
\end{eqnarray}
where ${\di \Pi^{\uparrow\downarrow}_{\lambda\mu}(\br,t)=\int\frac{d^3p}{(2\pi\hbar)^3}
\{p\otimes p\}_{\lambda\mu}f^{\uparrow\downarrow}(\br,\bp,t)}$ is the pressure tensor.
The variation of this integral reads:
\begin{eqnarray}
\label{VarI'}
\delta I'_{ss4}
=\int\! d^3r\left[n^{\downarrow\uparrow}({\rm eq})\delta\Pi^{\uparrow\downarrow}_{\lambda\mu}(\br,t)
+\Pi^{\uparrow\downarrow}_{\lambda\mu}({\rm eq})\delta n^{\downarrow\uparrow}(\br,t)\right].
\end{eqnarray}
The pressure tensor variation is defined in appendix~\ref{AppD}.
With formula (\ref{VarP1}) one finds for the first part of 
(\ref{VarI'}):
\begin{eqnarray}
\label{VarI'1}
I'_1
=\int\! d^3r n^{\downarrow\uparrow}({\rm eq})\delta\Pi^{\uparrow\downarrow}_{\lambda\mu}(\br,t)
\simeq
T^{\uparrow\downarrow}_{\lambda\mu}(t)\int\! d^3r n^{\downarrow\uparrow}({\rm eq})n^+(r)
\simeq
T^{\uparrow\downarrow}_{\lambda\mu}(t)n_0\int\! d^3r n^{\downarrow\uparrow}({\rm eq})=0.\quad
\end{eqnarray}
The last equality follows obviously from the definition of $n^{\downarrow\uparrow}$ (\ref{neq}).

The second part of (\ref{VarI'}) reads:
\begin{eqnarray}
\label{VarI'2}
I'_2
&=&\int\! d^3r\Pi^{\uparrow\downarrow}_{\lambda\mu}({\rm eq})\delta n^{\downarrow\uparrow}(r,t)
\nonumber\\
&=&-\sum_{\beta}(-1)^{\beta}\int\! d^3r\Pi^{\uparrow\downarrow}_{\lambda\mu}({\rm eq})
\left\{
N^{\downarrow\uparrow}_{\beta,-\beta}(t)n^+
+\sum_{\gamma}(-1)^{\gamma}N^{\downarrow\uparrow}_{\beta,\gamma}(t)\frac{1}{r}
\frac{\partial n^+}{\partial r}r_{-\beta}r_{-\gamma}
\right\}.
\end{eqnarray}
Let us consider at first the simpler part of this integral
\begin{eqnarray}
\label{I'22}
-\sum_{\beta}(-1)^{\beta}N^{\downarrow\uparrow}_{\beta,-\beta}(t)
\int\! d^3r\Pi^{\uparrow\downarrow}_{\lambda\mu}({\rm eq})n^+(r).
\end{eqnarray}
The value of the last integral is determined by the angular structure
of the function $\Pi^{\uparrow\downarrow}_{\lambda\mu}(\br)$. We are interested in
$\lambda=2, \mu=1$. By definition
\begin{eqnarray}
\label{Pieq}
\Pi^{\uparrow\downarrow}_{21}(\br)=\int\frac{d^3p}{(2\pi\hbar)^3}
\{p\otimes p\}_{21}f^{\uparrow\downarrow}(\br,\bp)
=\sum_{\nu,\sigma}C^{21}_{1\nu,1\sigma}
\int\frac{d^3p}{(2\pi\hbar)^3}
p_{\nu}p_{\sigma}f^{\uparrow\downarrow}(\br,\bp)
\nonumber\\
=2C^{21}_{11,10}
\int\frac{d^3p}{(2\pi\hbar)^3}
p_{1}p_{0}f^{\uparrow\downarrow}(\br,\bp)
=-\frac{\hbar^2}{2\sqrt2}\left[(\nabla'_1-\nabla_1)(\nabla'_0-\nabla_0)
\rho(\br'\uparrow,\br\downarrow)\right]_{\br'=\br}
\nonumber\\
=-\frac{\hbar^2}{2\sqrt2}\sum_k v_k^2
\left\{
[\nabla_1\nabla_0\psi_k(\br,\uparrow)]\psi_k^*(\br,\downarrow)
-[\nabla_1\psi_k(\br,\uparrow)][\nabla_0\psi_k^*(\br,\downarrow)]
\right.\nonumber\\ \left.
-[\nabla_0\psi_k(\br,\uparrow)][\nabla_1\psi_k^*(\br,\downarrow)]
+\psi_k(\br,\uparrow)[\nabla_1\nabla_0\psi_k^*(\br,\downarrow)]
\right\}
\end{eqnarray}
with $\psi_k$ being defined by (\ref{phi}).
Taking into account formulae~\cite{Var}
$$\nabla_{\pm1}Y_{l\lambda}=
-\sqrt{\frac{(l\pm\Lambda+1)(l\pm\Lambda+2)}{2(2l+1)(2l+3)}}
\frac lrY_{l+1,\Lambda\pm1}
-\sqrt{\frac{(l\mp\Lambda-1)(l\mp\Lambda)}{2(2l-1)(2l+1)}}
\frac{l+1}rY_{l-1,\Lambda\pm1},
$$
$$\nabla_{0}Y_{l\lambda}=
-\sqrt{\frac{(l+1)^2-\Lambda^2}{(2l+1)(2l+3)}}
\frac lrY_{l+1,\Lambda}
+\sqrt{\frac{l^2-\Lambda^2}{(2l-1)(2l+1)}}
\frac{l+1}rY_{l-1,\Lambda}
$$
one finds that 
\begin{eqnarray}
\label{I'Pi}
&&\int\! d^3r\Pi^{\uparrow\downarrow}_{\lambda\mu}({\rm eq})n^+(r)=
\nonumber\\
&&\hbar^2\sum_{nljm} v_{nljm}^2
\int\!dr n^+(r)\R_{nlj}^2(r)
(\delta_{j,l+\frac12}-\delta_{j,l-\frac12})
\frac{l(l+1)[(l+\frac12)^2-m^2]}{(2l+3)(2l+1)(2l-1)}m=0\qquad
\end{eqnarray}
due to summation over $m$. The more complicated part of the integral 
(\ref{VarI'2}) is calculated in a similar way with the same result,
hence $I'_2=0$.

So, we have shown that $I'_1+I'_2\simeq 0$, therefore one can neglect 
by the contribution of integrals $I'_h$ (together with $I_h$) into 
equations of motion.

$\bullet$ And finally, just a few words about the integrals with the 
weight $\{r\otimes r\}_{\lambda\mu}$:
$$I''_h=\int\! d(\bp,\br)\{r\otimes r\}_{\lambda\mu}
\left[h^{\uparrow\downarrow}f^{\downarrow\uparrow}-h^{\downarrow\uparrow}f^{\uparrow\downarrow}\right]=I''_{so}
+I''_{ss}.$$
The spin-orbital part $I''_{so}$ is neglected and for the spin-spin part we have
\begin{eqnarray}
\label{I"ss4}
I''_{ss4}
=\int\! d(\bp,\br)\{r\otimes r\}_{\lambda\mu}n^{\downarrow\uparrow}(\br,t)
f^{\uparrow\downarrow}(\br,\bp,t)
=\int\! d^3r\{r\otimes r\}_{\lambda\mu}n^{\downarrow\uparrow}(\br,t)n^{\uparrow\downarrow}(\br,t).
\end{eqnarray}
The variation of this integral reads:
\begin{eqnarray}
\label{VarI"}
\delta I''_{ss4}
=\int\! d^3r\{r\otimes r\}_{\lambda\mu}[n^{\downarrow\uparrow}({\rm eq})\delta n^{\uparrow\downarrow}(\br,t)
+n^{\uparrow\downarrow}({\rm eq})\delta n^{\downarrow\uparrow}(\br,t)].
\end{eqnarray}
With the help of formulae (\ref{neq}) and (\ref{Varn1}) the subsequent
analysis becomes quite similar to that of the integral (\ref{Var41})
with the same result, i.e. $I''_h\simeq 0.$

$\bullet$ The integrals 
$\int\! d(\bp,\br)W_{\lambda\mu}
\left[h^-f^{\downarrow\uparrow}-h^{\downarrow\uparrow}f^-\right]$ and
$\int\! d(\bp,\br)W_{\lambda\mu}
\left[h^-f^{\uparrow\downarrow}-h^{\uparrow\downarrow}f^-\right]$,
where $W_{\lambda\mu}$ is any of the above mentioned weights, can
be analyzed in an analogous way with the same result.

\section{ }
\label{AppD}

According to the approximation suggested in~\cite{Bal}, the variations
of density, current, and pressure tensor are expanded as the series
\begin{eqnarray}
\label{Varn}
\delta n^{\varsigma}(\br,t)
&=&-\sum_{\beta}(-1)^{\beta}\nabla_{-\beta}\left\{n^+(\br)\left[N^{\varsigma}_{\beta}(t)
+\sum_{\gamma}(-1)^{\gamma}N^{\varsigma}_{\beta,\gamma}(t)r_{-\gamma}
\right.\right.\nonumber\\ &&\left.\left.
+\sum_{\lambda',\mu'}(-1)^{\mu'}N^{\varsigma}_{\beta,\lambda'\mu'}(t)
\{r\otimes r\}_{\lambda'-\mu'}+...\right]\right\},
\\
\label{VarJ}
\delta J^{\varsigma}_{\beta}(\br,t)
&=&n^+(\br)\!\!\left[K^{\varsigma}_{\beta}(t)
+\sum_{\gamma}(-1)^{\gamma}K^{\varsigma}_{\beta,-\gamma}(t)r_{\gamma}
\right.\nonumber\\ &&\left.
+\sum_{\lambda',\mu'}(-1)^{\mu'}K^{\varsigma}_{\beta,\lambda'-\mu'}(t)
\{r\otimes r\}_{\lambda'\mu'}+...\right],
\\
\label{VarP}
\delta \Pi^{\varsigma}_{\lambda\mu}(\br,t)
&=&n^+(\br)\!\!\left[T^{\varsigma}_{\lambda\mu}(t)
+\sum_{\gamma}(-1)^{\gamma}T^{\varsigma}_{\lambda\mu,-\gamma}(t)r_{\gamma}
\right.\nonumber\\ &&\left.
+\sum_{\lambda',\mu'}(-1)^{\mu'}T^{\varsigma}_{\lambda\mu,\lambda'-\mu'}(t)
\{r\otimes r\}_{\lambda'\mu'}+...\right].
\end{eqnarray}
Putting these series into the integrals (\ref{Var4},~\ref{VarI'}), one discovers
immediately that all terms containing expansion coefficients $N,\,K,\,T$
with odd numbers of indices disappear due to axial symmetry. Furthermore,
we truncate these series omitting all terms generating higher (than 
second) order moments. So, finally the following expressions are used:
\begin{eqnarray}
\label{Varn1}
\delta n^{\varsigma}(\br,t)
&\simeq &-\sum_{\beta}(-1)^{\beta}\nabla_{-\beta}\left\{n^+(\br)
\sum_{\gamma}(-1)^{\gamma}N^{\varsigma}_{\beta,\gamma}(t)r_{-\gamma}\right\}
\nonumber\\
&=&-\sum_{\beta}(-1)^{\beta}\left\{
N^{\varsigma}_{\beta,-\beta}(t)n^+
+\sum_{\gamma}(-1)^{\gamma}N^{\varsigma}_{\beta,\gamma}(t)\frac{1}{r}
\frac{\partial n^+}{\partial r}r_{-\beta}r_{-\gamma}
\right\},
\end{eqnarray}
\begin{eqnarray}
\label{VarJ1}
\delta J^{\varsigma}_{\beta}(\br,t)
\simeq n^+(\br)
\sum_{\gamma}(-1)^{\gamma}K^{\varsigma}_{\beta,-\gamma}(t)r_{\gamma}
\end{eqnarray}
and
\begin{eqnarray}
\label{VarP1}
\delta \Pi^{\varsigma}_{\lambda\mu}(\br,t)
\simeq n^+(\br)T^{\varsigma}_{\lambda\mu}(t).
\end{eqnarray}
The coefficients $N^{\varsigma}_{\beta,\gamma}(t)$ and $K^{\varsigma}_{\beta,-\gamma}(t)$
are connected by the linear relations with collective variables $\R^{\varsigma}_{\lambda\mu}(t)$
and $\L^{\varsigma}_{\lambda\mu}(t)$ respectively.
\begin{eqnarray}
\R_{\lambda\mu}^\varsigma
&=&\int\! d^3r \{r\otimes r\}_{\lambda\mu}\delta n^\varsigma(\br)
\nonumber\\
&=&\frac{2}{\sqrt3}\left[{\cal A}_1 C_{1\mu,10}^{\lambda\mu}N_{\mu,0}^\varsigma-
{\cal A}_2 \left(C_{1\mu+1,1-1}^{\lambda\mu}N_{\mu+1,-1}^\varsigma+
C_{1\mu-1,11}^{\lambda\mu}N_{\mu-1,1}^\varsigma\right)\right],
\end{eqnarray}
where
\begin{eqnarray}
\label{CA} 
&&{\cal A}_1=\sqrt2\, R_{20}^{\rm eq}-R_{00}^{\rm eq}=\frac{Q_{00}}{\sqrt3}\left(1+\frac{4}{3}\delta\right),\quad
{\cal A}_2= R_{20}^{\rm eq}/\sqrt2+R_{00}^{\rm eq}=-\frac{Q_{00}}{\sqrt3}\left(1-\frac{2}{3}\delta\right),\quad
\\
&&R^{\rm eq}_{20}=Q_{20}/\sqrt6,\quad R^{\rm eq}_{00}=-Q_{00}/\sqrt3,\quad
 Q_{20}=\frac43\delta Q_{00},\quad Q_{00}=A\langle r^2\rangle =\frac35 AR_0^2.\nonumber
\end{eqnarray}\begin{eqnarray}
\label{G} 
\nonumber&&N_{-1,-1}^\varsigma=-{\frac{\sqrt3\,\R_{2-2}^\varsigma}{2{\cal A}_2}},\quad
N_{-1,0}^\varsigma={\frac{\sqrt6\,\R_{2-1}^\varsigma}{4{\cal A}_1}},\quad
N_{-1,1}^\varsigma=-{\frac{\R_{00}^\varsigma+\R_{20}^\varsigma/\sqrt2}{2{\cal A}_2}},\\
\nonumber&&N_{0,-1}^\varsigma=-{\frac{\sqrt6\,\R_{2-1}^\varsigma}{4{\cal A}_2}},\quad
N_{0,0}^\varsigma={\frac {\sqrt {2}\R_{2,0}^\varsigma-\R_{0,0}^\varsigma}{2{\cal A}_1}},\quad
N_{0,1}^\varsigma=-{\frac{\sqrt6\,\R_{21}^\varsigma}{4{\cal A}_2}},\\
&&N_{1,-1}^\varsigma=N_{-1,1}^\varsigma,\quad
N_{1,0}^\varsigma={\frac{\sqrt6\,\R_{21}^\varsigma}{4{\cal A}_1}},\quad
N_{1,1}^\varsigma=-{\frac{\sqrt3\,\R_{22}^\varsigma}{2{\cal A}_2}}.
\end{eqnarray}
\begin{eqnarray}
\L_{\lambda,\mu}^\varsigma
&=&\int\! d^3r \{r\otimes \delta J^\varsigma\}_{\lambda\mu}
\nonumber\\
&=&\frac{1}{\sqrt3}(-1)^\lambda 
\left[{\cal A}_1 C_{1\mu,10}^{\lambda\mu}K_{\mu,0}^\varsigma-
{\cal A}_2 \left(C_{1\mu+1,1-1}^{\lambda\mu}K_{\mu+1,-1}^\varsigma+
C_{1\mu-1,11}^{\lambda\mu}K_{\mu-1,1}^\varsigma\right)\right].\qquad
\end{eqnarray}
\begin{eqnarray}
\label{K} 
\nonumber&&K_{-1,-1}^\varsigma=-{\frac{\sqrt3\,\L_{2-2}^\varsigma}{{\cal A}_2}},
\quad
K_{-1,0}^\varsigma={\frac{\sqrt3\,(\L_{1-1}^\varsigma+\L_{2-1}^\varsigma)}{\sqrt2\,{\cal A}_1}},
\quad
K_{-1,1}^\varsigma=-{\frac{\sqrt3\, \L_{10}^\varsigma+\L_{20}^\varsigma+\sqrt2\, \L_{00}^\varsigma}
{\sqrt2\, {\cal A}_2}},\\
\nonumber&&K_{0,-1}^\varsigma={\frac{\sqrt3\,(\L_{1-1}^\varsigma-\L_{2-1}^\varsigma)}{\sqrt2\,{\cal A}_2}},
\quad
K_{0,0}^\varsigma={\frac {\sqrt {2}\L_{2,0}^\varsigma-\L_{0,0}^\varsigma}{{\cal A}_1}},
\quad
K_{0,1}^\varsigma=-{\frac{\sqrt3\,(\L_{11}^\varsigma+\L_{21}^\varsigma)}{\sqrt2\,{\cal A}_2}},\\
&&K_{1,-1}^\varsigma={\frac{\sqrt3\, \L_{10}^\varsigma-\L_{20}^\varsigma-\sqrt2\, \L_{00}^\varsigma}
{\sqrt2\, {\cal A}_2}},
\quad
K_{1,0}^\varsigma={\frac{\sqrt3\,(\L_{21}^\varsigma-\L_{11}^\varsigma)}{\sqrt2\,{\cal A}_1}},
\quad
K_{1,1}^\varsigma=-{\frac{\sqrt3\,\L_{22}^\varsigma}{{\cal A}_2}}.
\end{eqnarray}

The coefficient $T^{\varsigma}_{\lambda\mu}(t)$ is connected with 
$\P^{\varsigma}_{\lambda\mu}(t)$ by the relation 
$\P^{\varsigma}_{\lambda\mu}(t)=AT^{\varsigma}_{\lambda\mu}(t)$, $A$ being 
the number of nucleons. 

%

\section{ }
\label{AppE}

\begin{eqnarray}
&&I^{\kappa\Delta}_{pp}(\br,p)=
\frac{r_p^3}{\sqrt{\pi}\hbar^3}{\rm e}^{-\alpha p^2}
\int\!\kappa^r(\br,p')\left[\phi_0(x)
-4\alpha^2p'^4\phi_2(x)\right]
{\rm e}^{-\alpha p'^2}p'^2dp',
\\
&&I^{\kappa\Delta}_{rp}(\br,p)=
\frac{r_p^3}{\sqrt{\pi}\hbar^3}{\rm e}^{-\alpha p^2}
\int\!\kappa^r(\br,p')[\phi_0(x)
-2\alpha p'^2\phi_1(x)]{\rm e}^{-\alpha p'^2}p'^2dp',
\end{eqnarray}
where $x=2\alpha pp'$,

\begin{eqnarray}
&&\phi_0(x)=\frac{1}{x}\sinh(x),
\qquad\phi_1(x)=\frac{1}{x^2}\left[\cosh(x)-\frac{1}{x}\sinh(x)\right],
\nonumber \\
&&\phi_2(x)=\frac{1}{x^3}\left[\left(1
+\frac{3}{x^2}\right)\sinh(x)-\frac{3}{x}\cosh(x)\right].
\end{eqnarray}
The detailed derivation of these formulae can be found in~\cite{Urban}.

Anomalous density and semiclassical gap equation~\cite{Ring}:
\begin{eqnarray}
&&\kappa(\br,\bp)=\frac{1}{2}
\frac{\Delta(\br,\bp)}{\sqrt{h^2(\br,\bp)+\Delta^2(\br,\bp)}},
\\
&&\Delta(\br,\bp)=-\frac{1}{2}\int\!\frac{d^3\!p'}{(2\pi\hbar)^3}
v(|\bp-\bp'|)
\frac{\Delta(\br,\bp')}{\sqrt{h^2(\br,\bp')+\Delta^2(\br,\bp')}},
\end{eqnarray}
where $v(|\bp-\bp'|)=\beta {\rm e}^{-\alpha |\bp-\bp'|^2}\!$
with $\beta=-|V_0|(r_p\sqrt{\pi})^3$ and $\alpha=r_p^2/4\hbar^2$.

\end{document}